\documentclass[11pt,a4paper]{article}
\usepackage{jheppub}
\makeatletter
\def\@fpheader{\mbox{}}
\makeatother

\usepackage{amsthm}
\usepackage{dsfont}
\usepackage{tensor}
\usepackage{booktabs}
\usepackage{enumitem}
\usepackage[dvipsnames]{xcolor}
\usepackage{tikz}
\usetikzlibrary{arrows.meta}
\usepackage[most]{tcolorbox}
\usepackage{orcidlink}

\renewcommand\afterEmailSpace{\vskip50pt plus9pt minus10pt\filbreak}
\renewcommand\afterAbstractSpace{\vskip50pt plus9pt minus13pt}
\hypersetup{
    pdfencoding=unicode,
    colorlinks=true,
    urlcolor=Maroon,
    linkcolor=RoyalBlue,
    citecolor=Maroon,
    pdftitle={Classification of c=1 CFTs},
    pdfauthor={Lorenz Eberhardt, Nat Levine},
    pdfdisplaydoctitle=true,
    pdfstartview=FitH,
    linktocpage=true
}

\newtheorem{theorem}{Theorem}[section]

\newcommand{\be}{\begin{equation}}
\newcommand{\ee}{\end{equation}}

\newcommand{\ZZ}{\mathbb{Z}}
\newcommand{\QQ}{\mathbb{Q}}
\newcommand{\RR}{\mathbb{R}}
\newcommand{\CC}{\mathbb{C}}
\newcommand{\NN}{\mathbb{N}}
\newcommand{\TT}{\mathbb{T}}
\newcommand{\EE}{\mathbb{E}}
\newcommand{\PP}{\mathbb{P}}

\newcommand\SU{\mathrm{SU}}
\newcommand\SO{\mathrm{SO}}
\newcommand\Orth{\mathrm{O}}
\newcommand\Uone{\mathrm{U}(1)}
\newcommand{\Vir}{\mathrm{Vir}}
\let\Re\relax
\DeclareMathOperator{\Re}{Re}
\DeclareMathOperator{\Aut}{Aut}
\DeclareMathOperator{\ord}{ord}
\DeclareMathOperator{\sgn}{sign}
\DeclareMathOperator{\SL}{SL}
\newcommand{\AP}{\operatorname{AP}}
\newcommand{\covol}{\operatorname{covol}}
\newcommand{\Fun}{\operatorname{Fun}}
\newcommand{\Hom}{\operatorname{Hom}}
\newcommand{\Irr}{\operatorname{Irr}}
\newcommand{\rank}{\operatorname{rank}}
\newcommand{\Rep}{\operatorname{Rep}}
\newcommand{\supp}{\operatorname{supp}}

\newcommand\dd{\mathrm{d}}
\newcommand{\e}{\mathrm{e}}
\newcommand{\id}{\mathds{1}}
\newcommand{\ket}[1]{\lvert #1\rangle}
\newcommand{\bra}[1]{\langle #1\rvert}

\newtcbox{\eqbox}[1][]{%
  enhanced, nobeforeafter, tcbox raise base,
  math upper,
  boxrule=0.8pt, arc=2pt,
  left=6pt, right=6pt, top=6pt, bottom=6pt,
  colback=gray!5, colframe=black, #1}
\newtcolorbox{axiombox}{%
  enhanced,
  boxrule=0.8pt, arc=2pt,
  left=6pt, right=6pt, top=6pt, bottom=6pt,
  colback=gray!5, colframe=black}

\title{Classification of $\boldsymbol{c=1}$ CFTs}

\author{Lorenz Eberhardt \orcidlink{0000-0003-1912-2211} and  Nat Levine \orcidlink{0000-0002-1885-4958}}
\affiliation{Institute for Theoretical Physics, University of Amsterdam, Amsterdam, 1098XH, NL}
\emailAdd{l.eberhardt@uva.nl}
\emailAdd{n.j.levine@uva.nl}

\abstract{
We completely classify unitary 2d CFTs with central charge $c=1$ and discrete spectrum. We confirm the folklore that the only such theories are the compact free boson, its $\ZZ_2$-orbifold, and three exceptional theories obtained as orbifolds of the $\SU(2)_1$ WZW model.
The proof is by recasting the $c=1$ modular bootstrap as a \emph{crystalline measure} problem for the density of states, and analytically solving it.
We also show that, once the density of states is specified, the full CFT can be uniquely reconstructed.
}

\begin{document}
\maketitle

\section{Introduction}

Unitary 2d CFTs are some of the best-studied physical theories. Much is known about them in the rational case, i.e.\ with finitely many primary states. For example, unitary CFTs with $c<1$ are completely classified: their modular-invariant partition functions in~\cite{Cappelli:1986hf, Cappelli:1987xt, Gannon:1999cp} and the corresponding theories in~\cite{Kawahigashi:2002px, Kawahigashi:2003gi}. Chiral CFTs at $c=8$ and $c=16$ have also been classified~\cite{Dong:2004aa}. A complementary line of attack organizes rational theories by their number of characters~\cite{Mathur:1988na, Mukhi:2022bte}.

However, there is widespread belief that such theories are quite atypical and we are missing many generic and irrational CFTs. These are very far from being classified. Most of the general results for such CFTs originate from the modular bootstrap, which has been systematized in the language of linear functionals, leading to numerical bounds on spectral gaps for $c>1$ CFTs~\cite{Hellerman:2009bu, Friedan:2013cba, Collier:2016cls, Hartman:2019pcd}. Another class of results are various rigorously established universal behaviours in the spectra of 2d CFTs~\cite{Mukhametzhanov:2019pzy, Pal:2019zzr, Mukhametzhanov:2020swe, Pal:2022vqc}.
\bigskip

In this paper, we shall focus on the marginal case of $c=1$, which represents an intermediate complexity class. While such CFTs are typically not rational, they only have a polynomially growing number of primary states at large conformal weight, compared to the exponential growth in $c>1$ CFTs as dictated by Cardy's formula~\cite{Cardy:1986ie}.
Although there is a priori no reason to expect $c=1$ CFTs to exhibit rational or close-to-rational behavior, a list of $c=1$ CFTs was proposed almost 40 years ago~\cite{Ginsparg:1987eb}, and this list has been conjectured to be a complete classification~\cite{Kiritsis:1988et, Kiritsis:1988es, Dijkgraaf:1987vp}. This statement has been partially proven for the subclass of rational CFTs~\cite{Kiritsis:1988et, Kiritsis:1988es},\footnote{The proof of~\cite{Kiritsis:1988et, Kiritsis:1988es} is not complete, even under the additional hypothesis of rationality~\cite{Gannon:1999cp}. In particular, only uniqueness of the partition function, not of the theories themselves, is claimed.}\textsuperscript{,}\footnote{While this paper was nearing completion, a proof for the rational subclass was also announced in~\cite{Gannon:2026ihc}.} and the general case remains a longstanding open problem.

In this work, we shall prove this conjecture in general, without any assumption of rationality. In other words, we prove the following theorem.
\begin{theorem}\label{thm:c1 classification}
The following are the only unitary $c=1$ CFTs with discrete spectrum:
\begin{enumerate}
    \item Compact free boson theory with radius $R$.
    \item $\ZZ_2$-orbifold of the compact free boson with radius $R$.
    \item Three isolated theories given by orbifolding the $\SU(2)_1$ WZW model by one of the three exceptional discrete subgroups of $\SO(3)$: the tetrahedral, octahedral or icosahedral group.
\end{enumerate}
\end{theorem}
\noindent Note that T-duality identifies the theories of radius $R$ and $R^{-1}$ for the free boson and its $\ZZ_2$-orbifold.\footnote{In our conventions, the self-dual radius is $R=1$. In much of the literature the radius is rescaled by a factor of $\sqrt{2}$, e.g.\ $R=\sqrt{2}\, R_\text{G}$ in terms of the radius $R_\text{G}$ of~\cite{Ginsparg:1988ui}, whose self-dual value is $R_\text{G}=\frac{1}{\sqrt{2}}$; and $R=\frac{1}{\sqrt{2}}\, R_\text{FMS}$ in terms of the radius $R_\text{FMS}$ of~\cite{DiFrancesco:1997nk}, whose self-dual value is $R_\text{FMS}=\sqrt{2}$.} Moreover, the $\ZZ_2$-orbifold theory at $R=1$ is equivalent to the free boson at $R=2$. The corresponding densities of states $\rho_R$, $\rho_R^\mathrm{orb}$, $\rho^\mathrm{tet}$, $\rho^\mathrm{oct}$ and $\rho^\mathrm{ico}$ are listed in equations~\eqref{eq:rho free bos}, \eqref{eq:orbifold solution} and \eqref{eq:exceptional solutions}.

Let us comment on the level of rigor with which we prove this theorem. The largest part of our proof is classifying the possible torus partition functions using the modular bootstrap, which is completely rigorous. It then remains to show that, given a partition function, the corresponding theory is uniquely determined. For this we need the full axioms of 2d CFT, including the operator algebra. We formulate our proof in physics language, but everything can in principle also be formulated in mathematical language and turned into completely rigorous statements. We have also chosen to write our proof as a continuous text, without employing lemmas etc., in order to make it more readable for physicists.

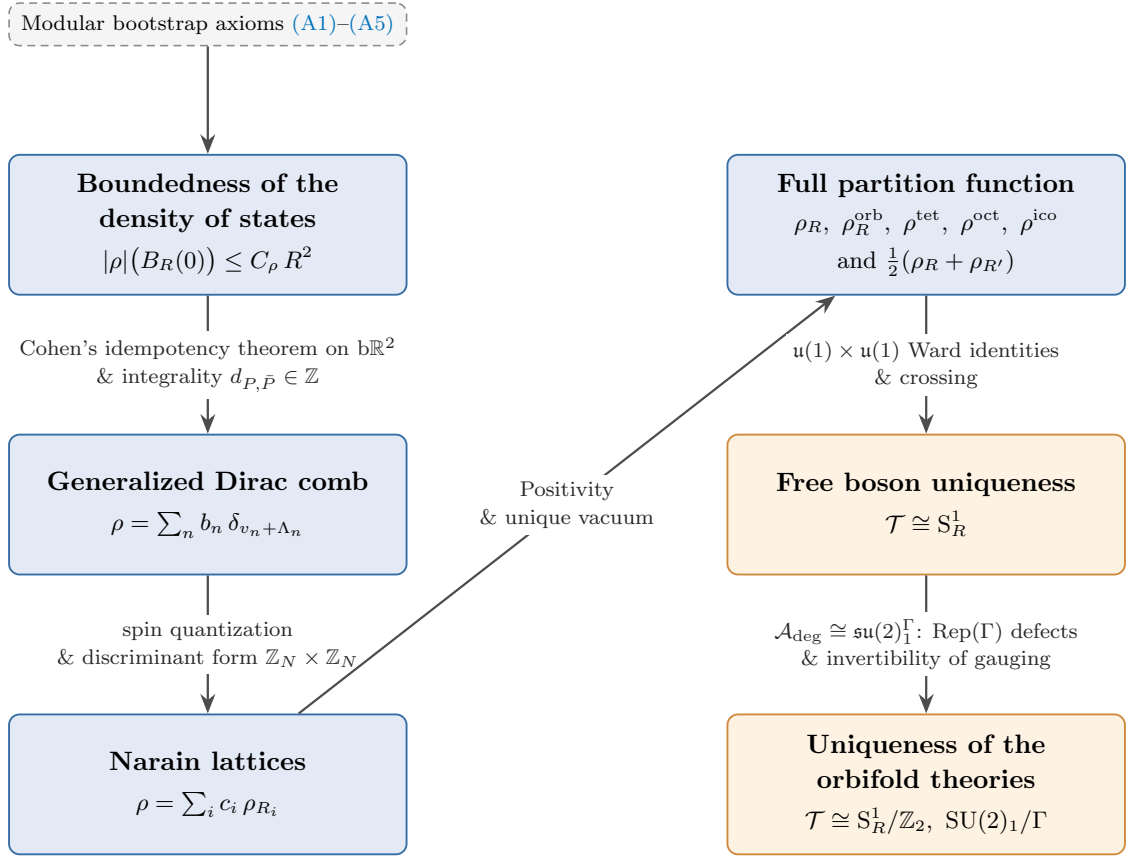
\begin{figure}[htbp]
\centering
\begin{tikzpicture}[
  x=1cm, y=1cm, font=\small,
  box/.style ={draw=#1!80!black, fill=#1!10, rounded corners=4pt, line width=0.7pt,
               text width=5.0cm, minimum height=1.85cm, align=flush center, inner ysep=5pt},
  box/.default=NavyBlue,
  ax/.style  ={draw=black!45, densely dashed, fill=black!4, rounded corners=4pt,
               line width=0.6pt, text width=5.0cm, align=flush center, inner ysep=4pt},
  arr/.style ={-{Stealth[length=2.6mm,width=2.0mm]}, line width=0.85pt, draw=black!70},
  lab/.style ={font=\scriptsize, align=center, inner sep=2.5pt, fill=white, text=black!85}
]

\node[ax]              (ax) at (0, 2.65)  {\scriptsize Modular bootstrap axioms
                                           \ref{axiom:S invariance}--\ref{axiom:vacuum}};

\node[box]             (b1) at (0, 0)     {\textbf{Boundedness of the\\ density of states}\\[2pt]
                                           {\footnotesize $|\rho|\big(B_R(0)\big)\le C_\rho\, R^{2}$}};
\node[box]             (b2) at (0,-3.7)   {\textbf{Generalized Dirac comb}\\[2pt]
                                           {\footnotesize $\rho=\sum_n b_n\,\delta_{v_n+\Lambda_n}$}};
\node[box]             (b3) at (0,-7.4)   {\textbf{Narain lattices}\\[2pt]
                                           {\footnotesize $\rho=\sum_i c_i\,\rho_{R_i}$}};
\node[box]             (b4) at (9.5,0) {\textbf{Full partition function}\\[2pt]
                                           {\footnotesize $\rho_R,\ \rho^{\mathrm{orb}}_R,\ \rho^{\mathrm{tet}},\ \rho^{\mathrm{oct}},\ \rho^{\mathrm{ico}}$\\[1pt]
                                            and $\tfrac{1}{2}(\rho_R+\rho_{R'})$}};
\node[box=BurntOrange] (b5) at (9.5,-3.7) {\textbf{Free boson uniqueness}\\[2pt]
                                           {\footnotesize $\mathcal{T}\cong \mathrm{S}^1_R$}};
\node[box=BurntOrange] (b6) at (9.5, -7.4)   {\textbf{Uniqueness of the\\ orbifold theories}\\[2pt]
                                           {\footnotesize $\mathcal{T}\cong \mathrm{S}^1_R/\ZZ_2,\ \SU(2)_1/\Gamma$}};

\draw[arr] (ax) -- (b1);
\draw[arr] (b1) -- node[lab] {Cohen's idempotency theorem on $\mathrm{b}\RR^2$\\ \& integrality $d_{P,\bar P}\in\ZZ$} (b2);
\draw[arr] (b2) -- node[lab] {spin quantization\\ \& discriminant form $\ZZ_N\times\ZZ_N$} (b3);
\draw[arr] (b3) -- node[lab] {Positivity\\ \& unique vacuum} (b4);
\draw[arr] (b4) -- node[lab] {$\mathfrak{u}(1)\times\mathfrak{u}(1)$ Ward identities\\ \& crossing} (b5);
\draw[arr] (b5) -- node[lab] {$\mathcal{A}_{\mathrm{deg}}\cong\mathfrak{su}(2)_1^{\Gamma}$: $\Rep(\Gamma)$ defects\\ \& invertibility of gauging} (b6);
\end{tikzpicture}
\caption{The logic of the proof of Theorem~\ref{thm:c1 classification}. The blue boxes are established by the modular bootstrap alone. The orange boxes use the full set of CFT axioms. Each arrow is labeled by the crucial input entering the corresponding step. The spurious bootstrap solution $\frac{1}{2}(\rho_R+\rho_{R'})$ is excluded in Section~\ref{subsec:spurious solution}.}
\label{fig:proof-logic}
\end{figure}

The logic of our proof of Theorem~\ref{thm:c1 classification} is displayed in Figure~\ref{fig:proof-logic}.
At the heart of the proof is an analytic solution of the modular bootstrap. For this, we decompose the torus partition function into Virasoro characters $\chi_P^\Vir(\tau)$,
\be\label{eq:Z}
  Z(\tau,\bar \tau)= \sum_{P \ge 0,\, \bar P\ge 0} D_{P,\bar {P}}\ \chi_P^\Vir(\tau)\chi_{\bar P}^\Vir(-\bar \tau)  \ ,
\ee
where the non-negative integers $D_{P,\bar P}$ are the multiplicities of the  Virasoro primaries of (anti)holomorphic weights $(h,\bar h) = (P^2, \bar P^2)$.  Since Virasoro representations can have null vectors at $c=1$, the corresponding Virasoro characters with certain values of $(P,\bar P)$ are non-trivial (and non-positive) combinations of generic Fock characters $\chi_P(\tau)=q^{P^2}/\eta(\tau)$, where as usual $q=\e^{2\pi i \tau}$. For the modular bootstrap, it is also useful to write the alternative decomposition into Fock characters, $Z(\tau,\bar \tau) = \sum_{P \ge 0,\, \bar P\ge 0} d_{P,\bar {P}}\, \chi_P(\tau)\chi_{\bar P}(-\bar \tau)$, with degeneracies $d_{P,\bar P}$ that are not necessarily positive. Their precise relation to the Virasoro degeneracies $D_{P,\bar P}$ is discussed below, see \eqref{eq:Ddec}.
The decomposition \eqref{eq:Z} of the partition function can equivalently be written as the integral of the character against the following density of states $\rho(P,\bar P)$,
\begin{align}
    Z(\tau,\bar \tau)&=  \int \dd\rho(P,\bar P) \  \chi_P(\tau)\chi_{\bar P}(-\bar \tau) \ , \label{eq:intDensity} \\
    \rho &= \frac{1}{4}\sum_{P,\bar P \ge 0}d_{P,\bar P}\ \big(\delta_{(P,\bar P)}+\delta_{(-P,-\bar P)}+\delta_{(-P,\bar P)}+\delta_{(P,-\bar P)}\big) \ , \label{eq:definition density of states}
\end{align}
where we choose the density to be even in both $P$ and $\bar P$. The measure $\rho$ is atomic (a discrete sum of delta-functions) because the spectrum is assumed to be discrete. The advantage of passing to $\rho$ is that the constraint of modular S-invariance translates to $\rho$ being self-dual under the Fourier transform,
\be
\mathbb S  \rho = \rho \ ,
\ee
where $\mathbb S$ is an appropriately normalized \textit{Lorentzian Fourier transform} in the $(P,\bar P)$ variables, see \eqref{eq:FourierDef}. In particular, this means that both $\rho$ and its Fourier dual have discrete support. Such a measure $\rho$ (with other technical assumptions) is called a \textit{crystalline measure} in the mathematical literature.

As we shall discuss, crystalline measures are highly constrained thanks to powerful theorems from harmonic analysis.
With appropriate additional assumptions, they are forced to have a regular lattice-like structure,
\be
\rho = \sum_i c_i \, \delta_{\lambda_i+\Lambda_i} \ ,
\ee
where $\Lambda_i$ are lattices and $\lambda_i$ are offsets.
One such additional assumption that forces this type of behavior is control of the growth of $\mathbb{S}\rho$ near infinity. Intuitively, this is a very constraining property because the Fourier transform of a single atom is an oscillating phase: such phases now need to sum in a way that does not produce any constructive interference in any direction near infinity---which is very fine-tuned. Control of the growth of $\mathbb{S}\rho$ is not obvious in our case because of non-positivity of $\rho$ due to null states at $c=1$, so the usual Cardy argument does not straightforwardly apply.

A large portion of the technical work below goes into proving in Section~\ref{sec:growth} that the modular bootstrap axioms (modular invariance; positive integer degeneracy) force $\rho=\mathbb{S}\rho$ to have upper bounded density near infinity, which can be seen as a $c=1$ Cardy theorem.

With this property in hand, in Section~\ref{sec:modular bootstrap crystalline measures} we prove a new crystalline measure classification theorem for the $c=1$ modular bootstrap. The main machine behind this is Cohen's idempotency theorem, characterizing measures on compact abelian groups whose Fourier transform only takes integer values. In order to circumvent the unpleasant property that $\RR^2$ is not compact, we employ a standard trick to uplift the measure to the so-called Bohr compactification of $\RR^2$, where the theorem can be applied.
Without assuming rationality, this brings us to a point where Kiritsis' argument~\cite{Kiritsis:1988et, Kiritsis:1988es} can be carried out, proving the modular bootstrap classification. One also finds one spurious solution $\frac{1}{2}(\rho_R+\rho_{R'})$ to the bootstrap.

We finally prove in Section~\ref{sec:uniqueness} the uniqueness of the associated theories, as well as ruling out the spurious solution. For the free boson, this is simple, because the spectrum tells us that the theory has a conserved current, whose Ward identities determine the theory to be the free boson. For the other theories, we show that they have a non-anomalous symmetry. Gauging it leads back to the free boson theory, whose uniqueness was already established. By inverting the gauging, the uniqueness of the other theories also follows. For the three exceptional theories, the corresponding symmetry is a non-invertible symmetry.

\section{Asymptotic growth of states} \label{sec:growth}
We are studying 2d CFTs with central charge $c\equiv c_\mathrm{L}=c_\mathrm{R}=1$ and discrete spectrum. In this section and the next, we will solve the modular bootstrap for such theories. Let us first recall the setup.

A consistent CFT can be placed on the torus~\cite{Segal:2002ei, Friedan:1986ua, Moore:1988qv}, and its torus partition function $Z(\tau,\bar\tau)$ must be modular-invariant,
\be
Z(\tau,\bar \tau)=Z(\tau+1,\bar \tau+1)=Z(-\tfrac{1}{\tau},-\tfrac{1}{\bar \tau})\ .
\ee
It is convenient to use the `momentum' parametrization for the conformal weights,\footnote{For general values of $c$, the standard parametrization is $(h,\bar h)=(\tfrac{c-1}{24}+P^2,\tfrac{c-1}{24}+\bar P^2)$.}
\be
    h = P^2 \ , \qquad \bar h  = {\bar P}^2 \ ,
\ee
where we have specialized to $c=1$. These variables naturally identify $P \leftrightarrow-P$ and $\bar P \leftrightarrow-\bar P$ so, without loss of generality, we may take the Virasoro primaries with non-zero $D_{P,\bar P}$ appearing in \eqref{eq:Z} to be located at $P,\bar P\geq 0$.

For each primary, the corresponding character counts the contributions of all of its Virasoro descendents. However, at  $c=1$, the Virasoro algebra has degenerate representations at each half-integer $P\in \tfrac 12\NN_0$, in which a descendent at level $2P+1$ is null and hence absent. The physical \textit{Virasoro character} is therefore modified at those special values,
\be
    \chi_P^\Vir(\tau) = \begin{cases}
        \chi_P(\tau) - \chi_{P+1}(\tau) \ , &P\in \tfrac12\NN_0\ , \\
        \chi_P(\tau) \ , &P\notin \tfrac12\NN_0  \ .
    \end{cases}
\ee
One can then re-write the decomposition \eqref{eq:Z} in terms of Fock characters,
\be
     Z(\tau,\bar \tau)  = \sum_{P,\bar P\geq 0} D_{P,\bar {P}}\ \chi^\Vir_P(\tau)\chi^\Vir_{\bar P}(-\bar \tau) =\sum_{P,\bar P\geq 0} d_{P,\bar {P}}\ \chi_P(\tau)\chi_{\bar P}(-\bar \tau)  \ ,
\ee
where the Fock degeneracies $d_{P,\bar P}$ are related to the Virasoro degeneracies by
\be
    D_{P,\bar P} = \sum_{P'\in C(P),\bar P'\in C(\bar P)} \label{eq:Ddec}
    d_{P',\bar P'} \ ,
   \ee
   with
   \be
   C(P)=\begin{cases}
        \{P,P-1,\ldots,P- \lfloor P\rfloor\} \ , &P\in \tfrac12\NN_0 \ , \\
        \{P\} \ , &P\notin \tfrac12\NN_0   \ .
    \end{cases}
\ee
Here, $\lfloor x \rfloor$ denotes the floor function. The Virasoro degeneracies $D_{P,\bar P}$ are positive, since they actually count physical states, while the Fock degeneracies $d_{P,\bar P}$ may be negative.

The modular bootstrap consists of imposing that the torus partition function is modular-invariant, and that it counts positive integer numbers of Virasoro primaries, with a unique vacuum.
This can be stated as the following set of modular bootstrap axioms:
\newlength{\axiomlab}
\newlength{\axiomnum}
\settowidth{\axiomlab}{\textbf{Positive Virasoro degeneracies.}\qquad}
\settowidth{\axiomnum}{(A5)}
\begin{axiombox}
\begin{enumerate}[label=(A\arabic*),ref=(A\arabic*),align=left,
  labelindent=0pt,labelwidth=\axiomnum,labelsep=0.5em,
  leftmargin=!,topsep=0pt]
\item \label{axiom:S invariance} \makebox[\axiomlab][l]{\textbf{Modular $\boldsymbol{S}$ invariance.}}$Z(-1/\tau,-1/\bar\tau)=Z(\tau,\bar\tau)$
\item \makebox[\axiomlab][l]{\textbf{Modular $\boldsymbol{T}$ invariance.}}$Z(\tau+1,\bar \tau+1) =  Z(\tau,\bar \tau)$
\item \label{axiom:integrality} \makebox[\axiomlab][l]{\textbf{Integer degeneracies.}}$d_{P,\bar P} \in \mathbb{Z}$
\item \label{axiom:positivity} \makebox[\axiomlab][l]{\textbf{Positive Virasoro degeneracies.}}$D_{P,\bar P} \geq 0 $
\item \label{axiom:vacuum} \makebox[\axiomlab][l]{\textbf{Unique vacuum.}}$d_{0,0}=1$
\end{enumerate}
\end{axiombox}
We of course also assume that the sum defining the partition function converges absolutely, so that the axioms even make sense. Our assumption of a discrete spectrum means that $d_{P,\bar P}$ has discrete support.

We are free to define the density of states as in \eqref{eq:definition density of states}, i.e.\ symmetrized under $P \to -P$ and $\bar P\to-\bar P$, since the conformal weights (and hence the characters) are invariant under this. The partition function is then the integral of the density against Fock characters \eqref{eq:intDensity}

Let us make a few comments on the content of these axioms.

\paragraph{$\boldsymbol{S}$ invariance is Fourier self-duality of $\boldsymbol{\rho}$.} Modular $S$ transformation acts on Fock characters as a cosine transform or, by evenness in $P$, a Fourier transform
\be
\chi_P(-1/\tau)=2\sqrt{2}\int_{0}^{\infty}\dd P'\,\cos(4\pi P P')\,\chi_{P'}({\tau})
=\sqrt{2}\int_{-\infty}^{\infty}\dd P'\,\e^{4\pi i P P'}\,\chi_{P'}({\tau})\ .
\ee
The combination of holomorphic and anti-holomorphic characters therefore transforms to its Lorentzian Fourier dual,
\be
\chi_P(-1/\tau)\chi_{\bar P}(1/{\bar \tau}) = \mathbb{S} \,  \chi_P(\tau) \chi_{\bar P}(-{\bar \tau}) \label{eq:chars-transf-S}
\ee
defined by
\be
\mathbb{S}\, f(P,\bar P) = 2 \int_{\RR^2} \dd P' \dd\bar P' \,  \e^{2\pi i \langle (P,\bar P),(P',\bar P') \rangle_\mathrm{L}} \, f(P',\bar P')\ . \label{eq:FourierDef}
\ee
with a Lorentzian inner product
\be
\langle (P,\bar P),(P',\bar P') \rangle_\mathrm{L}=2(PP'-\bar P \bar P')\ .  \label{eq:LIP}
\ee
Substituting \eqref{eq:chars-transf-S} into the statement of $S$ invariance \ref{axiom:S invariance}, one finds that $\rho$ equals its own dual,\footnote{To deduce this, one can use the fact that the characters $\chi_{P}(\tau)\chi_{\bar P}(-\bar \tau)$ are complete in the space of Schwartz functions even in both $P$ and $\bar P$, since they include Gaussians of all widths centered at the origin.}
\be
\rho = \mathbb S \rho \ .
\ee
The Fourier transform $\mathbb S \rho$ can be defined here in the sense of tempered distributions, i.e.\ by its integral  $\int (\mathbb S\rho) \, f = \int \rho \, (\mathbb S f) $ against Schwartz functions $f$. Indeed, as we will explain below, it is easy to show from axioms \ref{axiom:S invariance} and \ref{axiom:positivity} that $|\rho|$ is a tempered measure,\footnote{The notation $|\rho|$ means that we take $|d_{P,\bar P}|$ instead of $d_{P,\bar P}$ in \eqref{eq:definition density of states}, i.e.\ all atoms are counted positively.} i.e.\ that $|\rho|(B_R(0)) = \mathcal O(R^\alpha)$ for some $\alpha>0$. With more work---and using all of the axioms except uniqueness of the vacuum---we will show that this is true with $\alpha=2$.

\paragraph{$\boldsymbol{T}$ invariance is spin quantization.}
Expanding the statement of modular $T$ invariance into Fock characters, one finds that
\be
\rho(P,\bar P)=\e^{\pi i \lVert (P,\bar P)\rVert^2_\mathrm{L}}\,  \rho(P,\bar P)\ .
\ee
This forces that $\rho(P,\bar P)$ is supported on states with
\be
 h-\bar h = P^2-\bar P^2 = \tfrac 12 \lVert (P,\bar P)\rVert^2_\mathrm{L} \in \mathbb{Z}\ ,
\ee
i.e.\ with integer spin.

\paragraph{Standard degeneracy assumptions.} Axioms \ref{axiom:integrality} and \ref{axiom:positivity} are the basic consistency requirement that counting the Virasoro primaries in each representation returns a non-negative integer. Axiom \ref{axiom:vacuum} is the standard assumption that there is a unique vacuum state with $(h,\bar h)=(0,0)$.

\paragraph{Growth.} An important input that does not readily follow from the bootstrap axioms \ref{axiom:S invariance}--\ref{axiom:vacuum} is precise control over the behavior of $\rho(P,\bar P)$ at large $P$ and $\bar P$---which is usually achieved by Cardy's formula~\cite{Cardy:1986ie}. For our purposes in Section~\ref{sec:modular bootstrap crystalline measures}, it will be enough to show that $\rho$ is of \textit{upper bounded density}, meaning that
\be\label{eq:bounded density}
  \eqbox{|\rho|(B_R(0)) \le C_\rho \, R^2 \qquad \text{for } R \ge 1}
\ee
for some constant $C_\rho$. This is essentially Cardy's formula in the case of $c=1$. However, there is a major catch in proving this. The problem is that, because of the null states at $c=1$, the multiplicities $d_{P,\bar P}$ that enter the measure $\rho$ in \eqref{eq:definition density of states} are not necessarily positive: only the Virasoro degeneracies $D_{P,\bar P}$ are positive. This makes \eqref{eq:bounded density} quite non-trivial to establish, and the current section is devoted to it. In fact, we view it as the most difficult technical step of this paper. As a byproduct, we will also show the much simpler result that $\rho$ is a tempered measure, meaning that $|\rho|(B_R(0))$ has at most polynomial growth. The reader that is willing to take the bound \eqref{eq:bounded density} for granted may proceed to Section~\ref{sec:modular bootstrap crystalline measures}, where this growth enters as an input for the classification problem.

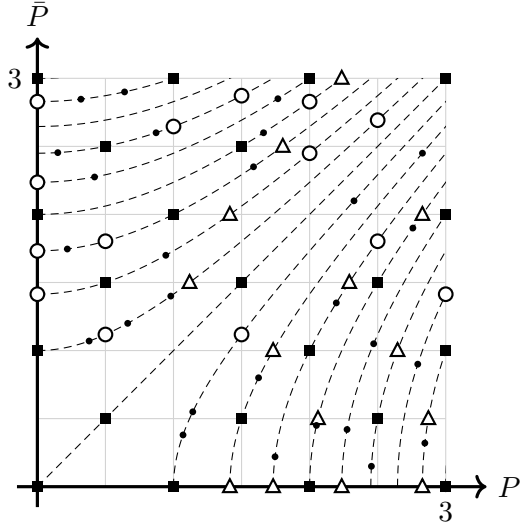
\begin{figure}[t]
\centering
\begin{tikzpicture}[scale=0.9]
  \foreach \x in {1,2,3,4,5,6} \draw[gray!35] (\x,0) -- (\x,6.0);
  \foreach \y in {1,2,3,4,5,6} \draw[gray!35] (0,\y) -- (6.0,\y);
  \begin{scope}
    \clip (0,0) rectangle (6.0,6.0);
    \draw[densely dashed] (0,0) -- (6.1,6.1);
    \foreach \n in {1,...,9} {
      \draw[densely dashed] plot[domain=0:2.0,samples=80,variable=\t]
        ({2*sqrt(\n)*0.5*(exp(\t)+exp(-\t))},{2*sqrt(\n)*0.5*(exp(\t)-exp(-\t))});
      \draw[densely dashed] plot[domain=0:2.0,samples=80,variable=\t]
        ({2*sqrt(\n)*0.5*(exp(\t)-exp(-\t))},{2*sqrt(\n)*0.5*(exp(\t)+exp(-\t))});
    }
  \end{scope}
  \draw[very thick, ->] (-0.3,0) -- (6.6,0) node[right] {$P$};
  \draw[very thick, ->] (0,-0.3) -- (0,6.6) node[above] {$\bar P$};
  \draw (6,-0.08) -- (6,0.08) node[below=4pt] {$3$};
  \draw (-0.08,6) -- (0.08,6) node[left=4pt] {$3$};
  \foreach \p in {(1.000,1.000),(3.000,3.000),(6.000,6.000),(1.000,3.000),(3.000,1.000),(2.000,4.000),(4.000,2.000),(3.000,5.000),(5.000,3.000),(1.000,5.000),(5.000,1.000),(2.000,6.000),(6.000,2.000),(4.000,6.000),(6.000,4.000),(0.000,0.000),(2.000,0.000),(4.000,0.000),(6.000,0.000),(0.000,2.000),(0.000,4.000),(0.000,6.000)}
    \fill \p ++(-0.075,-0.075) rectangle ++(0.15,0.15);
  \foreach \p in {(1.000,2.236),(1.000,3.606),(2.000,5.292),(3.000,2.236),(3.000,5.745),(4.000,4.899),(4.000,5.657),(5.000,3.606),(5.000,5.385),(6.000,2.828),(0.000,2.828),(0.000,3.464),(0.000,4.472),(0.000,5.657)}
    {\fill[white] \p circle (0.1); \draw[thick] \p circle (0.1);}
  \foreach \p in {(4.123,1.000),(5.745,1.000),(3.464,2.000),(5.292,2.000),(2.236,3.000),(4.583,3.000),(2.828,4.000),(5.657,4.000),(3.606,5.000),(4.472,6.000),(2.828,0.000),(3.464,0.000),(4.472,0.000),(5.657,0.000)}
    {\fill[white] \p ++(0,0.11) -- ++(-0.1,-0.19) -- ++(0.2,0) -- cycle;
     \draw[thick] \p ++(0,0.11) -- ++(-0.1,-0.19) -- ++(0.2,0) -- cycle;}
  \foreach \p in {(2.283,1.100),(3.250,1.600),(1.327,2.400),(4.392,2.700),(4.100,0.900),(1.887,3.400),(4.652,4.200),(3.176,4.700),(4.941,2.100),(3.323,5.200),(5.517,3.800),(3.534,5.700),(5.589,1.800),(5.658,4.900),(1.281,5.800),(1.744,5.200),(1.960,2.800),(0.640,5.693),(0.840,4.550),(0.440,3.492),(0.760,2.140),(0.300,4.908),(5.693,0.640),(4.550,0.840),(3.492,0.440),(2.140,0.760),(4.908,0.300)}
    \fill \p circle (0.05);
\end{tikzpicture}
\caption{A sketch of the type of spectrum that may appear in the modular bootstrap, drawn for $P,\, \bar P \le 3$. States can only appear on the dashed hyperbolas by spin quantization. The four different types of states are indicated as follows: squares are dd sites, circles dg sites, triangles gd sites and dots are gg sites.}
\label{fig:families}
\end{figure}

\paragraph{Notation.}
Our task is to bound
\be
V(R):=\sum_{0 \le P,\bar{P} \le R} |d_{P,\bar{P}}|\ ,
\ee
since the measure of a ball or a box is asymptotically the same, up to a constant.
We divide states into the generic (g) case $P \not\in \frac{1}{2} \NN_0$ and the degenerate (d) one $P \in \frac{1}{2}\NN_{0}$. We call a state gg if both $P$ and $\bar P$ are generic, dg if only $P$ is degenerate, gd if only $\bar P$ is degenerate and dd if both are degenerate. A sketch of the structure of the spectrum is given in Figure~\ref{fig:families}.
We write
\be
V(R)=V_{\mathrm{gg}}(R)+V_{\mathrm{gd}}(R)+V_{\mathrm{dg}}(R)+V_{\mathrm{dd}}(R)
\ee
for these four contributions. Notice that, for gg states, positivity means that $d_{P,\bar P} \ge 0$ and therefore bounding $V_{\mathrm{gg}}(R)$ will be simple; it is of order $R^2$. The dg and gd contributions will turn out to be the hardest. We will also need sometimes the following rectangle version,
\be
V_\mathrm{gg}(X,Y):=\sum_{0 \le P\le X,\, 0 \le \bar P \le Y,\text{ gg}} |d_{P,\bar{P}}|\ ,
\ee
so that $V_\mathrm{gg}(R)\equiv V_\mathrm{gg}(R,R)$, and similarly for the other families. We will always assume that $R,\, X,\, Y \ge 1$ in the following.

In this section, we use the Vinogradov notation frequently employed in analytic number theory. $\ll$ denotes $\le$ up to some constant depending only on $\rho$ itself and $\ll_\gamma$ (or $\ll_\beta$) denotes $\le$ up to some constant depending on $\gamma$ (or $\beta$). We also write $X \asymp Y$ when $X \ll Y \ll  X$.
In this notation, our task will be to show that $V(R) \ll R^2$.

\subsection{Theta function and the gg contribution}

\paragraph{Theta function.} We begin by introducing the Theta function. For $\Re \beta,\, \Re \bar \beta > 0$,
\be
\Theta(\beta,\bar \beta)=\sum_{P,\bar P \ge 0} d_{P,\bar P}\,  \e^{- (\beta P^2+\bar \beta\bar P^2)}= \eta\Big(\frac{i \beta}{2\pi}\Big)\,  \eta\Big(\frac{i \bar \beta}{2\pi}\Big) Z\Big(\tau=\frac{i \beta}{2\pi},\bar \tau=-\frac{i \bar \beta}{2\pi}\Big)\ . \label{eq:theta function definition}
\ee
This is a Lorentzian version of the partition function, with $\beta$ and $\bar \beta$ being the temperatures for the left- and the right-movers and descendants of the Fock characters removed.
As remarked below the axioms \ref{axiom:S invariance}--\ref{axiom:vacuum}, the sum defining the Euclidean partition function converges absolutely. Since
\be
\sum_{P,\bar P \ge 0} |d_{P,\bar P}|\,  \e^{- (\beta P^2+\bar \beta\bar P^2)} \le \sum_{P,\bar P \ge 0} |d_{P,\bar P}|\,  \e^{- \min(\beta,\bar \beta) (P^2+\bar P^2)}\ ,
\ee
the same follows also for $\Theta(\beta,\bar\beta)$.
$\Theta(\beta,\bar \beta)$ is an analytic function in both arguments. Thus modular invariance of the partition function can be extended away from the Euclidean regime to conclude that $\Theta$ satisfies
\be
\Theta(\beta,\bar \beta)=\frac{2\pi}{\sqrt{\beta \bar \beta}} \, \Theta\Big(\frac{4\pi^2}{\beta},\frac{4\pi^2}{\bar\beta}\Big)\ . \label{eq:Theta-function-modular-invariance}
\ee
We will in the following only make use of the Theta function for real $\beta,\, \bar \beta$.

\paragraph{Positivity.} The first step in our proof is to notice that $\Theta(\beta,\bar \beta)$ is a positive function, even though not all of the coefficients $d_{P,\bar P}$ are necessarily positive. Notice that
\begin{align}
    d_{P,\bar P}=\begin{cases}
        D_{P,\bar P}\ , &\mathrm{gg}\ , \\
        D_{P,\bar P}-D_{P-1,\bar P}\ , &\mathrm{dg}\ , \\
        D_{P,\bar P}-D_{P,\bar P-1}\ , &\mathrm{gd}\ , \\
        D_{P,\bar P}-D_{P-1,\bar P}-D_{P,\bar P-1}+D_{P-1,\bar P-1}\ , &\mathrm{dd}\ ,
    \end{cases} \label{eq:d-vs-D relation}
\end{align}
where $D_{P,\bar P}$ is the Virasoro multiplicity, which is by definition positive. We also have by convention $D_{P,\bar P}=0$ if either $P<0$ or $\bar P<0$, so that the corresponding terms in \eqref{eq:d-vs-D relation} are absent for small $P$ or $\bar P$. We can insert \eqref{eq:d-vs-D relation} into \eqref{eq:theta function definition} and reorganize the sum as follows,
\begin{align}
    \Theta(\beta,\bar \beta)&=\sum_{P,\bar P\, \mathrm{gg}} D_{P,\bar P}\, \e^{-(\beta P^2+\bar \beta \bar P^2)}+\sum_{P,\bar P\, \mathrm{dg}} (D_{P,\bar P}-D_{P-1,\bar P})\, \e^{-(\beta P^2+\bar \beta \bar P^2)} \nonumber \\
    &\qquad+\sum_{P,\bar P\, \mathrm{gd}} (D_{P,\bar P}-D_{P,\bar P-1})\, \e^{-(\beta P^2+\bar \beta \bar P^2)}\nonumber\\
    &\qquad+\sum_{P,\bar P\, \mathrm{dd}} (D_{P,\bar P}-D_{P-1,\bar P}-D_{P,\bar P-1}+D_{P-1,\bar P-1})\, \e^{-(\beta P^2+\bar \beta \bar P^2)} \nonumber\\
    &=\sum_{P,\bar P\, \mathrm{gg}} D_{P,\bar P}\, \e^{-(\beta P^2+\bar \beta \bar P^2)}+\sum_{P,\bar P\, \mathrm{dg}} D_{P,\bar P}\, \big(\e^{-\beta P^2}-\e^{-\beta (P+1)^2}\big)\e^{-\bar \beta \bar P^2}\nonumber\\
    &\qquad+\sum_{P,\bar P\, \mathrm{gd}} D_{P,\bar P}\, \e^{-\beta  P^2}\big(\e^{-\bar\beta \bar P^2}-\e^{-\bar\beta (\bar P+1)^2}\big)\nonumber\\
    &\qquad+\sum_{P,\bar P\, \mathrm{dd}} D_{P,\bar P}\, \big(\e^{-\beta P^2}-\e^{-\beta (P+1)^2}\big)\big(\e^{-\bar\beta \bar P^2}-\e^{-\bar\beta (\bar P+1)^2}\big)\ .
\end{align}
Here, we shifted the summation indices $P \to P+1$ or $\bar P \to \bar P+1$ in some of the terms. This is legal, since also the sum $\sum_{P,\bar P} D_{P,\bar P} \, \e^{-\beta P^2-\bar \beta \bar P^2}$ is absolutely convergent.
When written in this way, it becomes in particular obvious that $\Theta(\beta,\bar \beta)>0$ is strictly positive, since all terms are positive.
By modular invariance, we also have for $0< \beta,\,\bar \beta\le 1$,
\begin{align}
\Theta(\beta,\bar \beta) &=\frac{2\pi}{\sqrt{\beta \bar \beta}}\, \Theta\Big(\frac{4\pi^2}{\beta},\frac{4\pi^2}{\bar\beta}\Big) \nonumber\\
&\le \frac{2\pi}{\sqrt{\beta\bar \beta}}\, \sum_{P,\bar P} |d_{P,\bar P}|\, \e^{-4\pi^2(\beta^{-1} P^2+\bar \beta^{-1} \bar P^2)} \nonumber\\
&\le \frac{2\pi}{\sqrt{\beta \bar \beta}} \sum_{P,\bar P} |d_{P,\bar P}|\, \e^{-4\pi^2(P^2+\bar P^2)} \nonumber\\
&\ll \frac{1}{\sqrt{\beta \bar \beta}}\ .\label{eq:Theta small argument bound}
\end{align}

\paragraph{gg contribution.} We begin by bounding the generic contribution $V_\mathrm{gg}(R)$, for which we have thanks to $d_{P,\bar P} \ge 0$,
\begin{align}
V_\mathrm{gg}(X,Y)&=\sum_{0 \le P \le X,\, 0 \le \bar P \le Y\, \mathrm{gg}} d_{P,\bar P} \nonumber\\
&\ll \sum_{0 \le P \le X,\, 0 \le \bar P \le Y\, \mathrm{gg}} d_{P,\bar P}\, \e^{-X^{-2}P^2-Y^{-2}\bar P^2} \nonumber\\
&\le \Theta(X^{-2},Y^{-2}) \nonumber\\
&\ll  X Y\ , \label{eq:Vgg bound}
\end{align}
where we used \eqref{eq:Theta small argument bound} in the last step. Therefore, the gg contribution grows according to the rectangle size.
\subsection{Naive bounds and the dd contribution}
We will now work towards bounding the other contributions.
\paragraph{Naive dg contribution.}
For this we will first establish the simple bound
\be
    V_\mathrm{dg}(X,Y) \ll X^2 Y\ , \qquad V_\mathrm{gd}(X,Y) \ll X Y^2\ . \label{eq:Vdg Vgd naive bound}
\ee
This bound can be established with an argument similar to the computation in \eqref{eq:Vgg bound}. We focus on the first bound, since the proof for the second one is obtained by exchanging $P$ and $\bar P$. Let us begin by bounding the sum $\sum_{0 \le P \le X,\, 0 \le \bar P \le Y \ \mathrm{dg}} D_{P,\bar P}$.
The sum over $P$ can be further divided into the intervals $E_0=[0,1)$ and $E_i=[2^{i-1},2^i)$ with $1 \le i \le \lceil \log_2(X) \rceil$. We then have
\begin{align}
    \sum_{0 \le P \le X,\, 0 \le \bar P \le Y \ \mathrm{dg}} D_{P,\bar P} &= \sum_{i=0}^{\lceil \log_2(X) \rceil} \sum_{P \in E_i,\, 0 \le \bar P \le Y\, \mathrm{dg}} D_{P,\bar P} \nonumber\\
    &\ll \sum_{i=0}^{\lceil \log_2(X) \rceil} 2^{i-1} \sum_{P \in E_i,\, 0 \le \bar P \le Y\, \mathrm{dg}} D_{P,\bar P} \nonumber\\
    &\qquad\times \big(\e^{-2^{-2i} P^2}-\e^{-2^{-2i} (P+1)^2}\big) \e^{-Y^{-2} \bar P^2} \nonumber\\
    &\ll \sum_{i=0}^{\lceil \log_2(X) \rceil} 2^i \, \Theta(2^{-2i},Y^{-2}) \nonumber\\
    &\ll \sum_{i=0}^{\lceil \log_2(X) \rceil} 2^{2i} Y \nonumber\\
    &\ll X^2 Y\ . \label{eq:Vdg D bound}
\end{align}
Here we used that $\e^{-Y^{-2} \bar P^2}\ge \e^{-1}$ for $0 \le \bar P \le Y$ and
\be
\e^{-2^{-2i} P^2}-\e^{-2^{-2i} (P+1)^2}=\e^{-x^2}-\e^{-(x+2^{-i})^2} \gg \, 2^{-i}
\ee
for $2^{i-1} \le P <2^{i}$ with $x=2^{-i}P$ taking values in the interval $[\frac{1}{2},1)$. For $i=0$, this bound still holds.  We then again used the bound \eqref{eq:Theta small argument bound} on the Theta function and use that $\sum_{i=0}^{\lceil \log_2(X) \rceil} 2^{2i} \ll X^2$ by the geometric series.

We finally have
\begin{align}
V_{\mathrm{dg}}(X,Y) &= \sum_{0 \le P \le X,\, 0 \le \bar P \le Y\ \mathrm{dg}} |d_{P,\bar P}| \nonumber\\
&=\sum_{0 \le P \le X,\, 0 \le \bar P \le Y\ \mathrm{dg}} |D_{P,\bar P}-D_{P-1,\bar P}| \nonumber\\
&\leq \sum_{0 \le P \le X,\, 0 \le \bar P \le Y\ \mathrm{dg}} D_{P,\bar P}+D_{P-1,\bar P} \label{eq:triangle inequality}\\
&\ll \sum_{0 \le P \le X,\, 0 \le \bar P \le Y\ \mathrm{dg}} D_{P,\bar P} \ll X^2 Y\ .
\end{align}
Notice that all these inequalities are expected to be sharp (they are e.g.\ for the free boson case), with the only exception of \eqref{eq:triangle inequality}, where the triangle inequality is applied. That estimate removes all possible cancellations that can appear in the difference $D_{P,\bar P}-D_{P-1,\bar P}$.

\paragraph{Naive dd contribution.} An analogous proof shows also that
\be
V_\mathrm{dd}(X,Y) \ll X^2Y^2 \ .\label{eq:Vdd naive bound}
\ee
For this, we should break up the sum over $P$ and $\bar P$ into subintervals $P \in E_i=[2^{i-1},2^i)$, $i \ge 1$, with $E_0=[0,1)$ and $\bar P \in F_j=[2^{j-1},2^{j})$, $j\ge 1$, with $F_0=[0,1)$ with both elements degenerate, which leads to
\begin{align}
\sum_{0 \le P \le X,\, 0 \le \bar P \le Y\, \mathrm{dd}} D_{P,\bar P} &\ll \sum_{i=0}^{\lceil \log_2(X) \rceil} \sum_{j=0}^{\lceil \log_2(Y) \rceil } 2^{i+j} \, \Theta(2^{-2i},2^{-2j}) \nonumber\\
&\ll \sum_{i=0}^{\lceil \log_2(X) \rceil}\sum_{j=0}^{\lceil \log_2(Y) \rceil} 2^{2i+2j} \nonumber\\
&\ll X^2Y^2\ ,
\end{align}
and similarly by the triangle inequality also $V_\mathrm{dd}(X,Y) \ll X^2Y^2$.

\paragraph{Self-duality and temperedness.}
Our computations so far show that $V(R) \ll R^4$, which implies in particular that $\rho$ is a tempered measure. As discussed above, this means that self-duality of $\rho$ can be stated in the sense of tempered distributions: for any Schwartz function $F(P,\bar P) \in \mathcal{S}_\mathrm{even}(\RR^2)$ even in both arguments,
    \begin{align}
        \sum_{P,\bar P} d_{P,\bar P}\,  F(P,\bar P)=\sum_{P,\bar P} d_{P,\bar P} \, \mathbb{S}F(P,\bar P) \label{eq:measure self duality}
    \end{align}
    with
    \be
        \mathbb{S}F(P,\bar P)=8\int_0^\infty  \dd P' \, \dd \bar P'\, \cos(4\pi P P') \cos(4\pi \bar P \bar P')\, F(P',\bar P')\ , \label{eq:SF definition}
    \ee
    where we have folded the integral from \eqref{eq:FourierDef} onto the quadrant $P,\bar P\geq 0$ using evenness of $F$. The notation $\mathbb{S}F$ is borrowed from the fact that the integrand in \eqref{eq:SF definition} is $\mathbb{S}_{P,P'}\, \mathbb{S}_{\bar P,\bar P'}$ with $\mathbb{S}_{P,P'}=2\sqrt{2} \cos(4\pi P P')$ the modular S-kernel.

    The Theta function \eqref{eq:Theta-function-modular-invariance} is the special case of $F(P,\bar P)=\e^{-\beta P^2-\bar \beta \bar P^2}$. Since these functions for $\beta,\bar \beta>0$ span a dense subspace of the even Schwartz functions, the statement follows by linearity and continuity for all even Schwartz functions.

    To derive better bounds, we will apply \eqref{eq:measure self duality} to well-chosen test functions. Let us first illustrate this for the dd contribution, where it is quite simple to find a suitable test function.

    \paragraph{The dd contribution.} Assume for this paragraph that the bound $V_\mathrm{dg}(R) \ll R^\alpha$ and $V_\mathrm{gd}(R) \ll R^\alpha$ holds for some $\alpha \ge 2$. We proved it for $\alpha=3$ in \eqref{eq:Vdg Vgd naive bound} and the rest of the section below will be devoted to proving that it also holds for $\alpha=2$. Then we claim that also
    \be
        V_\mathrm{dd}(R) \ll R^\alpha \label{eq:Vdd bound}
    \ee
    follows. Thus, the dd contribution will follow automatically once we have bounded the dg and gd contributions.

To prove \eqref{eq:Vdd bound}, we apply \eqref{eq:measure self duality} for the bump function
\be
F_{P_0,\bar P_0}(P,\bar P)=\sum_{\sigma,\bar \sigma=\pm} \psi(\beta (\sigma P-P_0)) \, \psi(\beta(\bar \sigma \bar P-\bar P_0))\ , \label{eq:F bump function}
\ee
where $(P_0,\bar P_0)$ is part of the doubly degenerate spectrum. We'll assume that $P_0 \ge 1$ and $\bar P_0\ge 1$, since the remaining contribution to $V_\mathrm{dd}(R)$ where either $P_0 \le 1$ or $\bar P_0 \le 1$ is already known to grow at most as $V_\mathrm{dd}(R,1)+V_\mathrm{dd}(1,R) \ll R^2$ thanks to \eqref{eq:Vdd naive bound}.
Here, we fix once and for all
\be
\psi \in C_{\mathrm{c},\mathrm{even}}^\infty([-1,1]) \label{eq:psi choice}
\ee
with $\psi(0)=1$ and $0 \le \psi(x) \le 1$ a smooth, even bump function. $\beta$ is a parameter that can be adjusted in the proof. We have
\be
\mathbb{S}F_{P_0,\bar P_0}(P,\bar P)=8\beta^{-2} \cos(4\pi P P_0) \cos(4\pi \bar P \bar P_0) \widehat \psi \big(2 \beta^{-1} P\big)\widehat \psi \big(2 \beta^{-1} \bar P\big)\ ,
\ee
where
\be
\widehat{\psi}(x)=\int_{-\infty}^\infty \dd y\,\e^{2\pi i x y} \psi(y)  \label{eq:psi hat}
\ee
is the standard Fourier transform.

Now apply \eqref{eq:measure self duality}. We will choose $\beta$ large so that $F_{P_0,\bar P_0}$ is sharply peaked around $(P,\bar P)=(P_0,\bar P_0)$. It will be sufficient to choose $\beta \ge 4$. The main term that we are interested in is the term $P=P_0$ and $\bar P=\bar P_0$ around which $F_{P_0,\bar P_0}$ is centered. Notice that $F_{P_0,\bar P_0}(P_0,\bar P_0)=1$ due to the fact that $\supp \psi \subset [-1,1]$. Moreover, $F_{P_0,\bar P_0}$ vanishes on any other dd site, since they are separated by at least $\frac{1}{2}$ in the $P$ or the $\bar P$ direction and we chose $\beta \ge 4 \ge 2$. Terms with $\sigma=-$ or $\bar\sigma=-$ do not contribute, since they are centered at $-P_0\le -1$ and $-\bar P_0 \le -1$, and states are located at $P,\, \bar P \ge 0$.
By the triangle inequality, we can thus write
\be
|d_{P_0,\bar P_0}| \le \sum_{P,\bar P\, \text{not dd}} |d_{P,\bar P}\, F_{P_0,\bar P_0}(P,\bar P)|+8 \beta^{-2} \sum_{P,\bar P} \big|d_{P,\bar P}\, \widehat{\psi}\big(2 \beta^{-1} P\big) \widehat{\psi}\big(2 \beta^{-1} \bar P\big)\big|\ , \label{eq:dP0Pbar0 estimate}
\ee
Here, we bounded $|\cos(4\pi P P_0)| \le 1$.
Since $\widehat{\psi}$ is a Schwartz function and $\rho$ is tempered, the sum on the RHS converges.
We can then sum \eqref{eq:dP0Pbar0 estimate} over the doubly degenerate states in the box $1\le P_0, \bar P_0 \le R$. There are $\mathcal{O}(R^2)$ such states and thus the second term is bounded by $\ll_\beta R^2$. For the first term, we exchange the sum over $(P,\bar P)$ and $(P_0,\bar P_0)$. This leads to
\be
V_\mathrm{dd}(R) \ll_\beta R^2+\sum_{P,\bar P\text{ not dd}} |d_{P,\bar P}| \sum_{1 \le P_0,\bar P_0\le R\, \mathrm{dd}} |F_{P_0,\bar P_0}(P,\bar P) |\ .
\ee
Since $\psi$ has compact support and we chose $\beta \ge 4$, there is for a given $(P,\bar P)$ only at most one doubly degenerate state for which $F_{P_0,\bar P_0}(P,\bar P) \ne 0$. Moreover $|F_{P_0,\bar P_0}(P,\bar P)| \le 1$ and such a state can only exist when $0 \le P,\bar P\le R+\frac{1}{4}$ since the bump function forces it to also lie in the box up to a small error. Thus we have the bound
\be
V_\mathrm{dd}(R) \ll R^2+V_\mathrm{gg}(R+\tfrac{1}{4})+V_\mathrm{dg}(R+\tfrac{1}{4})+V_\mathrm{gd}(R+\tfrac{1}{4}) \ll R^\alpha\ .
\ee

\subsection{Choosing an appropriate test function}
We now work towards bounding the dg contribution (with the bound for the gd contribution following by an identical method). The crux of the argument is that $S$ invariance, combined with the other axioms, relates the dg contribution to the gg one, preventing it from growing too fast. To see this, we need to use $S$ invariance in a rather fine-grained way, by choosing a particular test function $F(P,\bar P)$ to smear against the statement of $S$ invariance in \eqref{eq:measure self duality}. To bound the dd contribution, we just chose a small bump function in \eqref{eq:F bump function}, which eliminated the other dd contributions. Here, for the dg contribution, this will not be enough, since we only have control over the gg contributions. Thus, we will need to choose it in such a way such that it can isolate dg contributions on one side and bound them in terms of the controlled gg contributions on the other side.

We will follow the convention that $d_{P,\bar P}=0$ when $(P,\bar P)$ does not appear in the spectrum. We call a site occupied when $d_{P,\bar P} \ne 0$, and axiom~\ref{axiom:integrality} then gives $|d_{P,\bar P}| \ge 1$.

\paragraph{Choice of test function.} Let $(P_0,\bar P_0)$ be an occupied dg site. We can assume that $P_0 \ge2$ and $\bar P_0 \ge 2$, since the contribution to $V_\mathrm{dg}(R)$ with either $P_0 \le 2$ or $\bar P_0 \le 2$ is already bounded by $R^2$ thanks to the naive bound \eqref{eq:Vdg Vgd naive bound}. Let
\be
G_{P_0,\bar P_0}(P,\bar P)=\cos\big(4\pi b(\bar P-\bar P_0)\big) \psi\big(\beta(P-P_0)\big)\psi\big(\beta(\bar P-\bar P_0)\big)\ , \label{eq:def G}
\ee
where $\psi$ is again a compactly supported bump function as in \eqref{eq:psi choice} and $F$ the following combination of this function
\be
F_{P_0,\bar P_0}(P,\bar P)=\sum_{\sigma,\bar \sigma=\pm} \sum_{s\in \{0,1\}} \sum_{t \in \{-1,0,1\}} a(s,t) \, G_{P_0,\bar P_0}(\sigma P-s n,\bar \sigma \bar P-t)\ , \label{eq:def F}
\ee
with $a(s,0)=(-1)^s$ and $a(s,\pm 1)=-\frac{1}{2} (-1)^s$. Here, $\beta>0$, $b>0$ and the positive integer $1 \le n \le R$ are parameters that we can freely choose.
We will set
\be
\beta=R^3\ , \label{eq:beta choice}
\ee
which sharply localizes $F$ around the positions $(P_0+s n,\bar P_0+t)$ with $s\in \{0,1\}$ and $t \in \{-1,0,1\}$, together with their reflected positions under the symmetrization $P \to -P$ and $\bar P \to -\bar P$.

It is also straightforward to compute the Fourier transform of \eqref{eq:def F}. The two symmetrized phase sums are
\begin{align}
\sum_{\sigma=\pm}\ \sum_{s \in \{0,1\}} (-1)^s\, \e^{4\pi i \sigma P (P_0+sn)}
&=4 \sin(2\pi n P)\, \sin\big(2\pi (2P_0+n) P\big)\ , \nonumber\\
\sum_{\bar \sigma=\pm}\ \sum_{t \in \{-1,0,1\}} c_t\, \e^{4\pi i \bar \sigma \bar P (\bar P_0+t)}
&=4\cos(4\pi \bar P \bar P_0)\, \sin^2(2\pi \bar P)\ ,
\end{align}
where $c_0=1$, $c_{\pm 1}=-\frac{1}{2}$, and we obtain
\begin{align}
\mathbb{S}F_{P_0,\bar P_0}(P,\bar P)=\frac{16}{\beta^{2}}\, &\sin(2\pi n P)\, \sin\big(2\pi (2P_0+n) P\big)\, \widehat{\psi}\big(2\beta^{-1}P\big) \nonumber \\
\times\ &\cos(4\pi \bar P \bar P_0)\, \sin^2(2\pi \bar P)\, \Big[\widehat{\psi}\big(2\beta^{-1}(\bar P-b)\big)+\widehat{\psi}\big(2\beta^{-1}(\bar P+b)\big)\Big]\ , \label{eq:SF}
\end{align}
where $\widehat{\psi}$ is the Fourier transform defined in \eqref{eq:psi hat}.

 Let us briefly comment why one would write down this test function. We will call the LHS of \eqref{eq:measure self duality} the `primal side', and the RHS the `Fourier side'. On the primal side, we wanted to isolate dg states. This is achieved because it is a sum of bump functions sharply localized enough that they will only catch dg atoms, up to some controllable error terms. On the Fourier side, by the antisymmetrized shift $P_0 \to P_0 +sn$ and the `doubly antisymmetrized' shift $\bar P_0 \to \bar P_0 \pm 1$ in
\eqref{eq:def F}, we have engineered zeros in \eqref{eq:SF} at all degenerate sites where $P \in \frac{1}{2}\ZZ$ or $\bar P \in \frac{1}{2} \ZZ$. The Fourier side is therefore entirely supported on gg sites, and this is the main reason why the construction is controllable. The additional sum over $\sigma,\bar \sigma$ simply makes the function even.

The particular symmetrization choices above in $P_0$ and $\bar P_0$ are made so that (i) only the unshifted term in $\bar P_0$ contributes to the primal side and (ii) after summing over different values of the integer $n$ and different sites $(P_0,\bar P_0)$, the primal side can be bounded below by the desired quantity $V_{dg}(R)$.  Since this still won't be enough to get the desired $R^2$ bound on the Fourier side, the $\cos\big(4\pi b(\bar P-\bar P_0)\big)$ factor in \eqref{eq:def G} preserves all of those properties and gives an additional handle $b$. Summing over $b$ will play an important role in the argument below.

\paragraph{Evaluation on the primal side.}
We now evaluate the primal side of \eqref{eq:measure self duality} for the choice \eqref{eq:def F}. As mentioned above, we'll assume that $2 \le P_0,\bar P_0\le R$.

We first notice that only $\sigma=\bar \sigma=+$ can contribute to this primal evaluation, since the terms with $\sigma=-$ or $\bar \sigma =-$ are supported for negative $P$ or $\bar P$. Let us also notice that the six terms in the sum over $s$ and $t$ in \eqref{eq:def F} have disjoint support, so at most one of them can be non-zero at any point. The primal side thus selects terms with $(P,\bar P)$ close to the centers of the bump functions located at $(P_0+s n,\bar P_0+t)$. The $t=0$ sites $(P_0,\bar P_0)$ and $(P_0+n,\bar P_0)$ are both dg sites.
On the other hand, the sites $(P_0+s n,\bar P_0+t)$ at the centers of the $t \in \{-1,1\}$ bump functions violate spin quantization, precisely because $\bar P_0 \not\in \frac{1}{2}\ZZ$ since the state is dg.

In fact, we claim that, for large enough $R$, there are no contributions at all on the primal side of \eqref{eq:measure self duality} from the $t\neq 0$ terms.  To see this, note that such contributions would be supported in the square with vertices $(P_0+s n\pm \beta^{-1},\bar P_0+t \pm \beta^{-1})$. We shall show that this square contains no state with integer spin. Indeed, the spin of a state in this square is an integer plus  the correction
\be
-2\bar P_0 t+ \mathcal{O}(R)\, \beta^{-1}=-2 \bar P_0 t+\mathcal{O}(R^{-2})\ , \label{eq:spin correction}
\ee
where the order $R$ terms come from the crossterms with $\beta^{-1}$ involving $P_0$, $\bar P_0$ or $n$. Now compute the distance of $2 \bar P_0$ to the nearest integer. Since $P_0^2-\bar P_0^2 \in \ZZ$ and $P_0 \in \frac{1}{2}\ZZ$, it follows that $\bar P_0=\frac{1}{2} \sqrt{m}$ for a non-square integer $m$. So
\begin{multline}
    \hspace{1cm}\mathrm{dist}(2 \bar P_0,\ZZ)=\mathrm{dist}(\sqrt{m},\ZZ) = \big|\sqrt{m}-\lfloor \sqrt{m} \rceil \big|\\
    = \frac{\big|m-\lfloor \sqrt{m} \rceil^2\big|}{\sqrt{m}+\lfloor\sqrt{m} \rceil} \ge \frac{1}{2 \sqrt{m}+1} \gg R^{-1}\ , \hspace{1cm}\label{eq:distance barP Z}
\end{multline}
with the last inequality following since we chose $\bar P_0 \le R$. Here, $\lfloor x \rceil$ is the closest integer to the real number $x$.  For large enough $R$, the $\mathcal{O}(R^{-1})$ distance \eqref{eq:distance barP Z} dominates over the error term in \eqref{eq:spin correction}, so there are no integer spins in the support and we may discard these contributions. This is the main reason why we chose $\beta$ as in \eqref{eq:beta choice}.

By similar reasoning, we can see that all states in the analogous square $(P_0+s n \pm \beta^{-1},\bar P_0 \pm \beta^{-1})$  with $t=0$ are gg, except for the central site $(P_0+s n,\bar P_0)$. Indeed, by \eqref{eq:distance barP Z},  $\bar P_0$ is far enough away from being half-integer that a small perturbation by $\beta^{-1}$ cannot make any state in the square degenerate for the right-moving variable. The left-moving variable is then tied to the right-moving one because of spin quantization. For a state $(P_0+s n+x,\bar P_0+y)$ with $|x|,\,|y| \le \beta^{-1}$, spin quantization tells us that
\be
2(P_0+n s)x-2 \bar P_0 y+x^2-y^2=0\ .
\ee
Spin quantization only forces the left hand side to be an integer, but since $|x|,\, |y| \le \beta^{-1}$ and $P_0,\,\bar P_0,\, n\le R$, the only attainable integer for large enough $R$ is 0.
If $x=0$, then also $y=0$ in the region $|y| \le \beta^{-1}$, and vice versa. Therefore, all states except for $(x,y)=(0,0)$ are gg.

Therefore, the primal evaluation can be written as
\be
\sum_{P, \bar P} d_{P,\bar P} F_{P_0,\bar P_0}(P,\bar P)=d_{P_0,\bar P_0}-d_{P_0+n,\bar P_0}+\sum_{P, \bar P\, \mathrm{gg}} \sum_{s \in \{0,1\}} d_{P,\bar P}\, (-1)^s G_{P_0,\bar P_0}(P-s n,\bar P)\ . \label{eq:primal evaluation}
\ee
We will consider the terms $d_{P_0,\bar P_0}-d_{P_0+n,\bar P_0}$ to be the main terms, while the other terms are error terms.

\paragraph{Assembly.} We thus see that the self-duality of $\rho$ \eqref{eq:measure self duality} with our choice \eqref{eq:def F} extracts dg atoms only in terms of gg atoms. By moving the error terms in \eqref{eq:primal evaluation} onto the Fourier side, we get
\be
d_{P_0,\bar P_0}-d_{P_0+n,\bar P_0}=-\sum_{P, \bar P\, \mathrm{gg}} \sum_{s \in \{0,1\}} d_{P,\bar P} (-1)^s G_{P_0,\bar P_0}(P-s n,\bar P)+\sum_{P,\bar P\,\mathrm{gg}} d_{P,\bar P} \mathbb{S}F_{P_0,\bar P_0}(P,\bar P) \ . \label{eq:self duality identity dg}
\ee
At this point, we have only assumed that $2 \le P_0, \bar P_0 \le R$ and we have specified $\beta$ as in \eqref{eq:beta choice}. We have also assumed that the integer $n$ satisfies $1 \le n \le R$. The parameter $b$ entering \eqref{eq:def G} is still undetermined.

\paragraph{Summation.} We want to make a statement about $V_\mathrm{dg}(R)$. Therefore, we will now multiply \eqref{eq:self duality identity dg} by $\sgn(d_{P_0,\bar P_0}-d_{P_0+n,\bar P_0})$ and sum over all occupied dg sites in $2 \le P_0,\bar P_0 \le R$. On the LHS side, this produces
\be
\sum_{\begin{subarray}{c} 2 \le P_0,\bar P_0\le R\, \mathrm{dg} \\ \text{occupied} \end{subarray}} \big| d_{P_0,\bar P_0}-d_{P_0+n,\bar P_0}\big|\ ,
\ee
which we now further bound from below. For this, we first fix $\bar P_0$, summing over $P_0$. Let us additionally average over the integers $1\leq n\leq R$:
\begin{align}
&\frac{1}{R}\sum_{n=1}^R \sum_{\begin{subarray}{c} 2 \le P_0 \le R \text{ d} \\ \text{occupied} \end{subarray}} \big| d_{P_0,\bar P_0}-d_{P_0+n,\bar P_0}\big| \nonumber\\
&\qquad\ge \frac{1}{R}\sum_{{\begin{subarray}{c} 2 \le P_0 \le R\text{ d} \\ \text{occupied} \end{subarray}}} \sgn(d_{P_0,\bar P_0}) \sum_{n=1}^R \big(d_{P_0,\bar P_0}-d_{P_0+n,\bar P_0}\big) \nonumber\\
&\qquad=\sum_{2 \le P_0 \le R\text{ d}} |d_{P_0,\bar P_0}|-\frac{1}{R}\sum_{\begin{subarray}{c} 2 \le P_0 \le R\text{ d} \\ \text{occupied} \end{subarray}} \sgn(d_{P_0,\bar P_0})\big(D_{P_0+R,\bar P_0}-D_{P_0,\bar P_0}\big) \nonumber\\
&\qquad\ge\sum_{2 \le P_0 \le R\text{ d}} |d_{P_0,\bar P_0}|-\frac{1}{R}\sum_{2 \le P_0 \le 2R} D_{P_0,\bar P_0}\ , \label{eq:primal side bound from below}
\end{align}
where $D_{P_0,\bar P_0}$ are the Virasoro multiplicities.

We can now reinstate the sum over $\bar P_0$. The first term in \eqref{eq:primal side bound from below} becomes the quantity $V_\mathrm{dg}(R)$ that we are after, while the second is an error term. We already know from \eqref{eq:Vdg D bound} that this error term is of order $\mathcal{O}(R^2)$. We thus have
\be
\frac{1}{R} \sum_{n=1}^R \sum_{\begin{subarray}{c} 2 \le P_0,\bar P_0\le R\, \mathrm{dg} \\ \text{occupied} \end{subarray}} \big| d_{P_0,\bar P_0}-d_{P_0+n,\bar P_0}\big| \ge V_\mathrm{dg}(R)-C R^2\ ,
\ee
for some constant $C$. Since this is true on average in $n$, there must be at least one $n$ for which this is also true separately without taking averages. We fix this $n$ in the following.
The self-duality equation \eqref{eq:self duality identity dg} thus gives the bound
\begin{multline}
    V_\mathrm{dg}(R) \le C R^2+\sum_{\begin{subarray}{c} 2 \le P_0,\bar P_0\le R\, \mathrm{dg} \\ \text{occupied} \end{subarray}}\sgn(d_{P_0,\bar P_0}-d_{P_0+n,\bar P_0})\sum_{P,\bar P\text{ gg}} d_{P,\bar P} \\
    \times \bigg[-\sum_{s \in \{0,1\}} (-1)^s G_{P_0,\bar P_0}(P-s n,\bar P)+\mathbb{S} F_{P_0,\bar P_0}(P,\bar P)\bigg]\ , \label{eq:Vdg bound from selfduality}
\end{multline}
where we also absorbed the $\mathcal{O}(R^2)$ contribution to $V_\mathrm{dg}(R)$ from the states with $P_0 \le 2$ or $\bar P_0 \le 2$ into the constant $C$.
\paragraph{Primal bound.} It remains to bound the various terms on the RHS of this inequality further. First, we claim that the first term in the parenthesis of \eqref{eq:Vdg bound from selfduality}, i.e.\ the remaining terms from the primal evaluation, is $\mathcal O(R^2)$.
To see this, we take absolute values and swap the order of summations of $(P,\bar P)$ and $(P_0,\bar P_0)$. For fixed $s$, the crucial observation is that every $P,\bar P$ can in fact only appear once in the double sum, since it is only close (on the scale $\beta^{-1}=R^{-3}$) to at most one dg site. This follows from the same computation as in \eqref{eq:distance barP Z}: dg states are separated in $P_0$ by $\gg 1$ (distance $\frac{1}{2}$ to be precise) and in $\bar P_0$ by $\gg R^{-1}$, much more than the scale $\beta^{-1}$. Furthermore, every $(P,\bar P)$ that appears has to lie in the rectangle $0 \le P-sn,\bar P \le R+\beta^{-1}$. In particular, all appearing $(P,\bar P)$ lie in a square of width $2R+1$ and appear at most twice, once for $s=0$ and once for $s=1$. Since, $G_{P_0,\bar P_0}$ is upper bounded by 1 by construction, then this term is bounded by $2 V_\mathrm{gg}(2R+1)$, which in turn is $\ll R^2$ by \eqref{eq:Vgg bound}.

We can therefore absorb the first term of the parenthesis into the constant $C$ in \eqref{eq:Vdg bound from selfduality}, giving
\be
V_\mathrm{dg}(R) \ll R^2+\bigg|\sum_{\begin{subarray}{c} 2 \le P_0,\bar P_0\le R\, \mathrm{dg} \\ \text{occupied} \end{subarray}} \sum_{P,\bar P\text{ gg}} d_{P,\bar P}\, \sgn(d_{P_0,\bar P_0}-d_{P_0+n,\bar P_0}) \, \mathbb{S} F_{P_0,\bar P_0}(P,\bar P) \bigg|\ ,\label{eq:Vdg bound from selfduality 2}
\ee
with $\mathbb{S}F_{P_0,\bar P_0}$ given in \eqref{eq:SF}. Notice also that we have now chosen $\beta$ and $n$, but not $b$ and the implied constant in \eqref{eq:Vdg bound from selfduality 2} is independent of $b$.

\subsection{Localizing the Fourier side}
It remains to bound the Fourier side of \eqref{eq:Vdg bound from selfduality 2}. For this, it is useful to swap the sum over $(P,\bar P)$ and $(P_0,\bar P_0)$. Since $\widehat \psi$ is Schwartz, it satisfies the bound
\be
\big| \widehat \psi(x) \big| \ll_\gamma (1+\tfrac{1}{2} |x|)^{-\gamma} \label{eq:Schwartz decay}
\ee
for any $\gamma>0$. For the argument below, it will be enough to choose $\gamma=5$.
Substituting \eqref{eq:Schwartz decay} into \eqref{eq:SF} and using $|\bar P+b| \ge |\bar P-b|$, we immediately find that we can remove the second term in the parenthesis in \eqref{eq:SF} at the cost of a factor of 2. The inequality then gives the following bound on the Fourier side,
\be
    V_\mathrm{dg}(R) \ll R^2+\beta^{-2} \sum_{P,\bar P\text{ gg}} d_{P,\bar P}\, \big(1+\beta^{-1} P\big)^{-5}\big(1+\beta^{-1}|\bar P-b|\big)^{-5} \big| \Psi(P,\bar P) \big|\ ,\label{eq:Vdg bound from selfduality 3}
\ee
where we define
\be
\Psi(P,\bar P)=\sum_{\begin{subarray}{c} 2 \le P_0,\bar P_0 \le R\text{ dg} \\ \text{occupied} \end{subarray}} c_{P_0,\bar P_0} \sin\big(2\pi (2 P_0+n) P\big) \cos(4\pi \bar P \bar P_0)\ , \label{eq:Psi definition}
\ee
and $c_{P_0,\bar P_0}=\sgn(d_{P_0,\bar P_0}-d_{P_0+n,\bar P_0})$ is a sign.
Let us remark that the bounds on the primal side do not depend on $b$, while the Fourier side is sensitive to $b$. In fact, the sum is localized around $\bar P \sim b$.

If we naively eliminate all the trigonometric functions in the definition of $\Psi(P,\bar P)$, it is bounded by the total number of occupied dg sites, which in turn is bounded by $V_\mathrm{dg}(R)$. Thus, we get the trivial bound
\be
|\Psi(P,\bar P)| \le V_\mathrm{dg}(R)\ . \label{eq:Psi trivial inequality}
\ee
Let us note that the simplest possible approach using this will not work. If we choose $b=0$ and use the bound \eqref{eq:Psi trivial inequality}, the remaining sum over $(P,\bar P)$ can be bounded by $\ll \beta^{-2}V_\mathrm{gg}(\beta)\ll 1$. This means that the Fourier side is bounded by $\ll V_\mathrm{dg}(R)$, which is useless since we did not control the relative constant.\footnote{Let us note that the scale $\beta$ canceling out on the Fourier side is why the choice \eqref{eq:beta choice} did not have to be particularly sharp.}

In the following, we will show that we can use $b$ to our advantage.

\paragraph{Summing over $b$.} We will now sum the Fourier side over $b=B\ell \beta$ with $\ell \in \{1,\dots, L\}$ for some $L$, with $B\ge 1$ a fixed $\mathcal{O}(1)$ constant. We momentarily ignore the fact that $P$ and $\bar P$ also appear in the inner sum in \eqref{eq:Vdg bound from selfduality 3}. We  claim that
\be
\sum_{\ell=1}^L \sum_{P, \bar P\text{ gg}} d_{P,\bar P} \big(1+\beta^{-1} P\big)^{-3} \big(1+ |\beta^{-1}\bar P-B\ell |\big)^{-3} \ll LB \beta^2\ . \label{eq:summed widehat psi bound}
\ee
We will later use the remaining two powers $(1+\beta^{-1} P)^{-2} (1+ |\beta^{-1}\bar P-B\ell |)^{-2}$ from \eqref{eq:Vdg bound from selfduality 3}. The implicit constant is in particular independent of $B,\, L \ge 1$.
This makes sense since, after doing the sum over $\ell$, the functions $(1+|x|)^{-3}$ localize the sum over gg atoms approximately to the rectangle $[0,\beta] \times [0,BL\beta]$. The bound \eqref{eq:Vgg bound} then motivates the RHS of \eqref{eq:summed widehat psi bound}.

To demonstrate it precisely, we shall control the tails. To begin with, let us note that
\be
\sum_{\ell=1}^{L} \big(1+|x-B\ell|\big)^{-3} \ll \big(1+\max\big(x-BL,0\big)\big)^{-2}\ , \label{eq:grid sum}
\ee
with the implicit constant independent of $B$ and $L$. To see this, note that for any $x$ the LHS is bounded by $\sum_{ \ell \in \ZZ} \big(1+|x-B\ell|\big)^{-3} \le 2+2 \sum_{k \ge 1} (1+k B)^{-3} \ll 1$. This follows because the points $x-B \ell$ meet each of the intervals $[Bk,B(k+1))$ once.
Now write $x=BL+u$ and assume $u>0$. Then
\be
\sum_{\ell=1}^L \big(1+|x-B\ell|\big)^{-3} = \sum_{k=0}^{L-1} \big(1+u+Bk\big)^{-3} \le \sum_{k=0}^\infty \big(1+u+Bk\big)^{-3}\ ,
\ee
where $k=L-\ell$. Since the function we are summing is monotonically decreasing, we can bound $\sum_{k=0}^\infty f(k) \le f(0)+\int_0^\infty \dd k\, f(k)$. Both terms are bounded by the RHS of \eqref{eq:grid sum}, which demonstrates its validity.

Thanks to \eqref{eq:grid sum}, it is sufficient for \eqref{eq:summed widehat psi bound} to prove that
\be
\sum_{P,\bar P\ \mathrm{gg}} d_{P,\bar P}\, \big(1+\beta^{-1}P\big)^{-3}\Big(1+\max\big(\beta^{-1}\bar P-BL,0\big)\Big)^{-2}\ \ll\ L B\beta^2\ . \label{eq:envelope mass sum}
\ee
To carry this out, we partition the sum over $P$ into the intervals $E_0=[0,\beta]$, $E_i=(2^{i-1}\beta,2^i \beta]$ and the sum over $\bar P$ into the intervals $F_0=[0,LB \beta]$, $F_j=(2^{j-1} LB\beta,2^j LB \beta]$ with $i,\, j \in \NN$. We apply the bound \eqref{eq:Vgg bound} on each $E_i \times F_j \subset [0,2^i \beta] \times [0,2^j LB\beta]$. On $E_i \times F_j$, we have furthermore $(1+\beta^{-1} P)^{-3} \ll 2^{-3i}$ and $(1+\max(\beta^{-1}\bar P-BL,0))^{-2} \ll 2^{-2j}$. Therefore, we can bound the LHS of \eqref{eq:envelope mass sum} by
\be
\ll \  L B \beta^2 \bigg(\sum_{i \ge 0} 2^i \times 2^{-3i}\bigg)\bigg(\sum_{j \ge 0} 2^j \times 2^{-2j}\bigg)\ \ll\ L B \beta^2\ ,
\ee
since both geometric series converge. This proves \eqref{eq:summed widehat psi bound}.

\paragraph{Good $\boldsymbol{\ell}$'s.} We now apply the pigeonhole principle to \eqref{eq:summed widehat psi bound}. Since the whole sum of $L$ terms is bounded by $\ll LB\beta^2$, there must be at least $\frac{L}{2}$ terms that are bounded by twice the average value $B\beta^2$. We call values \textit{good values} of $\ell$, for which we have
\be
 \sum_{P, \bar P\text{ gg}} d_{P,\bar P} \big(1+\beta^{-1} P\big)^{-3} \big(1+\beta^{-1} |\bar P-B\ell \beta|\big)^{-3} \  \ll \  B \beta^2\ . \label{eq:good l bound}
\ee
Notice that it doesn't matter that $\ell \le L$ since we could choose $L$ arbitrarily large above.
As we mentioned, at least half of the values of $\ell$ are good and we can take $\ell$ arbitrarily large below.

\paragraph{Sharp localization.} Let us come back to the RHS of \eqref{eq:Vdg bound from selfduality 3} and take $b=B \ell \beta$ for a good value of $\ell$. We will eventually take $\ell$ very large. Similarly to \eqref{eq:summed widehat psi bound} above, the sum over $P$ and $\bar P$ is heuristically localized around $(0,b)$. We shall now make this precise and turn it into a sharp localization.

Split the sum over $\bar P$ in \eqref{eq:Vdg bound from selfduality 3} into a `localized' region $|\bar P-b| \le \frac{1}{4} B \beta$ and an `error' region $|\bar P-b| > \frac{1}{4} B \beta$. The constant of $\frac{1}{4}$ is for convenience. We now estimate the error contribution.

Notice that
\be
\big(1+|\beta^{-1} \bar P-B \ell|\big)^{-2} \ll B^{-2}\ .
\ee
Thus, we can use the two additional powers in \eqref{eq:Vdg bound from selfduality 3} to gain an extra power of $B^{-2}$. Then combining \eqref{eq:good l bound} and the trivial bound \eqref{eq:Psi trivial inequality}, the error term is bounded by
\be
\ll \beta^{-2} \times B^{-2}\times B\, \beta^2 \, V_\mathrm{dg}(R) \ll B^{-1} V_\mathrm{dg}(R)\ .
\ee
Even though the RHS contains the quantity $V_\mathrm{dg}(R)$ we are trying to bound, we have the additional parameter $B$ to adjust. We can choose $B$ large enough, so that $B^{-1}$ times the implicit constant in the bound \eqref{eq:Vdg bound from selfduality 3} is smaller than unity. We can then take the error term on the RHS and absorb it into the LHS.
This gives the following bound on the Fourier side. For at least half of the $\ell$'s and for sufficiently large $B$,
\be
    V_\mathrm{dg}(R) \ll R^2+\beta^{-2} \sum_{\begin{subarray}{c} P,\bar P\text{ gg},\\ |\bar P-B \ell\beta | \le \frac{1}{4} B \beta \end{subarray}} d_{P,\bar P}\, \big(1+\beta^{-1} P\big)^{-5}\big(1+\beta^{-1}|\bar P-b|\big)^{-5} \big| \Psi(P,\bar P) \big|\label{eq:Vdg bound from selfduality 4}\ ,
\ee
i.e.\ we have localized $\bar P$.

Now notice that $\Psi(P,\bar P)$ is by definition \eqref{eq:Psi definition} 1-periodic in $P$. Moreover, it is continuous and thus attains a maximum on the rectangle $[0,1] \times [B\ell \beta-\frac{1}{4} B \beta,B\ell \beta+\frac{1}{4} B \beta]$. We will denote the location of this maximum by $(P_\ell,\bar P_\ell)$. We estimate \eqref{eq:Vdg bound from selfduality 4} with the help of \eqref{eq:good l bound}. For this term we can just waste the extra powers $(1+\beta^{-1} P)^{-2} (1+ |\beta^{-1}\bar P-B\ell |)^{-2}$ and have
\be
V_\mathrm{dg}(R) \ll R^2+B\, |\Psi(P_\ell,\bar P_\ell)|\ . \label{eq:Vdg above}
\ee
It is useful to give names to the constants in this inequality and write
\be
2\delta\,  V_\mathrm{dg}(R) \le c R^2 +|\Psi(P_\ell,\bar P_\ell)|\ , \label{eq:Vdg bound from selfduality 5}
\ee
where $c \asymp\delta\asymp B^{-1}$.
\subsection[Bounding \texorpdfstring{$\Psi$}{Psi}]{Bounding $\boldsymbol{\Psi}$}
It remains to bound $|\Psi(P_\ell,\bar P_\ell)|$ with $\Psi$ defined in \eqref{eq:Psi definition} by a stronger bound than the trivial bound \eqref{eq:Psi trivial inequality}. We don't do this directly since we haven't found a way to bound it for a single $\ell$. Instead, we say something about the quantity
\be
\frac{1}{T}\, \Big|\Big\{(P,\bar P) \in [0,1]\times [0,T]\ :\ \big|\Psi(P,\bar P)\big| \ge \delta\, V_\mathrm{dg}(R)-c R^2\Big\}\Big| \label{eq:average extreme}
\ee
as $T \to \infty$ and fixed $\delta$. The bound \eqref{eq:Vdg bound from selfduality 5} implies a lower bound on this quantity. We will derive an upper bound by studying the statistics for large $T$ and applying equidistribution theorems. Comparing the two bounds will yield the desired statement.

\paragraph{Lower bound.} Let us first work out the lower bound. Consider now $(P_\ell,\bar P_\ell)$ as defined above \eqref{eq:Vdg above}, and the small square $[P_\ell- \frac{\delta}{100 R},P_\ell+\frac{\delta}{100 R}]\times [\bar P_\ell - \frac{\delta}{100 R},\bar P_\ell + \frac{\delta}{100 R}]$ around it, possibly reduced mod 1 in $P$ if the square goes outside of the strip $[0,1] \times [0,T]$.
Differentiate the expression \eqref{eq:Psi definition} for $\Psi$ term-by-term and use that there are at most $V_\mathrm{dg}(R)$ terms. Since $2P_0+n \le 3R$ and $\bar P_0 \le R$, then we find
\be
\bigg|\frac{\partial \Psi}{\partial P}\bigg| \le 6\pi R\, V_\mathrm{dg}(R)\ , \qquad \bigg|\frac{\partial \Psi}{\partial \bar P}\bigg| \le 4\pi R\, V_\mathrm{dg}(R)\ ,
\ee
i.e.\ $\Psi$ does not vary very quickly.  On the small square, we have by the triangle inequality from \eqref{eq:Vdg bound from selfduality 5}
\be
|\Psi(P,\bar P)| \ge \delta\Big(2-\frac{6\pi}{100}-\frac{4\pi}{100}\Big) V_\mathrm{dg}(R) -c R^2\ge \delta V_\mathrm{dg}(R)-c R^2\ .
\ee
These squares around $(P_\ell,\bar P_\ell)$ for $\ell$ good are all disjoint since good heights $\bar P= B \ell \beta$ are at least $B\beta$ apart and the height of the $\bar P_\ell$ is chosen within distance $\frac{1}{4}B \beta$ from a good height, so the $\bar P_\ell$'s are at least $\frac{1}{2}B \beta$ apart. Thus, disjointness requires $\frac{\delta}{50R} \le \frac{1}{2}B \beta$, which is true for large enough $R$. Thus, $|\Psi(P,\bar P)| \ge \delta V_\mathrm{dg}(R)-c R^2$ holds on the union of all these squares, which makes a finite proportion of $[0,1] \times [0,T]$. The proportion is at least
\begin{multline}
\liminf_{T \to \infty} \frac{1}{T}\, \Big|\Big\{(P,\bar P) \in [0,1]\times [0,T]\ :\ \big|\Psi(P,\bar P)\big| \ge \delta\, V_\mathrm{dg}(R)-c R^2\Big\}\Big| \\
\ge \frac{1}{2B \beta}\Big(\frac{\delta}{100R}\Big)^2 \gg \frac{\delta^2}{B R^5}\ . \label{eq:average extreme density lower bound}
\end{multline}
Even though this is a very small proportion, it will be enough to conclude below.

\paragraph{Decomposition into squarefree parts.} We now start to work out the upper bound. First, some elementary arithmetic observations.
Since $(P_0,\bar P_0)$ is of type dg, we can write $P_0=\frac{k}{2}$ for $k$ integer. Spin quantization imposes $\bar P_0=\frac{1}{2}\sqrt{m}$ for a non-square $m$ integer, as well as $m \equiv k^2 \bmod 4$.
Write $m=sj^2$ for $s\geq 2$ a squarefree integer. Since $\bar P_0 \le R$ then we have $j \ll R$.
Thus we have
\be
\Psi(P,\bar P)=\sum_{s \in \mathcal S}  \Phi_s(P,\sqrt{s} \bar P)\ , \label{eq:Psi Phi sum}
\ee
where
\be
\Phi_s(P, x):=\sum_{k,j}c_{k/2,\sqrt{s} j/2} \,  \sin\big(2\pi (k+n) P\big) \cos\big(2\pi jx\big)\ , \label{eq:Phi definition}
\ee
Here $k$ and $j$ run over the allowed finite sets, and $s$ over the finite set of squarefree parts $\mathcal S$, such that $(P_0,\bar P_0)=(\frac{k}{2},\frac{j}{2}\sqrt{s})$ is an occupied dg site in the range $2 \le P_0,\bar P_0 \le R$.

\paragraph{Equidistribution. } The main idea is now the following. Only $P \bmod 1$ and $\sqrt{s} \bar P \bmod 1$ enter the formula \eqref{eq:Psi Phi sum} and, for large values of $\bar P$, the collection of values $\{\sqrt{s} \bar P \bmod 1\}_{s \in \mathcal{S}}$ equidistributes. We will study the limiting distribution
\be
\lim_{T \to \infty} \frac{1}{T} \int_0^1 \dd P \int_0^T \dd \bar P\, f\big(\Psi(P,\bar P)\big)\ ,
\ee
where $f$ will be a continuous function chosen to bound the quantity \eqref{eq:average extreme}.

To do so, we will apply Weyl's equidistribution theorem~\cite{Weyl:1916}. It says that for irrational numbers $\alpha_1,\dots,\alpha_N \in \RR$ that are linearly independent over $\QQ$, we have
\be
\lim_{T \to \infty}\frac{1}{T}\int_0^T \dd y\, G(\alpha_1 y,\dots,\alpha_N y)=\int_{\mathbb{T}^N} \dd^N x\ G(x_1,\dots,x_N)\ ,
\ee
where $G$ is a continuous function that is 1-periodic in every entry, i.e.\ a continuous function on $\mathbb{T}^N$, where $\mathbb{T}=\RR/\ZZ$.

We will now apply this to the case of $\{\alpha_1,\dots,\alpha_N\}=\{\sqrt{s}\}_{s \in \mathcal{S}}$. This is possible because the real numbers $\sqrt{s}$ for $s$ squarefree are linearly independent over $\QQ$, which is known as Besicovitch's theorem~\cite{Besicovitch:1940}.
We will furthermore choose
\be
G:\ \mathbb{T}^{\mathcal{S}} \longrightarrow \RR\ , \qquad
G(x)=\int_0^1 \dd P\ f\bigg(\sum_{s \in \mathcal{S}} \Phi_s(P,x_s)\bigg)\ ,
\ee
which is continuous because $(P,x) \mapsto f\big(\sum_{s} \Phi_s(P,x_s)\big)$ is continuous on the
compact set $[0,1] \times \mathbb{T}^{\mathcal{S}}$ and thus uniformly continuous. Equidistribution gives that
\be
\lim_{T \to \infty} \frac{1}{T} \int_0^1 \dd P \int_0^T \dd \bar P\, f\big(\Psi(P,\bar P)\big)
=\int_{\mathbb{T}^{\mathcal{S}}} \dd^{\mathcal{S}} x\ G(x)
=\int_0^1 \dd P\  \EE \big( f(\mathcal X(P))\big)\ , \label{eq:equidistribution theorem}
\ee
where in the last step we used Fubini's theorem. Here we are regarding $X_s=\sqrt{s} \bar P \bmod 1$ as a collection of independent random variables on $\RR/\ZZ$, and we define a real-valued random variable
\be
\mathcal X(P):=\sum_{s \in \mathcal{S}} \Phi_s(P,X_s)\ .
\ee
Since the $X_s$ are independent random variables, then $\Phi_s(P,X_s)$
for $s \in \mathcal{S}$ are also independent.
Let us also note that since only positive modes $j>0$ appear in \eqref{eq:Phi definition}, these variables have zero mean,
\be
\EE \Phi_s(P,X_s)=0 \ , \qquad \EE \mathcal{X}(P)=\sum_{s \in \mathcal{S}}\EE \Phi_s(P,X_s)=0 \ .
\ee
A simple integral on the circle gives
\begin{align}
    \sigma^2(P)&:=\EE \mathcal{X}(P)^2 \nonumber\\
    &=\sum_{s \in \mathcal S}\EE \big(\Phi_s(P,X_s)^2\big)\nonumber\\
    &=\frac{1}{2}\sum_{s \in \mathcal{S}}\sum_{j} \bigg(\sum_k c_{k/2,\sqrt{s} j/2} \sin\big(2\pi (k+n) P \big) \bigg)^2 \nonumber\\
    &\le \sum_m \bigg(\sum_k 1 \bigg)^2 \ll R \sum_{m,k} 1 \ll R\, V_\mathrm{dg}(R)\ . \label{eq:X variance}
\end{align}
Here, we used that $c_{k/2,\sqrt{s} j/2}$ is a sign and we changed variables back from $j$ to $m=s j^2$. In the last step, we used that the sum only runs over occupied sites, and we have $|d_{P,\bar P}| \ge 1$ for such sites.
Finally, we will also need the simple fact that
\be
|\Phi_s(P,x)| \ll R^2\ , \label{eq:Phi sup bound}
\ee
since both $k$ and $j$ take at most $\mathcal{O}(R)$ values in the definition \eqref{eq:Phi definition}.

\paragraph{Extreme values.} Recall that our goal was to bound the function $\Psi(P,\bar P)$ defined in \eqref{eq:Psi definition}. In the worst case, all the signs from the trigonometric terms in \eqref{eq:Psi definition} line up, in which case $\Psi(P,\bar P)$ is of order the number of occupied dg sites in the box $2 \le P_0,\bar P_0 \le R$. This is much larger than the typical size, which is of order the standard deviation $\sigma$ that we bounded by $\sqrt{R \, V_{\mathrm{dg}}(R)}$ in \eqref{eq:X variance}. For any constant $\delta>0$, let us estimate the likelihood that
\be
\PP\Big(|\mathcal X(P)| \ge \frac{1}{2}\delta \, V_\mathrm{dg}(R)-c R^2 \Big)\ .
\ee
This is precisely the setup of Bennett's inequality. It states that for $Y_1,\dots,Y_N$ independent random variables with $\EE Y_i=0$ and $|Y_i| \le a$ almost surely for all $i$ and $\sigma^2=\sum_{i=1}^N \EE (Y_i^2 )$, we have
\be
\PP \Bigg(\bigg|\sum_{i=1}^N Y_i\bigg| \ge t \Bigg) \le 2\, \exp\bigg(-\frac{\sigma^2}{a^2}\, h\Big(\frac{at}{\sigma^2}\Big)\bigg) \ , \label{eq:Bennett inequality}
\ee
where $h(u)=(1+u)\log(1+u)-u$~\cite{Bennett:1962}.

We apply this with $Y_s=\Phi_s(P,X_s)$ and $t=\frac{1}{2}\delta V_\mathrm{dg}(R)-c R^2$.
In our case, $|\Phi_s(P,X_s)| \le C R^2$ by \eqref{eq:Phi sup bound} for some constant $C$ and thus we can apply Bennett's inequality for $a=CR^2$.
Since the RHS of \eqref{eq:Bennett inequality} is monotonically increasing in $\sigma^2(P)$, we may thus insert the upper bound \eqref{eq:X variance}, $\sigma^2(P) \le C'\,R\, V_\mathrm{dg}(R)$, and apply \eqref{eq:Bennett inequality} with $\sigma^2=C'\,R\, V_\mathrm{dg}(R)$. Here $C'$ is another universal constant.

Assume first $t>0$. We can write $V_\mathrm{dg}(R)=\frac{2}{\delta}(t+cR^2)$, so that
\be
\frac{a t}{\sigma^2}=\frac{C R\, t}{C' V_\mathrm{dg}(R)}=\frac{C \delta R}{2C'}\cdot
\frac{t}{t+c R^2}\ .
\ee
Using $h(u) \ge \frac{1}{2}u\log u$ and then $\log(1+x) \le x$, the exponent in
\eqref{eq:Bennett inequality} is bounded below by
\begin{align}
\frac{\sigma^2}{a^2}\, h\Big(\frac{at}{\sigma^2}\Big)
&\ge \frac{t}{2a}\log\Big(\frac{at}{\sigma^2}\Big)
\nonumber\\
&=\frac{t}{2C R^2}\bigg[\log\Big(\frac{C\delta R}{2C'}\Big)-\log\Big(1+\frac{cR^2}{t}\Big)\bigg]
\nonumber \\
&\ge \frac{t}{2C R^2}\,\log\Big(\frac{C\delta R}{2C'}\Big)-\frac{c}{2C}
\nonumber \\
&\ge\ \frac{1}{8C}\,\frac{\delta\, V_\mathrm{dg}(R)-2cR^2}{R^2}\,\log R-\frac{c}{2C}\ ,
\end{align}
where the last step holds for $R \ge (\frac{2C'}{C \delta})^2$ and we substituted back $t=\frac{1}{2}\delta V_\mathrm{dg}(R)-c R^2$.
We therefore have
\be
\PP\Big(|\mathcal X(P)| \ge \frac{1}{2}\delta \, V_\mathrm{dg}(R) -c R^2\Big) \ll \exp\Bigg(-c' \, \frac{\delta\, V_\mathrm{dg}(R)-2c R^2}{R^2}\log R \Bigg)\ , \label{eq:probability extreme values}
\ee
with the constant $c':=\frac{1}{8C}$ and the implied constant independent of $\delta$. The same bound \eqref{eq:probability extreme values} holds trivially for the $t \le 0$ case, as the inequality becomes vacuous. Therefore, we see that if $V_\mathrm{dg}(R)\ll R^2$ is \emph{violated}, extreme values become very rare.
Also note that the bound \eqref{eq:probability extreme values} holds uniformly in $P$.

We can turn the probability \eqref{eq:probability extreme values} into a bound on the desired quantity \eqref{eq:average extreme}. We are only allowed to use continuous functions in \eqref{eq:equidistribution theorem}, so let us choose an approximate indicator function: a continuous even function $f$ with $0 \le f \le 1$, with $f=1$ on $\{|x| \ge \delta V_\mathrm{dg}(R)-c R^2\}$ and $f=0$ on $|x| < \frac{1}{2}\delta V_\mathrm{dg}(R)-cR^2$. Thus
\begin{align}
&\limsup_{T \to \infty}\frac{1}{T}\, \Big|\Big\{(P,\bar P) \in [0,1]\times [0,T]\ :\ \big|\Psi(P,\bar P)\big| \ge \delta\, V_\mathrm{dg}(R)-c R^2\Big\}\Big| \nonumber\\%
&\qquad\le \limsup_{T \to \infty}\frac{1}{T} \int_0^1 \dd P \int_0^T \dd \bar P\, f(\Psi(P,\bar P))\\
&\qquad=\int_0^1 \dd P \, \EE\big(f(\mathcal{X}(P))\big) \nonumber\\
&\qquad\le \int_0^1 \dd P\ \PP\Big(|\mathcal X(P)| \ge \frac{1}{2}\delta \, V_\mathrm{dg}(R) -cR^2\Big) \nonumber\\
&\qquad\ll \exp\Bigg(-c' \, \frac{\delta\, V_\mathrm{dg}(R)-2c R^2}{R^2}\log R \Bigg)\ . \label{eq:average extreme density upper bound}
\end{align}
Thus, the proportion of values for which $|\Psi(P,\bar P)|$ is large becomes very small.

\paragraph{Comparison and conclusion.} To complete the proof, let us compare the upper bound \eqref{eq:average extreme density upper bound} with the lower bound \eqref{eq:average extreme density lower bound} above. Upon taking the logarithm, this gives
\be
\log\Big(\frac{\delta^2}{BR^5}\Big)\le -c' \, \frac{\delta\, V_\mathrm{dg}(R)-2c R^2}{R^2}\log R+\mathcal{O}(1)\ . \label{eq:exponent comparison}
\ee
Recall that $\delta$, $B$, $c$ and $c'$ are all positive $\mathcal{O}(1)$ quantities. Solving the inequality \eqref{eq:exponent comparison} for $V_\mathrm{dg}(R)$ immediately demonstrates that
\be
V_\mathrm{dg}(R)  \ll R^2\ , \label{eq:Vdg bound}
\ee
where we dropped the implicit dependence of the implicit constant on $B$ (which is independent of $R$).
This establishes the desired bound on the dg family.

An identical argument with left- and right-movers exchanged demonstrates also the same conclusion for the gd family.
Let us also recall that, given \eqref{eq:Vdg bound} and its gd analog, equations \eqref{eq:Vgg bound} and \eqref{eq:Vdd bound} then establish the main claim of this section, namely \eqref{eq:bounded density}.

\section{Modular bootstrap using crystalline measures}\label{sec:modular bootstrap crystalline measures}
Our new line of attack for the $c=1$ bootstrap problem is to reformulate it as a classification problem for the density of states $\rho$ on $\RR^2$, which is a \textit{crystalline measure}. That means it is a tempered distribution such that both itself and its Fourier dual are atomic measures with discrete support.

\subsection{Crystalline measures}
The prototypical example of a crystalline measure is the Dirac comb on $\RR$,
\be
\rho=\sum_{n \in \ZZ} \delta_n\ ,
\ee
where $\delta_n$ is the Dirac measure supported at $n$. This example may suggest that crystalline measures are very orderly objects, but this is far from the truth. Some rather non-trivial examples of crystalline measures were found by Guinand and later discussed by Meyer~\cite{Guinand:1959, Meyer:2016}, as well as recent examples due to Kurasov and Sarnak~\cite{Kurasov:2020}.

However, under various additional assumptions on the asymptotic behavior of the density of states, it has been possible to classify crystalline measures. For example, on $\RR^d$, under the additional assumptions that (i) $\rho$ and its Fourier dual are tempered  measures, (ii) both have uniformly discrete support\footnote{A set $S$ is \textit{uniformly discrete} if
\be
\inf_{\begin{subarray}{c} v,\, v' \in S \\ v \ne v' \end{subarray}} |v-v'|>0\ ,
\ee
i.e.\ the distance between different elements in the set is bounded from below.} and (iii) $\rho$ is positive if $d\geq 2$, Lev and Olevskii~\cite{Lev:2015} showed that  $\rho$ must be a \textit{generalized Dirac comb},
\be
\rho = \sum_i c_i \sum_{v \in \Lambda_i} \e^{2\pi i \langle v,w_i \rangle_\mathrm{L}} \ \delta_{v_i+v} \ , \qquad v_i,w_i\in \RR^d \ , \ c_i\in \CC \ , \label{eq:genDirac}
\ee
that is, a finite linear combination of Dirac combs on rank-$d$ lattices $\Lambda_i$, possibly translated in position and frequency domain.\footnote{In fact, Lev-Olevskii showed more strongly all of the lattices can be taken to be the same, $\Lambda_i=\Lambda$, under these assumptions.} The class of generalized combs is closed under Fourier transformation, with the Fourier dual of \eqref{eq:genDirac} as defined in \eqref{eq:FourierDef} being
\be
  \mathbb{S}\rho = \sum_i \frac{c_i}{\covol(\Lambda_i)}
  \sum_{u \in \Lambda_i^\vee} \e^{2\pi i \langle -w_i+ u,v_i \rangle_\mathrm{L}} \, \delta_{-w_i+u} \ ,
\ee
where the dual lattice $\Lambda_i^\vee:=\{x\in\RR^d \ | \ \langle x,y\rangle_\mathrm{L} \in \ZZ\text{ for all }y\in\Lambda_i\}$ and $\covol(\Lambda_i)$ is the volume of the lattice's unit cell as measured by the pairing $\langle\bullet,\bullet\rangle_\mathrm{L}$.
In our case of $d=2$, the pairing will be taken to be the Lorentzian one defined in \eqref{eq:LIP}, but in general it may be an arbitrary non-degenerate bilinear form.

In our case, we are in very good shape to get analytic control over the $c=1$ modular bootstrap, as we have established in Section~\ref{sec:growth} that $\rho$ has upper bounded density. Although this is strictly weaker than uniform discreteness assumption of Lev-Olevskii (which is not satisfied by the symmetrized free boson CFT density of states in the irrational cases $R^2 \notin \QQ$),  it will still be sufficient to prove a similar classification. This weakened assumption, and the lack of positivity, is compensated here by the additional assumption of integer degeneracies, which will play a crucial role. The rest of this section is devoted to proving and applying the following theorem:
\begin{theorem} \label{thm:improved Meyer theorem}
    Let $\rho$ be a tempered, atomic measure $\rho=\sum_{v \in \supp(\rho)} a_v\, \delta_v$ on $\RR^d$. Assume furthermore that $\{a_v\}_{v \in \supp(\rho)}$ are integers and that the Fourier transform $\mathbb{S}\rho$ is a measure of upper bounded density. Then $\rho$ is a finite sum of combs
    \begin{align}
        \rho=\sum_{i} b_i \, \delta_{v_i+\Lambda_i} \label{eq:rho-Dirac-comb-sum}
    \end{align}
    with $\Lambda_i$ lattices of rank $\rank \Lambda_i \in \{0,\dots,d\}$, $v_i \in \RR^d$ and $b_i \in \ZZ$.
\end{theorem}
Here and in the following, we use the notation
\be
\delta_{v+\Lambda}:=\sum_{w \in v+\Lambda} \delta_w \label{eq:comb-def}
\ee
for the formal sum of unit atoms on a coset $v+\Lambda$ of a subgroup $\Lambda \subset \RR^d$. Upper boundedness of a measure $\mu$ on $\RR^d$ means $|\mu|(B_R(0))\leq C_\mu R^d$, generalising the $d=2$ case \eqref{eq:bounded density}.

This is a variation of a theorem of Meyer~\cite[pp.~25--26]{Meyer:1970} in the one-dimensional case and of C\'ordoba \cite{Cordoba:1989} in $d$ dimensions, both with stronger assumptions. The special case of unit masses and positive Fourier transform, where the conclusion is that the support is a single lattice, was treated in~\cite{Cordoba:1988}.
The version of the theorem above is essentially discussed in~\cite{Kolountzakis:1996,Kolountzakis:2000}, except that we assume that $\{a_v\}_{v \in \supp(\rho)}$ take integer values, while they assumed that they take values in a finite set (which follows a posteriori from the theorem). The proof for both versions goes through the same steps.

Let us also mention that the assumption of upper bounded density on $\mathbb{S}\rho$ is necessary, e.g.\ the examples of Kurasov and Sarnak \cite{Kurasov:2020} are crystalline measures that are not finite sums of Dirac combs. They satisfy every assumption of the theorem, except that $\mathbb{S}\rho$ is not of upper bounded density.

We will apply this theorem below to the modular bootstrap problem. Let us note that the theorem only assumes temperedness, atomicity of $\rho$, upper bounded density of $\mathbb{S} \rho$ and integrality of the degeneracies, i.e.\ only part of the axioms \ref{axiom:S invariance}--\ref{axiom:vacuum} are necessary to deduce this. It is surprising to us that invariance under S-transformations is only used in a relatively weak form, while positivity and spin quantization play no role in this proof. Those assumptions are only used via \eqref{eq:bounded density}, which we showed in Section~\ref{sec:growth} to follow from the axioms.

While the proof techniques of this theorem are rather standard in mathematics, to our knowledge, they have not previously appeared in the physics literature. We thus give an account of the proof strategy below in Section~\ref{subsec:Reduction to lattices}.

\subsection{Reduction to lattices (proof of Theorem \ref{thm:improved Meyer theorem})}\label{subsec:Reduction to lattices}
The proof of the theorem relies on harmonic analysis on compact abelian groups, specifically Cohen's idempotency theorem, which characterizes measures whose Fourier transform only takes integer values. Throughout the proof we normalize the bilinear form $\langle \bullet,\bullet \rangle_\mathrm{L}$ entering \eqref{eq:FourierDef} such that its Gram matrix has determinant of absolute value $4$, as is the case for \eqref{eq:LIP}. This is precisely what makes $\mathbb{S}$ square to the reflection $x \mapsto -x$, and it is no restriction: rescaling the bilinear form amounts to a dilation of $\mathbb{S}$, under which neither the hypotheses nor the conclusion of Theorem~\ref{thm:improved Meyer theorem} change.

\paragraph{Measures on compact abelian groups.} Let $G$ be a compact abelian group. Let us recall some basic definitions. We denote by $M(G)$ the space of all complex-valued regular Borel measures.\footnote{That is, the underlying $\sigma$-algebra of measurable sets is the Borel $\sigma$-algebra of the topology. Regularity of the measure $\mu$ means both inner regularity, $|\mu|(E)=\sup\{|\mu|(K)\mid K \subset E \text{ compact}\}$ and outer regularity, $|\mu|(E)=\inf\{|\mu|(U)\mid U \supset E \text{ open}\}$, where $|\mu|$ is the total variation measure of $\mu$, defined by $|\mu|(E)=\sup \sum_i |\mu(E_i)|$, where $E=\sqcup_i E_i$ and the supremum runs over all such finite disjoint partitions.} For two measures $\mu$, $\nu$ on $G$ we can define addition $\mu+\nu$ as well as convolution $\mu*\nu$ by
\be
\int_G f \,  \dd (\mu * \nu)=\int_G \int_G f(x+y) \, \dd \mu(x) \, \dd \nu(y)\ .
\ee
$M(G)$ becomes a Banach algebra with the norm
\be
\lVert \mu \rVert=|\mu|(G)\ .
\ee
We can also define the Fourier transform $\hat{\mu}$ of a measure $\mu \in M(G)$. $\hat{\mu}$ is a continuous function on the Pontryagin dual group $\hat{G}=\Hom(G,\Uone)$.\footnote{For a compact group, the Pontryagin dual is discrete and continuity of a function on it is automatic.} For $\gamma \in \hat{G}$ a character, one defines
\be
\hat{\mu}(\gamma)=\int_G \bar{\gamma} \, \dd \mu\ . \label{eq:FTtransf}
\ee
The complex conjugation is conventional and mimics the standard conventions of the Fourier transform on $\RR^d$. A simple calculation gives
\be
\widehat{\mu * \nu}=\hat{\mu} \hat{\nu}\ , \label{eq:convolution-to-product Fourier transform}
\ee
i.e.\ convolution gets translated to multiplication of continuous functions under Fourier transformation.

\paragraph{Cohen's idempotency theorem.} Cohen's idempotency theorem~\cite{Cohen:1960} characterizes measures $\mu$ whose Fourier transforms $\hat{\mu}$ takes only integer values.\footnote{A slightly weaker formulation that is also often found in the literature treats the case where $\hat{\mu}$ only takes the values $0$ and $1$, in which case $\hat{\mu}^2=\hat{\mu}$ and via \eqref{eq:convolution-to-product Fourier transform} also $\mu * \mu=\mu$. Such measures are called idempotent, which gives the theorem its name~\cite{Rudin:1962, Cohen:1960}.}\textsuperscript{,}\footnote{The theorem also extends to the case of finite measures on a locally compact abelian groups, but we will not need it. The reduction is described in~\cite{Rudin:1959}.} The standard reference on the subject is~\cite{Rudin:1962} and a short proof can be found in~\cite{Ito:1964}. The simplest such measure is the Haar measure of a compact subgroup $H \subset G$, normalized such that $m_H(G)=m_H(H)=1$. The corresponding Fourier transform is\footnote{We use mathematics conventions in which the inner product $\langle\bullet,\bullet\rangle_H$ is antilinear in its second argument.}
\be
\widehat{m_H}(\gamma)=\int_G \bar{\gamma} \, \dd m_H=\int_H \bar{\gamma}|_H \, \dd m_H=\langle 1, \gamma|_H \rangle_H\ ,
\ee
the standard inner product of characters on $H$. Thus $\widehat{m_H}=\id_{H^\perp}$,
where the annihilator of $H$,
\be
H^\perp=\{\gamma \in \hat{G} \mid \gamma|_H\equiv 1\} \subset \hat{G}
\ee
is an open subgroup of $\hat G$.
$\widehat{m_H}$ therefore only takes the values 0 and 1. We can also multiply $m_H$ by a character $\gamma$, which in Fourier space translates the support set $H^\perp$ of the indicator function by $\gamma$ and thus preserves idempotency.
We can consider the following finite linear combination of measures
\be
\mu=\sum_{i} b_i \, \gamma_i \,  m_{H_i} \label{eq:mu canonical form Cohen}
\ee
with $b_i \in \ZZ$, $\gamma_i \in \hat{G}$ a character and $m_{H_i}$ the Haar measure of a compact subgroup. Linearity of the Fourier transform implies that
\be
\hat{\mu}=\sum_i b_i \, \id_{\gamma_i+H_i^\perp}
\label{eq:muHatCohen}
\ee
and thus $\hat{\mu}$ only takes integer values.

Cohen's theorem states the converse: \emph{Every} measure such that $\hat{\mu}$ takes only integer values is of the form \eqref{eq:mu canonical form Cohen}. This is very non-trivial, since one might have thought that any integer linear combination of arbitrary indicator functions would work in place of \eqref{eq:muHatCohen}. But in fact, unless they are of this form, they are not realized as the Fourier transform of a measure on $G$.

\paragraph{The Bohr compactification.} This is close to the statement we wanted. However, we want to apply this to $G=\RR^d$ and the crystalline measure $\rho$. But $G$ is not compact and $\rho$ is not a finite measure.
To apply Cohen's theorem, we first have to pass to a different \emph{compact} abelian group: the Bohr compactification $\mathrm{b}\RR^d$.

For a \emph{locally compact} abelian group $G$ (such as $\RR^d$), the Bohr compactification is defined abstractly as
\be
\mathrm{b}G=\widehat{\hat{G}_\mathrm{disc}} \ ,
\ee
i.e.\ the double dual of the group, but where we use the discrete topology on $\hat{G}_\mathrm{disc}$. Since the Pontryagin dual of a discrete group is compact, this defines a compact group. As we shall discuss below, the resulting group for $G=\RR^d$ is huge. This means it is much more difficult for a function on $\mathrm{b}G$ to be continuous: as we explain below, the continuous functions on $\mathrm{b}G$ are in one-to-one correspondence with \textit{almost periodic} functions on $G$, which are determined by a discrete set of Fourier modes. The dual space of finite measures on $\mathrm{b}G$ is accordingly much bigger, and in fact an infinite measure on $G$ can uplift to a finite one on $\mathrm{b}G$. Indeed, we shall show in our case, with $\rho$ atomic and $\mathbb{S}\rho$ of upper bounded density, that $\rho$ uplifts canonically to a finite measure on $\mathrm{b}G$---to which Cohen's theorem applies.

Let us disentangle the definition of $\mathrm{b}G$ to make more sense of it. An element of $\mathrm{b}G$ is an arbitrary homomorphism $\varphi: \hat{G} \to \Uone$ (with continuity being automatic in the discrete topology on $\hat G$). Notice that the set of functions
\be
\mathrm{Map}(\hat{G},\Uone)=\prod_{\gamma \in \hat{G}} \Uone_\gamma
\ee
is an infinite (possibly uncountable) product of compact spaces and thus compact by Tychonoff's theorem. The Bohr compactification is the closed subgroup defined by $\varphi \in \mathrm{Map}(\hat{G},\Uone)$ with $\varphi(\gamma+\eta)=\varphi(\gamma)\varphi(\eta)$ for $\gamma,\, \eta \in \hat{G}$.

There is a canonical dense injective homomorphism $\mathrm{b}: G \to \mathrm{b}G$, defined by evaluation $g \mapsto e_g$ with $e_g(\gamma)=\gamma(g)$. In other words, points $g\in G$ are mapped to the values of all characters at $g$. Now consider continuous functions on $\mathrm{b}G$: a large family are just the characters themselves. By the Stone-Weierstrass theorem, any continuous function on $\mathrm{b}G$ is in fact a uniform limit of linear combinations of the characters. This is essentially the statement that there is a correspondence $f=(\mathrm{b}f) \circ \mathrm{b}$ between continuous functions $\mathrm{b}f:\mathrm{b}G\to\mathbb{C}$ and almost periodic functions $f:G\to\mathbb{C}$, i.e.\ $f$ factors through $\mathrm{b}G$. The almost periodic functions $\AP(G)$ on $G$ can be defined as exactly the uniform limits of trigonometric polynomials (finite linear combinations of characters),
\be
\sum_{n=1}^N a_n \gamma_n\ ,
\ee
where $\gamma_n \in \hat{G}$ are characters and $a_n \in \CC$. Intuitively, such functions repeat arbitrarily well---but not necessarily exactly. A non-trivial example on $\RR$ is $f(x)=\e^{2\pi i x}+\e^{2\pi i\sqrt{2}x}$. Since we may approximate $\sqrt{2}$ by rational numbers $\frac{p}{q}$ to arbitrary precision, the function is almost periodic under $x \mapsto x+q$. The main point is that almost periodic functions can be specified by the collection of their frequencies, which is captured by the Bohr compactification.

\paragraph{Uplifting measures.}
The final ingredient is to show that $\rho$ uplifts to a finite measure on the Bohr
compactification $\mathrm{b}\RR^d$, so that the version of Cohen's theorem just discussed becomes
applicable. We closely follow the argument of Kolountzakis and Lagarias
\cite{Kolountzakis:1996, Kolountzakis:2000}.

Let $\rho=\sum_{v \in \supp(\rho)} a_v\, \delta_v$ be an atomic tempered measure on $\RR^d$
whose Fourier transform $\mathbb{S}\rho$ is again a measure of upper bounded density,
$|\mathbb{S}\rho|(B_R(0)) \le C_\rho\, R^d$ for $R \ge 1$.
Fix a bump function $\phi \in C_\mathrm{c}^\infty(B_1(0))$ normalized to $\phi(0)=1$. Its
Fourier transform is a Schwartz function and hence decays faster than any power,
$|\mathbb{S}\phi(\xi)| \le C_\alpha\, \lVert \xi\rVert^{-\alpha}$ for every $\alpha>0$.
We then smear the measure $\rho$ with such bump functions to produce a sequence of smooth functions on $\RR^d$. For $n \in \NN$, set
\be
\rho_n(x):=\int_{\RR^d} \phi\big(n(x-v)\big)\, \dd\rho(v)=\sum_{v \in \supp(\rho)} a_v\, \phi\big(n(x-v)\big)\ . \label{eq:rho-n-def}
\ee
Only atoms with $\lVert x-v \rVert<\frac{1}{n}$ contribute, so that the sum converges absolutely. The same argument holds for arbitrary derivatives so that $\rho_n$ is smooth. Since $|\rho_n(x)| \le \lVert \phi \rVert_\infty\, |\rho|\big(B_{1/n}(x)\big)$ grows at most polynomially in $x$, $\rho_n \equiv \rho_n(x)\, \dd^d x$ defines a tempered measure on $\RR^d$.
Thus we can consider the Fourier transform $\mathbb{S}\rho_n$, which is a priori defined as a tempered distribution. Since convolution is translated to multiplication in Fourier space, $\mathbb{S}\rho_n$ can be checked to equal
\be
\mathbb{S}\rho_n=\frac{1}{2n^d}\, \mathbb{S}\phi\big(\tfrac{\bullet}{n}\big)\, \mathbb{S}\rho\ .
\ee
The factor of $\frac{1}{2}$ originates from the conventional factor of $2$ in \eqref{eq:FourierDef}---and is immaterial for the following proof. Thus $\mathbb{S}\rho_n$ is actually a tempered measure on $\RR^d$.

The purpose of the smearing is that $\mathbb{S}\rho_n$ are \emph{finite} measures, with a
bound uniform in~$n$:
\be
|\mathbb{S}\rho_n|(\RR^d) \le C\ . \label{eq:uniform-bound-mun}
\ee
To see this, fix any $\alpha>d$ and split $\RR^d$ into the ball $B_n(0)$ and the shells $S_k=\{2^k n \le \lVert \xi\rVert < 2^{k+1}n\}$, $k \ge 0$. On
$B_n(0)$ we bound $\mathbb{S}\phi$ by its supremum and use upper bounded density. On $S_k$,
where $\lVert \xi/n\rVert \ge 2^k$, we use the decay of $\mathbb{S}\phi$ and upper boundedness:
\be
|\mathbb{S}\rho_n|(B_n(0)) \le C_\rho\lVert \mathbb{S}\phi\rVert_\infty\ ,\qquad
|\mathbb{S}\rho_n|(S_k) \le C_\rho C_\alpha\, 2^d\, 2^{k(d-\alpha)}\ ,
\ee
and summing the convergent geometric series over $k$ yields
\eqref{eq:uniform-bound-mun}. This is the only step in which upper bounded
density enters.
    
Since $\mathrm{b}\RR^d$ is compact, the Riesz representation theorem identifies
$M(\mathrm{b}\RR^d)$ with the dual of $C(\mathrm{b}\RR^d)=\AP(\RR^d)$
\cite{Rudin:1991}. Each finite measure $\mathbb{S}\rho_n$ defines a functional on
$C(\mathrm{b}\RR^d)$ by
\be
L_n(F):=2\int_{\RR^d} (F \circ \mathrm{b})\, \dd (\mathbb{S}\rho_n)\ , \qquad F \in C(\mathrm{b}\RR^d)\ , \label{eq:L-n}
\ee
of operator norm at most $2|\mathbb{S}\rho_n|(\RR^d)\le 2C$. The factors of 2 match our normalization of the Fourier transform in \eqref{eq:FourierDef}.

We can then define a limiting measure as $n\to\infty$ as follows. Recall that characters $\chi_v = \e^{2\pi i \langle v,\bullet\rangle_\mathrm{L}}$ on $\RR^d$ have a canonical embedding into $C(\mathrm{b}\RR^d)$, and their span is dense. First consider the action of $L_n$ on a character $F = \bar{\chi}_v=\chi_{-v}$ (with the complex conjugation for later convenience). 
Using that $\mathbb{S}$ squares to the reflection $x \mapsto -x$, we have
\begin{multline}
L_n(\bar{\chi}_v)=2\int_{\RR^d} \e^{-2\pi i \langle v,\xi\rangle_\mathrm{L}}\,
\dd (\mathbb{S}\rho_n)(\xi)=\big(\mathbb{S}^2\rho_n\big)(-v)=\rho_n(v)\\
=a_v\, \id_{\supp(\rho)}(v) \ , \quad  n \gg 1\ .\label{eq:L limit}
\end{multline}
In the last step we used that $\supp(\rho)$ is locally finite, and assumed that $n$ is sufficiently large, so that \eqref{eq:rho-n-def} only selects the single atom at $v$.\footnote{Local finiteness follows from temperedness, and the assumption in Theorem~\ref{thm:improved Meyer theorem} that the coefficients are integers. \label{footnotes:local finiteness}}
Thus, $L_n(\bar{\chi}_v)$ is eventually constant and in particular convergent.

The finite span of the characters is the subspace of trigonometric polynomials inside $C(\mathrm{b}\RR^d)$. By linearity, we may therefore define
\begin{equation}
    L(T) := \lim_{n\to\infty} L_n(T)
\end{equation}
for any trigonometric polynomial $T$. By the uniform bound on the norms of $L_n$, this defines a bounded linear functional on this subspace. As stated above, the trigonometric polynomials are dense in $C(\mathrm{b}\RR^d)$. Hence, $L$ extends uniquely to a bounded linear functional on all of $C(\mathrm{b}\RR^d)$,\footnote{The unique extension of continuous functionals from dense subspaces is true in any topological vector space.}  equivalently characterized by a finite measure $\mu \in M(\mathrm{b}\RR^d)$.

The limiting measure $\mu$ is indeed an uplift of $\rho$ to the Bohr compactification, in the sense that its values on characters \eqref{eq:L limit} read off the coefficients of the atoms of $\rho$. More directly, by a similar computation, its Fourier transform in the sense of \eqref{eq:FTtransf} is
\be
\hat{\mu}(v)=\int_{\mathrm{b}\RR^d} \bar{\chi}_v \, \dd \mu=\lim_{n \to \infty}L_n(\bar{\chi}_v)
=a_v\, \id_{\supp(\rho)}(v) = \rho(\{v\})\ . \label{eq:hat mu Fourier}
\ee
This is a continuous function on $\RR^d_\mathrm{disc}=\widehat{\mathrm{b}\RR^d}$ whose values are the same as the coefficients of atoms of $\rho$.

Since $\hat{\mu}$ only takes integer values (as $a_v\in \ZZ$ by assumption), then Cohen's theorem applies to the measure $\mu$ on $G=\mathrm{b}\RR^d$. It states that
\be
\hat{\mu}=\sum_i b_i \, \id_{v_i+\Lambda_i}\ , \qquad b_i\in \ZZ \ , \quad  v_i \in \RR^d \label{eq:bExp}
\ee
is a finite sum of coset indicator functions of open subgroups $\Lambda_i =H_i^\perp \subset \RR^d_\mathrm{disc}$, i.e.\ arbitrary subgroups $\Lambda_i\subset \RR^d$. By \eqref{eq:hat mu Fourier}, we have thus fixed the form of $\rho$ to
\be
\rho(\{v\})=\sum_i b_i \, \id_{v_i+\Lambda_i}(v)  \ .\label{eq:rho-abstract-combs}
\ee

\paragraph{From arbitrary subgroups to lattices.} Equation \eqref{eq:rho-abstract-combs} is almost the statement of Theorem~\ref{thm:improved Meyer theorem}. However, $\Lambda_i$ in the theorem should be lattices, i.e.\ discrete subgroups of $\RR^d$, while the subgroups appearing in \eqref{eq:rho-abstract-combs} are \emph{arbitrary} subgroups, including $\QQ$ or other non-discrete groups.

It remains to show that only discrete subgroups can appear, which follows from temperedness of the measure $\rho$. As remarked in footnote~\ref{footnotes:local finiteness}, temperedness implies local finiteness. A coset $v+\Lambda$ is  locally finite precisely when $\Lambda$ is discrete, i.e.\ a lattice. The non-lattice terms in \eqref{eq:rho-abstract-combs} must therefore conspire to cancel out.

In fact, one may choose the representation in \eqref{eq:rho-abstract-combs} such that only lattices appear to begin with. This was established in~\cite[Theorem 3]{Kolountzakis:2000}. The proof follows from elementary group theory and we therefore do not repeat it here. This implies that $\rho$ is a sum of combs on lattice cosets~\eqref{eq:rho-Dirac-comb-sum}, completing the proof of Theorem~\ref{thm:improved Meyer theorem}.

\subsection{Commensurability classes} \label{subsec:commensurability}
We now turn back to the modular bootstrap problem and will use Theorem~\ref{thm:improved Meyer theorem} to further constrain $\rho$. Strictly speaking, the multiplicities of $\rho$ are only quarter integers, see \eqref{eq:definition density of states}. Thus we should apply the theorem to $4\rho$ and the conclusion of the theorem holds with $b_i \in \frac{1}{4}\ZZ$. We will now use other axioms of the bootstrap to further constrain the possible form of the density of states.

So far, we have not even assumed that $\mathbb{S}\rho$ is atomic, so we cannot do much better. We will now make the first improvement by imposing atomicity of $\mathbb{S}\rho$.

\paragraph{Commensurability.} We first recall the notion of \emph{commensurability}. Let $\Lambda_1$ and $\Lambda_2$ be two full-rank lattices in $\RR^d$. Then $\Lambda_1$ and $\Lambda_2$ are called commensurable if $\Lambda_1 \cap \Lambda_2$ also has full rank. This is an equivalence relation and the equivalence classes are called commensurability classes. Intuitively, two lattices are commensurable if they only differ by a finite amount.

We can then refine Theorem~\ref{thm:improved Meyer theorem} as follows. Under the assumptions of the theorem and the additional assumption that $\mathbb{S}\rho$ is atomic, $\rho$ takes the form of a finite sum
\be
\rho=\sum_{\mathrm{c}} \sum_{n} b_{\mathrm{c},n} \, \delta_{v_{\mathrm{c},n}+\Lambda_\mathrm{c}}\ ,\label{eq:rhoComb}
\ee
where each $\rank \Lambda_c=d$, and $\Lambda_\mathrm{c}$ and $\Lambda_{\mathrm{c}'}$ are incommensurable for $\mathrm{c} \ne \mathrm{c}'$. Moreover, we can assume that $\bigcup_\mathrm{c}\bigcup_n (v_{\mathrm{c},n}+\Lambda_\mathrm{c}) \setminus \supp(\rho)$ is contained in a finite union of cosets of rank at most $d-1$.

Note that these assumptions are satisfied for the modular bootstrap on $\RR^2$ for the rescaled density of states $4\rho$ with the axioms \ref{axiom:S invariance}--\ref{axiom:vacuum}. In particular, (i) $\rho$ is atomic by assumption of discrete spectrum, (ii) $\mathbb{S}\rho=\rho$ is also atomic by \ref{axiom:S invariance}, (iii) the degeneracies of $4\rho$ are integer by \ref{axiom:integrality} and \eqref{eq:definition density of states} and (iv) the axioms together imply upper bounded density (by Section~\ref{sec:growth}).

\paragraph{Proof of \eqref{eq:rhoComb}.} To prove \eqref{eq:rhoComb}, we first show that $\rank \Lambda_i=d$ in the representation \eqref{eq:rho-Dirac-comb-sum} for all $\Lambda_i$.
Write $\rho=\rho_d+\rho_{<d}$, where $\rho_d$ contains the full-rank pieces of \eqref{eq:rho-Dirac-comb-sum} and $\rho_{<d}$ all lower-rank pieces. We then have
\be
\mathbb{S}\rho_{<d}=\mathbb{S}\rho-\mathbb{S}\rho_d\ . \label{eq:Fourier-transform-atomic-split}
\ee
The Fourier transform of a full-rank Dirac comb is atomic. In contrast, the Fourier transform of a lower-rank Dirac comb is atomless, meaning that it does not contain any point masses. This also carries over to finite sums of lower-rank Dirac combs. Thus the RHS of \eqref{eq:Fourier-transform-atomic-split} is atomic, while the LHS is atomless. This implies that both the left and right-hand side of \eqref{eq:Fourier-transform-atomic-split} have to vanish separately. Thus $\mathbb{S}\rho_{<d}=0$ and therefore also $\rho_{<d}=0$.

Let us now partition the representation \eqref{eq:rho-Dirac-comb-sum} into commensurability classes. For each appearing commensurability class $\mathrm{c}$, consider $\Lambda_c=\bigcap_{i \in \mathrm{c}} \Lambda_i$, which has finite index in each $\Lambda_i$, so that the cosets of $\Lambda_i$ can be represented by finite unions of cosets of $\Lambda_\mathrm{c}$. This gives the representation \eqref{eq:rhoComb}.

For the statement on the support, notice that we may assume that $(v_{\mathrm{c},n}+\Lambda_{\mathrm{c}})\cap (v_{\mathrm{c},n'}+\Lambda_{\mathrm{c}})=\varnothing$ for $n \ne n'$, since otherwise we could merge the terms in \eqref{eq:rhoComb}. By definition also $(v_{\mathrm{c},n}+\Lambda_\mathrm{c})\cap (v_{\mathrm{c}',n'}+\Lambda_{\mathrm{c}'})$ has at most rank $d-1$ for $\mathrm{c} \ne \mathrm{c}'$, since they are incommensurable. Thus $(v_{\mathrm{c},n}+\Lambda_\mathrm{c})\cap (v_{\mathrm{c}',n'}+\Lambda_{\mathrm{c}'})$ has at most rank $d-1$ unless $(\mathrm{c},n)=(\mathrm{c}',n')$.
For fixed $(\mathrm{c},n)$ let $x \in (v_{\mathrm{c},n}+\Lambda_\mathrm{c}) \setminus \bigcup_{(\mathrm{c}',n') \ne (\mathrm{c},n)} (v_{\mathrm{c}',n'}+\Lambda_{\mathrm{c}'})$. Since $x$ appears only in one of the terms in the sum \eqref{eq:rhoComb}, we necessarily have $x \in \supp(\rho)$. Equivalently,
\be
(v_{\mathrm{c},n}+\Lambda_\mathrm{c}) \setminus \supp(\rho) \subset  \bigcup_{(\mathrm{c}',n') \ne (\mathrm{c},n)} (v_{\mathrm{c}',n'}+\Lambda_{\mathrm{c}'}) \cap (v_{\mathrm{c},n}+\Lambda_\mathrm{c})\ .
\ee
Thus it follows that $\bigcup_{(\mathrm{c},n)} (v_{\mathrm{c},n}+\Lambda_\mathrm{c}) \setminus \supp \rho$ is contained in the union of all pairwise intersections, which is indeed contained in a finite union of cosets of rank at most $d-1$.

\subsection{Integral lattices}
We now continue to restrict the lattices appearing in \eqref{eq:rhoComb} further for the case of the modular bootstrap. Thus we will set $d=2$ in the following. Recall that for the modular bootstrap, the coefficients $b_{\mathrm{c},n}$ in \eqref{eq:rhoComb} take values in $\frac{1}{4}\ZZ$.
In this subsection, we will show that the problem can be treated separately for each commensurability class and that the corresponding lattices must be integral lattices.

First note that for any rank-2 coset $x+\Lambda$ appearing in the above sum, and any finite set of vectors $u_1,\ldots, u_r$ such that the cosets $x+u_i+\Lambda$ also appear in the sum, one can shift $x$ by an element of $\Lambda$ so that
\be
x+u_1, \, \ldots\, ,\, x+u_r \in \supp(\rho) \ .
\ee
This is because almost all of each such coset is contained in $\supp(\rho)$ by what we explained in Section~\ref{subsec:commensurability}. We will call this a `generic' choice for the point $x$.

\paragraph{Integrality.} We claim that the lattices $\Lambda_\mathrm{c}$ in \eqref{eq:rhoComb} are integral,
\be
\langle \Lambda_\mathrm{c},\Lambda_\mathrm{c} \rangle_\mathrm{L} \subset \ZZ\ ,
\ee
where $\langle \bullet, \bullet \rangle_\mathrm{L}$ is the Lorentzian inner product defined in \eqref{eq:LIP}.

To see this, for any $v,\, v'\in\Lambda_\mathrm{c}$, choose a generic point $x$ so that  $x$, $x+v$, $x+v'$ and $x+v+v'$ all lie in $\supp(\rho)$. By spin quantization, each of these vectors' norms lie in $2\ZZ$, and hence
\be
2\langle v,v' \rangle_\mathrm{L}=\lVert x+v+v' \rVert_\mathrm{L}^2-\lVert x+v \rVert_\mathrm{L}^2-\lVert x+v' \rVert_\mathrm{L}^2+\lVert x \rVert_\mathrm{L}^2 \in 2\ZZ\
\ee
as required.

In terms of the dual lattice
\be
\Lambda^\vee:=\{v \in \RR^2 \mid \langle v,\lambda \rangle_\mathrm{L} \in \ZZ \text{ for all $\lambda\in \Lambda$}\} \ ,
\ee
an equivalent statement is that $\Lambda_\mathrm{c} \subset \Lambda_\mathrm{c}^\vee$. In particular, this means $\Lambda_\mathrm{c}$ and $\Lambda_\mathrm{c}^\vee$ are commensurable, since they are lattices of the same rank and one is contained in the other.

\paragraph{Self-duality of each commensurability class.}
We claim that the contribution in each commensurability class
\be
\rho = \sum_\mathrm{c} \rho_\mathrm{c} \ , \qquad \rho_\mathrm{c} = \sum_n b_{\mathrm{c},n} \, \delta_{v_{\mathrm{c},n}+\Lambda_\mathrm{c}}
\ee
is individually self-dual,
\be
\mathbb{S} \rho_\mathrm{c} = \rho_\mathrm{c}  \ . \label{eq:rhocSD}
\ee
Here we take each $\rho_\mathrm{c}$ to be even under $(P,\bar P)\to(-P,-\bar P)$, replacing it by its symmetrization if it is not.\footnote{Notice that the chiral reflection $(P,\bar P) \to (P,-\bar P)$ does in general not preserve the commensurability class and thus $\rho_\mathrm{c}$ is in general not separately even in $P \to -P$ and $\bar P \to - \bar P$.} To show this, define $\Delta_\mathrm{c}:= \mathbb{S}\rho_\mathrm{c} - \rho_\mathrm{c}$. These sum to zero, $\sum_\mathrm{c}\Delta_\mathrm{c} = 0$, and we will show that each $\Delta_c=0$.

First, $\rho_\mathrm{c}$ has support contained in $\cup_n(v_{\mathrm{c},n}+\Lambda_\mathrm{c})$, and its dual $\mathbb{S}\rho_\mathrm{c}$ has support contained in $\Lambda_\mathrm{c}^\vee$. The support of $\Delta_\mathrm{c}$ is therefore contained in $S_\mathrm{c}:= \cup_n(v_{\mathrm{c},n}+\Lambda_\mathrm{c}) \cup \Lambda_\mathrm{c}^\vee$. This is a union of cosets of lattices commensurable with $\Lambda_\mathrm{c}$, and is in particular atomic.

Now consider any atom where $\Delta_\mathrm{c}$ has support. It would have to be canceled by an atom of $\Delta_{\mathrm{c}'}$ for some other $\mathrm{c}'\neq \mathrm{c}$.  Hence it actually belongs to $S_\mathrm{c}\cap S_{\mathrm{c}'}$. But since the cosets in $S_\mathrm{c}$ and $S_{\mathrm{c}'}$ are incommensurable with each other, then their intersections are either empty or cosets of rank at most 1. Hence the support of $\Delta_\mathrm{c}$ is contained in a finite union of cosets of rank at most 1.  $\Delta_\mathrm{c}$ is a finite sum of generalized Dirac combs since $\rho_\mathrm{c}$ is a finite sum of combs.  Hence it can be written as a finite sum of generalized Dirac combs of rank at most 1.\footnote{To see this, one can consider the restrictions of $\Delta_\mathrm{c}$ to the different rank-1 components of its support, which are rank-1 generalized combs. The full $\Delta_\mathrm{c}$ is almost given by the sum of these restrictions, but one has to cancel any intersections by subtracting or adding rank-0 generalized combs (single atoms).}

Since $\mathbb S^2f(x)=f(-x)$ and $\rho_\mathrm{c}$ is even, then $\mathbb{S}^2\rho_c=\rho_\mathrm{c}$. Hence we have $\mathbb{S} \Delta_\mathrm{c} = -\Delta_\mathrm{c}$. But the Fourier transform of generalized Dirac combs of non-maximal rank are atomless, so $\mathbb{S} \Delta_\mathrm{c}$ is atomless. On the other hand $\Delta_\mathrm{c}$ contains atoms if it is non-zero, since it is a finite sum of generalized combs. This gives a contradiction unless $\Delta_c=0$ as desired.

\paragraph{The lattice of periods.} From now on, we focus on a single commensurability class $\rho_\mathrm{c}$, which we just proved is individually self-dual. Define its lattice of periods $K=\{v \in \RR^2 \mid \rho_\mathrm{c}(\bullet+v)=\rho_\mathrm{c}(\bullet)\}$. Also define $M\subset \RR^2$ to be the subgroup  generated by $\supp(\rho_\mathrm{c})$.

We claim that $K$ and $M$ are dual rank-2 lattices:
\be
K = M^\vee \ , \qquad \quad M=K^\vee \ .\label{eq:MKdual}
\ee
To see this, note that $x+K\subset \supp(\rho_\mathrm{c})$  for any $x\in \supp(\rho_\mathrm{c})$,  implying that $K$ is discrete.  Subtracting $x$, it also implies that $K\subset M$. Since $K$ contains $\Lambda_\mathrm{c}$, then it must be a rank-2 lattice. Now since $\rho_\mathrm{c}$ is $K$-periodic, then it is a finite\footnote{Finiteness follows because a fundamental domain of $K$ is finite and since $\rho_\mathrm{c}$ is locally finite then it has finitely many atoms inside it.} linear combination of Dirac combs on $K$-cosets. Hence $\rho_c=\mathbb{S} \rho_\mathrm{c}$ has support inside $K^\vee$, implying $M\subset K^\vee$. Since $K\subset M\subset K^\vee$ and $M$ is a subgroup, then it is a rank-2 lattice.
Now since $\rho_\mathrm{c}$ is supported in $M$, then it follows that $M^\vee$ are periods of $\mathbb{S}\rho_c=\rho_\mathrm{c}$, i.e.\ $M^\vee\subset K$. Dualizing we get $K^\vee\subset (M^\vee)^\vee=M$. In total we have $M\subset K^\vee \subset M$, proving~\eqref{eq:MKdual}.

A consequence is that $K$ is an \textit{even} lattice, i.e.\ $\|k\|_\mathrm{L}^2 \in 2\ZZ$ for all $k\in K$. Indeed, pick a generic point $x\in\supp(\rho)$ so that also $x+k\in\supp(\rho)$. This is possible since $\rho_\mathrm{c}$ is $K$-periodic, so that the coset of $\Lambda_\mathrm{c}$ containing $x+k$ also appears in \eqref{eq:rhoComb}. By spin quantization their norms are even. We have $x\in M$ and $k\in M^\vee$ so that $\langle x,k\rangle_\mathrm{L}\in \ZZ$, and hence
\be
\lVert k \rVert_\mathrm{L}^2=\lVert x+k \rVert_\mathrm{L}^2-\lVert x \rVert_\mathrm{L}^2-2\langle x,k\rangle_\mathrm{L} \in 2\ZZ\ .
\ee

A similar argument shows that spin quantization is satisfied on all of $\supp(\rho_\mathrm{c})$ (not just the part in $\supp(\rho)$). This is because $K$ is a rank-2 lattice of periods of $\supp(\rho_\mathrm{c})$, so only rank-1-sized parts of it can fail to be in the support of $\rho$. Hence for any $y\in \supp(\rho_\mathrm{c})$, there exists $k\in K$ such that $y+k\in \supp(\rho)$. Since $y\in M$ and $k\in M^\vee$ then $\lVert y \rVert_\mathrm{L}^2=\lVert y+k \rVert_\mathrm{L}^2-\lVert k \rVert_\mathrm{L}^2-2\langle y,k\rangle_\mathrm{L} \in 2\ZZ$ as desired.

\subsection{A finite classification problem} We shall now reduce the $c=1$ modular bootstrap to a finite classification problem. The trick is to consider the quotient $G = M/K$. This is a finite abelian group of order $|G| = [M:K]<\infty$, since $K\subset M$ are both rank-2 lattices. Since $M$ has rank 2 then $G$ is generated by at most 2 elements.  For $x,y\in G$, the quantities $\lVert x \rVert_\mathrm{L}^2 \bmod 2$ and $\langle x,y \rangle_\mathrm{L} \bmod 1$ are well-defined (i.e.\ $K$-invariant). This follows because $K$ is even and $\langle K,K^\vee\rangle_\mathrm{L} \subset \ZZ$ in general. In fact $\langle\bullet,\bullet\rangle_\mathrm{L}$ defines a non-degenerate symmetric bi-additive pairing $G\times G \to \QQ/\ZZ$.\footnote{Non-degeneracy follows because if $\langle x,y\rangle\in \ZZ$ for all $y\in K^\vee$, then $x\in (K^\vee)^\vee=K$.}

 Since the measure $\rho_\mathrm{c}$ is supported inside $M$ and is $K$-periodic, then it is captured by a function $f$ on $G$:
\be
\rho_\mathrm{c} = \sum_{x\in G} f(x) \, \delta_{x+K} \ . \label{eq:rho from f}
\ee
The bootstrap problem can now be stated in terms of $f$. Spin quantization says that $f$ is supported on the set of isotropic elements $\mathcal{I}=\{x \in G \mid \lVert x \rVert_\mathrm{L}^2  \equiv 0 \bmod 2\}$. Since $\supp(\rho_\mathrm{c})$ generates $M$, it follows that $\supp(f)$ generates $G$. Finally, $S$ invariance says that
\be
f(y)=\frac{1}{\sqrt{|G|}}\sum_{x \in G} f(x)\, \e^{2\pi i \langle x,y\rangle_\mathrm{L}}\ . \label{eq:finiteS}
\ee
This formula, which is invariance under the standard \textit{finite} Fourier transform on the group $G$, follows from \eqref{eq:rhocSD} and $\mathbb{S} \, \delta_{x+K} =\frac{1}{\sqrt{|G|}} \sum_{u\in K^\vee} e^{2\pi i\langle x,u\rangle_\mathrm{L}}\, \delta_u =\frac{1}{\sqrt{|G|}} \sum_{y\in G} e^{2\pi i\langle x,y\rangle_\mathrm{L}}\, \delta_{y+K}$.

This is the promised finite problem: to classify such group $G$, norms $\|\bullet\|_\mathrm{L}^2$ and functions $f$. This is a special case of a general problem. In standard terminology, the group and $\QQ/2\ZZ$-valued quadratic form $(G,\|\bullet\|_L^2)$ are called the \emph{discriminant form} of the even lattice $K$~\cite{Nikulin:1980}. Discriminant forms of lattices of even signature have a canonical representation of $\SL(2,\ZZ)$ on the complex-valued functions $\CC[G]$ on $G$, called the Weil representation. As we shall discuss, modular $S$ and $T$ invariance forces the function $f$ to be an invariant of the Weil representation. These invariants have been determined in general~\cite{Nebe:2006,Skoruppa:2008,Bieker:2023,Mueller:2025}. In what follows, we shall give a self-contained classification in our case, which is elementary because the group $G$ needs at most two generators.

\paragraph{The group and pairing are determined.} This structure in fact forces the group $G$ and its pairing $\langle \bullet,\bullet\rangle_\mathrm{L}$ to take the form
\be
G\cong \ZZ_N\times \ZZ_N \ , \qquad \langle (a,b),\,(c,d)\rangle_\mathrm{L} = \frac{1}{N}(ad+bc)\ .\label{eq:Gcanon}
\ee
Its only remaining freedom is the natural number $N$. To see this, let us prove two intermediate claims.

First, $G$ can be generated by two isotropic elements, $x_1,x_2\in \mathcal I$. (A priori, it is generated by two elements, and it is generated by the isotropic elements $\supp(f)$, but it could have required more than two isotropic generators.) This follows by the prime decomposition of the abelian group $G=\bigoplus_p G_p$, where each factor satisfies $p^{r_p} G_p=\{0\}$ for $p$ prime and $r_p>0$ integer. One can check that the factors are orthogonal and the squared-norm factorizes as $\|x\|_\mathrm{L}^2 = \sum_p\|x_p\|_\mathrm{L}^2$. It then follows that $x$ is isotropic iff each $x_p$ is isotropic. For each factor $G_p$, one can show that a set is generating if its image in the quotient  $G_p/pG_p$ is generating. Since the original group $G$ is a quotient of rank-2 lattices, then it is a product of at most 2 cyclic groups: hence $G_p/pG_p$ is a vector space over the finite field $\ZZ/p\ZZ$ of dimension at most 2. Since $\supp(f)$ consists of isotropic elements and generates all of $G$, then the images of isotropic elements span the whole vector space. It can therefore be generated by at most 2 of them, since it has dimension at most 2. The sum of isotropic generators in each factor then provides the two isotropic generators of $G$.

Second, the two isotropic generators $x_1,x_2 \in G$ and their inner product $\lambda:=\langle x_1,x_2\rangle_\mathrm{L} \in\QQ/\ZZ$ all have the same order $N$. This follows because, on the one hand, $\langle x_1,x_1\rangle_\mathrm{L}=0 \in \QQ/(2\ZZ)$ so that $\langle \ord(\lambda)x_1,x_i\rangle_\mathrm{L}=0 \in \QQ/\ZZ$ for $i=1,2$ implying $\ord(\lambda)x_1=0$ (and similarly $\ord(\lambda)x_2=0$). Conversely, $\ord(x_1) \lambda = \langle \ord(x_1)x_1,x_2\rangle_\mathrm{L}=0$ (and similarly  $\ord(x_2) \lambda =0$).

We can then prove the claim \eqref{eq:Gcanon} as follows. We have the relations $N x_1=Nx_2=0$. There are no other relations because any relation $n_1 x_1+n_2 x_2=0$ gives $n_1\lambda = \langle n_1x_1,x_2\rangle_\mathrm{L}=- n_2\langle x_2,x_2\rangle_\mathrm{L}=0$ and similarly $n_2\lambda=0$: hence $N$ divides $n_1$ and $n_2$ so it is not an independent relation. We have therefore shown that $G\cong \ZZ_N\times \ZZ_N$. To fix the pairing, we can always write $\lambda=\ell/N$ for $\ell$ an integer coprime to $N$. The generator $x_1$ can be replaced by $\ell^{-1}x_1$, where $\ell^{-1}$ denotes the inverse mod $N$ (an integer with $\ell^{-1}\ell\equiv 1\bmod N$), so that we replace $\lambda \to 1/N$. Then we have derived the above pairing with the coordinates $x\equiv(a,b)\equiv a\, x_1+b\, x_2$.

\paragraph{$\boldsymbol{\SL(2,\ZZ)}$ invariance.} A convenient way to characterize the constraints of modular $S$ and $T$ invariance is to take the finite Fourier transform of $f$ in only the second coordinate:
\be
F(a,b):= \sum_{w\in \ZZ_N} f(a,w)\, \e^{2\pi i bw/N} \ , \qquad (a,b)\in\ZZ_N\times \ZZ_N \ .
\ee
For this function on $\ZZ_N\times \ZZ_N$, $S$ invariance \eqref{eq:finiteS} and $T$ invariance $\e^{\pi i \|x\|_\mathrm{L}^2}f(x) = f(x)$ say that
\begin{subequations}
\begin{align}
&S: \qquad F(b,-a) = F(a,b)  \ , \\
&T: \qquad F(a,a+b) = F(a,b)  \ .
\end{align}
\end{subequations}
In other words, $F$ must be invariant under the right-action of $\SL(2,\ZZ)$ on the row-vector $(a,b)$, generated by
\be
S=\begin{pmatrix} 0 & -1 \\ 1 & 0\end{pmatrix} \ , \qquad
T=\begin{pmatrix} 1 & 1 \\ 0 & 1\end{pmatrix}\ .
\ee
 The labels $(a,b)$ and their modular transformation are familiar from $\ZZ_N$ orbifolds~\cite{Vafa:1986wx}, with $F(a,b)$ playing the role of the contribution of the sector twisted by $(a,b)$.

Here $\SL(2,\ZZ)$ acts only through its quotient $\SL(2,\ZZ_N)$. We claim that the orbits of its action on $\ZZ_N \times \ZZ_N$ are labeled by $g:=\gcd(a,b,N)$. To see this, note on one hand that $g$ is clearly $\SL(2,\ZZ)$-invariant. Conversely, consider any element $(a,b)\in \ZZ_N\times \ZZ_N$ with $\gcd(a,b,N)=g$: we will show that this is in the same orbit as $(g,0)$. We have $a=g a'$, $b=g b'$ and $N=g N'$ for integers $a'$, $b'$ and $N'$ with $\gcd(a',b',N')=1$. Then there is an element $\gamma\in \SL(2,\ZZ)$ whose top row is $(a',b')\bmod N'$.\footnote{To see this, note that since $\gcd(a',b',N')=1$, then $a'$ and $b'$ can be shifted by multiples of $N'$ to make them coprime. Any coprime pair of integers form the top row of some $\SL(2,\ZZ)$ element.} Then we have $(a,b)=(g,0)\cdot \gamma$ in $\ZZ_N\times \ZZ_N$ as claimed.

In terms of the function $F$ on $G=\ZZ_N\times \ZZ_N$, we have found the most general $\SL(2,\ZZ)$-invariant solution:
an arbitrary function of $\gcd(a,b,N)$,
\be
F(a,b) = \sum_{g|N} d_g \, \mathds{1}_{g=\gcd(a,b,N)} \ ,\qquad d_g\in\CC \ .
\ee
Equivalently, acting on these fundamental indicator functions with an invertible triangular matrix, we may instead use the indicators $\mathds{1}_{g|\gcd(a,b,N)}$ as a basis.

Let us study these elementary functions $\mathds{1}_{g|\gcd(a,b,N)}$ of $(a,b)$, indexed by numbers $g$ that divide $N$. Translating this back to the function $f$ on $G=\ZZ_N\times \ZZ_N$ by taking the inverse Fourier transform in the $b$-variable, we find for $g|N$,
\be
f_g(a,w) = \frac{1}{N} \, \mathds{1}_{g|a}  \sum_{b\in\ZZ_N} \mathds{1}_{g|b}\, \e^{-\frac{2\pi ibw}{N}}  = \frac{1}{N} \, \mathds{1}_{g|a}  \sum_{b'\in\ZZ_{N/g}} \, \e^{-\frac{2\pi iwgb'}{N}} =\frac{1}{g}\, \mathds{1}_{g|a}  \, \mathds{1}_{(N/g)|w} \ .
\ee
Up to a harmless rescaling by $1/g$, these are the indicator functions for the sets $g\ZZ  \times (N/g)\ZZ $ in $G=\ZZ_N\times \ZZ_N$. The most general solution for $f$ is therefore a complex-linear combination of such functions,
\be
f(a,w) = \sum_{g|N} c_g \, \mathds{1}_{g\ZZ  \times (N/g)\ZZ}\, (a,w) \ , \qquad c_g\in\CC\ ,
\ee
where we have absorbed the factor of $1/g$ into $c_g$. Finally translating back to the $\rho_\mathrm{c}$ with \eqref{eq:rho from f}, we have found that the most general solution is
\be
\rho_\mathrm{c} = \sum_{g|N} c_g \, \delta_{\Lambda^{(g)}} \ , \qquad c_g\in\CC \ ,
\ee
where we define the lattices
\be
\Lambda^{(g)}:= \ZZ \, (g \, \tilde  x_1)+ \ZZ \, \big(\tfrac Ng \, \tilde x_2\big) + K
\ee
and $\tilde x_1,\tilde x_2$ are any representatives of $x_1,x_2\in M/K$ in $M$. Since it is a subgroup of the rank-2 lattice $M$,  then $\Lambda^{(g)}$ is a rank-2 lattice.

\paragraph{Narain lattices.} The lattices $\Lambda^{(g)}$ are in fact even and self-dual.

To see evenness, consider any element $x=m \, g \,  \tilde x_1 + n \, (N/g) \,  \tilde x_2 + k \in \Lambda^{(g)}$  where $m,n\in\ZZ$ and $k\in K$. We know the inner product \eqref{eq:Gcanon}, so we can simply write it down $\|x\|^2_{\mathrm{L}} \equiv  2 (mg) (n N/g)/N \equiv 0 \bmod 2$ and see that it is even.

Self-duality follows from counting. The subgroup $\Lambda^{(g)}/K=g\ZZ_N\times\tfrac{N}{g}\ZZ_N\subset G$ has order $\tfrac{N}{g}\, g=N$, while $|G|=N^2$. On the other hand, $\Lambda^{(g)}{}^\vee/K$ is the orthogonal complement of $\Lambda^{(g)}/K$ with respect to the non-degenerate pairing on $G$ and thus has order $|G|/N=N$. Since $\Lambda^{(g)}/K \subset \Lambda^{(g)}{}^\vee/K \subset G$, it follows by comparing orders that $\Lambda^{(g)}/K=\Lambda^{(g)}{}^\vee/K$ and thus $\Lambda^{(g)}=\Lambda^{(g)}{}^\vee$.

The only even self-dual lattices in $\RR^{1,1}$ are the Narain lattices, labeled by the radius $R>0$,\footnote{$\Lambda$ is an even self-dual lattice of signature $(1,1)$, hence isometric to the
hyperbolic plane. Thus it has a basis $x_1,\,x_2$ with $\lVert x_1\rVert^2_\mathrm{L}=\lVert x_2
\rVert^2_\mathrm{L}=0$ and $\langle x_1,x_2\rangle_\mathrm{L}=1$~\cite[Ch.~V, Thm.~6]{Serre:1973}. In terms of the basis $(P,\bar P)$ where the Lorentzian inner product takes the form \eqref{eq:LIP}, the most general basis satisfying these requirements can be parametrized as $x_1=a(1,1)$, $x_2=b(1,-1)$ with $4ab=1$. We can also choose $a>0$ and $b>0$ by possibly flipping the sign of the generators. To get the form \eqref{eq:Narain lattice new}, we set $a=\frac{R}{2}$ and $b=\frac{1}{2R}$.}
\be
\Lambda_R=\big\{\big(\tfrac{1}{2}(m R+n R^{-1}),\,\tfrac{1}{2}(m R-n R^{-1})\big) \mid m,\, n \in \ZZ\big\}\ . \label{eq:Narain lattice new}
\ee
What we have shown is that the modular bootstrap axioms imply that $\rho$ is a finite linear combination of Dirac combs on Narain lattices
    \be
        \rho=\sum_{R} c_R \, \rho_{R}\ , \qquad c_R \in \CC\ . \label{eq:rho Narain decomposition}
    \ee
The sum goes over some finite set of radii.

We are now very close to our goal, since each of these summands $\rho_R$ is itself the density of states of the free boson CFT with radius $R$,
\be
\rho_R = \delta_{\Lambda_R} \ . \label{eq:rho free bos}
\ee
Let us make a comment on conventions. Since the characters $\chi_P(\tau)\chi_{\bar P}(-\bar \tau)$ are even in both $P$ and $\bar P$, only the completely even part of $\rho$ is physical, in the sense that it appears in the partition function \eqref{eq:intDensity}. The antisymmetric components of $\rho$ are unphysical, and until now  we had fixed them to zero by explicitly symmetrizing $\rho$. This is not mandatory, and asymmetric densities like \eqref{eq:rho free bos} are perfectly healthy. In this case, the reflection $\sigma\cdot\rho(P,\bar P):=\rho(P,-\bar P)$ tells us that the densities $\rho_R\leftrightarrow \rho_{R^{-1}}$ related by T-duality are physically equivalent. From now on, we shall fix this ambiguity in $\rho$ differently, by insisting that the values of $R$ appearing in \eqref{eq:rho free bos} are in the range $R\geq 1$. One can return to the symmetrized density discussed above by simply symmetrizing (which preserves the general form of \eqref{eq:rho Narain decomposition}).

\subsection{Kiritsis' classification} \label{subsec:Kiritsis}
Up to this point, the positivity axiom~\ref{axiom:positivity} was only used to derive upper bounded density in Section~\ref{sec:growth}, and was not directly used in this section. The unique vacuum axiom~\ref{axiom:vacuum} has never been used.
The remaining task is to analyze which values of $c_R$ are compatible with these additional axioms. This problem was solved by Kiritsis in the rational case~\cite{Kiritsis:1988et, Kiritsis:1988es} and some of his proofs can be transferred to the irrational case. We will therefore follow Kiritsis' argument closely.

We should notice that we seemingly made a step back in \eqref{eq:rho Narain decomposition} since we only asserted that $c_R \in \CC$, whereas we already showed in \eqref{eq:rhoComb} that the coefficients of the combs must take values in $\frac{1}{4}\ZZ$. This step back does not make a difference in the remaining part of the proof, where we will establish that $c_R \in \frac{1}{2}\ZZ$ and $c_R \ge 0$ for $R>1$.

\paragraph{Special primaries.} The first step in the proof is to notice that, for every $R$ with $R>1$ in the sum \eqref{eq:rho Narain decomposition}, there is at least one primary $(P,\bar P)$ in $\rho$ that (i) only appears in $\rho_{R}$ and (ii) is non-degenerate in both $P$ and $\bar P$.

To see this, consider the state $P=\frac{1}{2}(R+pR^{-1})$, $\bar{P}=\frac{1}{2}(R-pR^{-1})$ appearing in \eqref{eq:Narain lattice new} for a prime $p$. This state is non-degenerate, unless $R+pR^{-1} \in \ZZ$ or $R-p R^{-1} \in \ZZ$. Each of these conditions can happen for at most one prime $p$.\footnote{Let us give the argument for the first case (the second case is identical).
Assume that $R+p_1 R^{-1} \in \ZZ$ and $R+p_2 R^{-1} \in \ZZ$ for two primes $p_1, p_2$. Assume that $p_1 \ne p_2$. Taking the difference shows that $R$ must be rational, $R=\frac{a}{b}$ with $a$ and $b$ coprime, i.e.\ $a^2+p_i b^2=k_i a b$ for $k_i \in \ZZ$ and $i=1,2$. Rearranging gives $a(k_ib-a)=p_ib^2$. But $\gcd(a,b)=1$ and $\gcd(k_i b-a,b)=1$, which means that the LHS has no common factor with $b$ and thus $b=1$. This means that $R=a$ must be an integer satisfying $R(k_i-R)=p_i$. Since $p$ is prime, then either $R=1$ or $R=p_i$. As $R>1$ by assumption then $p_1=p_2=R$.}
Next, we need to show that the state doesn't appear in $\rho_{R'}$ for any other $R'$, up to potentially finitely many exceptions.
If the state were to appear in another $\rho_{R'}$ with $R' \ne R$, then
\be
R+pR^{-1}=\pm (m R'+n R'^{-1})\ , \qquad R-pR^{-1}=\pm(m R'-n R'^{-1}) \label{eq:momentum matching}
\ee
for two integers $m,\, n \in \ZZ$. By flipping jointly the signs of $m$ and $n$ or applying $R' \to R'^{-1}$ we can assume both prefactors to be $+$. Taking sums and differences gives
\be
R=m R'\ , \qquad p R^{-1}=n R'^{-1}\ .
\ee
By multiplying the equations, we see that the solutions must satisfy $mn=p$. Since $p$ is prime then either $(m,n)=(p,1)$ and $R=pR'$, or $(m,n)=(1,p)$ and $R=R'$. Thus, as long as we choose the prime $p$ avoiding the two values where the state is degenerate and the finitely many values $p=\frac{R}{R'}$ (and by T-duality also $p=RR'$) for $R'$ appearing in $\rho$ in \eqref{eq:rho Narain decomposition}, then this non-degenerate state only appears in the single term $\rho_R$.

Choose such a state in each $\rho_{R}$ with $R>1$. It appears precisely twice in $\rho_R$ (with the second copy related by $(P,\bar P) \to (-P,-\bar P)$).
Also the state with $(P,\bar P) \to (P,-\bar P)$ carries the same conformal weight, but it doesn't appear in $\Lambda_R$ unless $R^2=p$, in which case $\bar P=0$ and this is the same state.\footnote{Indeed, if it does appear, then $R+p R^{-1}=m R+n R^{-1}$ and $-R+pR^{-1}=m R-n R^{-1}$ for some integers $m$ and $n$. Taking sums and differences gives $n=R^2=\frac{p}{m}$. Since $R \ne 1$ by assumption and $p$ is prime, the only solution is $n=p=R^2$ and $m=1$. Thus indeed, $\bar P=0$ for this state and its conformal weight is two-fold degenerate.}

Thus, positivity, integrality  and the non-degeneracy of that state tell us that the Virasoro multiplicity satisfies $D_{P,\bar P}=d_{P,\bar P}=2c_R \in \NN_0$ and thus $c_R \in \frac{1}{2}\NN$. We can of course assume that $c_R \ne 0$, since we just omit that term from \eqref{eq:rho Narain decomposition}.

\paragraph{The case $\boldsymbol{R=1}$.} We have to treat the case $R=1$ that may appear in the decomposition \eqref{eq:rho Narain decomposition} separately. All the states appearing in $\Lambda_1$ are degenerate. We will consider the special state $(P,\bar P)=(p,0) \in \Lambda_1$, obtained by setting $m=n=p$ in \eqref{eq:Narain lattice new}. For a prime $p$, the state again does not appear in any other $\rho_{R}$ with $R\ne 1$, since otherwise
\be
2p=mR+n R^{-1}\ , \qquad 0=m R-n R^{-1}
\ee
for $m,\, n \in \ZZ$, which implies
\be
p^2=m n\ , \qquad R^2=\frac{n}{m}\ .
\ee
Since $p$ is prime, the only solutions are $(m,n)=(p^2,1),\, (p,p),\, (1,p^2)$ which only leaves $R=p$ in the range $R>1$. Thus, as long as $p$ avoids the finitely many radii appearing in \eqref{eq:rho Narain decomposition}, the state only appears in $\rho_1$.

This state only appears twice in $\rho_1$ at $(p,0)$ and $(-p,0)$ (while a general state would appear four times). Therefore, integrality of $\rho$ implies that the coefficient $c_1$ of $\rho_{R=1}$ takes values in $\frac{1}{2}\ZZ$. It may be negative, since the corresponding Virasoro representation is degenerate.

Thus, we have now shown that
\be
\rho=\frac{1}{2} \sum_R m_R \,  \rho_{R} \label{eq:rho integer free boson decomposition}
\ee
with $m_R \in \NN$, except for $m_1 \in \ZZ$. Imposing uniqueness of the vacuum, axiom~\ref{axiom:vacuum}, for the first time gives
\be
1= D_{0,0} = \frac{1}{2}\sum_R m_R \ . \label{eq:to-combine}
\ee

\paragraph{Solutions with spin-1 currents.}
To continue, we split into cases, depending on the number of Virasoro holomorphic currents, i.e.\ $D_{n,0}$ for $n \in \NN$. Let us first assume that $D_{1,0}\ge 1$, i.e.\ the theory has holomorphic currents of spin 1. We have $D_{1,0}=d_{0,0}+d_{1,0}$. Every $\rho_R$ contributes $1$ to $d_{0,0}$, but only $\rho_1$ contributes $2$ to $d_{1,0}$.\footnote{Indeed, $\frac{1}{2}(mR+n R^{-1})=1$ and $\frac{1}{2}(m R-n R^{-1})=0$ imply $R^2=\frac{n}{m}$ and $\sqrt{mn}=1$, i.e.\ $m=n=\pm 1$ and hence $R=1$.}
Thus, we have
\be
D_{1,0}=m_1+\frac{1}{2}\sum_R m_R=m_1+1 \ge 1\ , \label{eq:number of spin-1 currents}
\ee
where we used the uniqueness of the vacuum \eqref{eq:to-combine}. Therefore, also $m_1 \ge 0$ and all coefficients appearing in \eqref{eq:rho integer free boson decomposition} are positive. Solutions of \eqref{eq:to-combine} are therefore decompositions of $1$ into positive half-integers. There are only two ways to do this: $m_R=2$ for some radius $R$, or $m_R=m_{R'}=1$ for two different radii $R \ne R'$. The corresponding densities of states are
\be
\rho=\rho_R\qquad \text{or}\qquad \rho=\frac{1}{2}(\rho_R+\rho_{R'})\ . \label{eq:free boson and spurious solution}
\ee
The first is the free boson density of states and we refer to the second as the spurious solution.

\paragraph{Solutions without spin-1 current.} We have found all solutions with a holomorphic spin-1 current. Thus, let us now assume that $D_{1,0}=0$. We still have $D_{1,0}=m_1+1$ as in \eqref{eq:number of spin-1 currents}, which forces $m_1=-1$. We find the remaining possibilities by computing $D_{n,0}=\sum_{k=0}^n d_{k,0}$ for $n\ge 2$. Let us first compute the contribution of $\rho_R$ to $d_{k,0}$ for $k\geq 2$. A state contributing to $d_{k,0}$ must satisfy $k=\pm \frac{1}{2}(m R+n R^{-1})$ and  $m R-n R^{-1}=0$.
If the state appears, it appears with two-fold degeneracy, corresponding to the two choices of sign.
Assuming without loss of generality the $+$ sign, so that $m,\, n \ge 0$, this implies
\be
k^2=mn\ , \qquad R^2=\frac{n}{m}\ .
\ee
For $k=2$, only $R=1$ and $R=2$ contribute. For $k=3$, only $R=1$ and $R=3$ contribute.
For $k=4$, only $R=1$, $R=2$ and $R=4$ contribute.
For $k=5$, only $R=1$ and $R=5$ contribute. For $k\ge 6$, also non-integer radii can contribute.
Thus, we have
\be
d_{2,0}=m_1+m_2\ , \quad d_{3,0}=m_1+m_3\ , \quad d_{4,0}=m_1+m_2+m_4\ , \quad d_{5,0}=m_1+m_5\ .
\ee
Summing up, and using that $d_{0,0}=1$, $d_{1,0}=-1$ and $m_1=-1$, we get
\begin{subequations}
\begin{align}
    D_{2,0}&=m_2-1\ , \\
    D_{3,0}&=m_2+m_3-2\ , \\
    D_{4,0}&=2m_2+m_3+m_4-3\ , \\
    D_{5,0}&=2m_2+m_3+m_4+m_5-4\ .
\end{align} \label{eq:D20-D50 degeneracies}
\end{subequations}
Positivity in particular implies that $D_{n,0} \ge 0$ for $n \in \{2,\dots,5\}$.

It is now simple to find all the integer solutions to these constraints, given that $\sum_{R>1} m_R=3$ from \eqref{eq:to-combine}. First suppose that $m_2 \ge 2$, so that each of \eqref{eq:D20-D50 degeneracies} are automatically positive as needed. The most general solution is
\be
\rho=\rho_R^\mathrm{orb}=\frac{1}{2}(-\rho_1+2\rho_2+\rho_R) \label{eq:orbifold solution}
\ee
for $R>1$. This is because $\sum_{R>1} m_R=3$ with $m_2\geq 2$ forces $m_2=2$ or $3$. The $m_2=3$ case coincides with \eqref{eq:orbifold solution} with $R=2$. In the $m_2=2$ case there is room for only one more $\rho_R$ with $R>1$, giving the above solution. We refer to it as the orbifold solution. For $R=1$, the orbifold solution coincides with the free boson solution \eqref{eq:free boson and spurious solution} with $R=2$.

Since $D_{2,0}=m_2-1 \ge 0$, the only remaining case is $m_2=1$. Imposing positivity of each line of \eqref{eq:D20-D50 degeneracies}, with $\sum_{R \ne 1,\, 2} m_R=2$ from \eqref{eq:to-combine}, only has the following three exceptional solutions $\rho=\rho^\mathrm{tet}$, $\rho=\rho^\mathrm{oct}$ and $\rho=\rho^\mathrm{ico}$:
\begin{subequations}
\begin{align}
\rho^\mathrm{tet}&=\frac{1}{2}(-\rho_1+\rho_2+2\rho_3)\ , \\
\rho^\mathrm{oct}&=\frac{1}{2}(-\rho_1+\rho_2+\rho_3+\rho_4)\ , \\
\rho^\mathrm{ico}&=\frac{1}{2}(-\rho_1+\rho_2+\rho_3+\rho_5)\ .
\end{align}\label{eq:exceptional solutions}%
\end{subequations}
It remains to show that $D_{P,\bar P} \ge 0$ for the solutions \eqref{eq:free boson and spurious solution}, \eqref{eq:orbifold solution} and \eqref{eq:exceptional solutions}. One can demonstrate this directly, but we will not carry this out here, since this follows from the fact that we can realize them as partition functions of physical theories (apart from the spurious solution, which is the average of the partition functions of two physical theories and thus still satisfies positivity).
This completes the classification of the solutions to the modular bootstrap.

\section{Uniqueness of the \texorpdfstring{$\boldsymbol{c=1}$}{c=1} theories}\label{sec:uniqueness}
So far, we have shown that the density of states of a compact unitary $c=1$ CFT must take the form of the free boson \eqref{eq:rho free bos}, the $\ZZ_2$-orbifold of the free boson \eqref{eq:orbifold solution}, or one of the three exceptional theories \eqref{eq:exceptional solutions}---or the spurious average $\frac{1}{2}(\rho_R+\rho_{R'})$ of two free boson densities \eqref{eq:free boson and spurious solution}. We will now show that these are the unique $c=1$ theories with those partition functions, and the spurious solution is not realized by a physical theory. This requires us to move beyond the modular bootstrap. Thus, we assume now the full axioms of a unitary 2d CFT with central charge $c=1$.

\subsection{The spurious solution} \label{subsec:spurious solution}
We begin by ruling out the spurious solution $\frac{1}{2}(\rho_R+\rho_{R'})$ for $R' \ne R,R^{-1}$. The density of states shows that $D_{1,0}\ge 1$ and thus any such putative theory would have a state of dimension $(h,\bar h)=(1,0)$. Call the state $\ket{j}$ and corresponding field from the operator-state correspondence $j$. Since $\lVert \bar L_{-1} \ket{j} \rVert^2=\bra{j} [\bar L_1,\bar L_{-1}] \ket{j}=2 \bra j \bar L_0 \ket j=0$, unitarity forces $\bar L_{-1} \ket j=0$ and thus $\bar \partial j=0$. Thus, $j$ is a conserved current. This means that the theory has an enhanced chiral symmetry. In particular, the associated Ward identities must hold inside correlation functions. The same argument works with the corresponding right-moving current $\bar \jmath$ satisfying $\partial \bar \jmath=0$.

Next, study the OPE $j(z)j(w)$, which is constrained by conformal invariance to take the form
\be
j(z) j(w) \sim \frac{k}{(z-w)^2}+\frac{a(w)}{z-w}\ ,
\ee
where $a(w)$ is also a weight-one holomorphic field. Exchange symmetry under $z \leftrightarrow w$ requires the simple pole to vanish. Unitarity requires $k=\langle j \,|\, j \rangle>0$ and thus we may rescale $j$ to assume that $k=1$.

The existence of $j$ also allows us to construct its corresponding Sugawara tensor $T_j(z)=\frac{1}{2} \, (j\,j)(z)$, where the parenthesis denote normal ordering. Unitarity imposes $T\equiv T_j$, i.e.\ the Sugawara stress-tensor coincides with the actual stress tensor of the theory. To see this, form the coset stress tensor $T-T_j$, which satisfies the Virasoro algebra of central charge $1-1=0$. Therefore, the state corresponding to $T-T_j$ is null, which implies that $T=T_j$ as claimed.

A state with conformal weight $(h,\bar h)=(P^2,\bar P^2)$ with respect to $T=T_j$ therefore has a momentum assignment $(P,\bar P)$ with a definite sign (at least after diagonalizing in case of degeneracy). However, for $R'\neq R,R^{-1}$, the argument around \eqref{eq:momentum matching} shows that there are generic states whose conformal weight appears with multiplicity $d_{P,\bar P}=D_{P,\bar P}=1$ in $\frac{1}{2}(\rho_R+\rho_{R'})$ (since either $R$ or $R' \ne 1$). For such a state with momentum $(P,\bar P)$, a charge conjugate state would have momentum $(-P,-\bar P)$ and hence the same conformal weight, but this is impossible since $D_{P,\bar P}=1$. Thus, all two-point functions of this operator vanish by momentum conservation, so the CFT does not carry a non-degenerate inner product, in contradiction with the theory being a unitary CFT.

Let us remark that unitarity is necessary to eliminate this modular invariant. It was shown in~\cite{Flohr:1992wi} that this modular invariant \emph{can} instead be realized in non-unitary models with $c_\mathrm{eff}=1$, at least for certain choices of $R$ and $R'$. Let us also mention that similar spurious solutions have been also been discussed in \cite{Dymarsky:2020bps, Dymarsky:2022kwb}.

\subsection{The free boson} \label{subsec:free boson}
Next, let us show that a theory with density of states $\rho_R$ has to be a free boson theory.

First, since $D_{1,0}\ge 1$, the same argument as in the previous Subsection~\ref{subsec:spurious solution} shows that unitarity implies the existence of a holomorphic current $j$ and an anti-holomorphic current $\bar \jmath$. Moreover, $T=T_j$ as before. We therefore can again write $(h,\bar h)=(P^2,\bar P^2)$ and we can assign to every state a momentum $(P,\bar P)$ with definite signs.

Let us denote vertex operators with momentum $(P,\bar P)$ by $V_{P,\bar P}$. Unitarity implies that it must have non-zero two-point function with a state of momentum $(-P,-\bar P)$.
In the OPE, the momentum behaves additively by momentum conservation. Unitarity implies that the OPE cannot be empty, since otherwise the sphere four-point function
\be
\langle V_{(P_1,\bar P_1)}(z_1,\bar z_1)\, V_{(P_2,\bar P_2)}(z_2, \bar z_2)\, V_{(-P_1,-\bar P_1)}(z_3, \bar z_3)\, V_{(-P_2,-\bar P_2)}(z_4, \bar z_4) \rangle \label{eq:sphere four-point function}
\ee
would need to vanish, which is inconsistent with the OPE in the $V_{(P_1,\bar P_1)}(z_1,\bar z_1)\times V_{(-P_1,-\bar P_1)}(z_3,\bar z_3)$ channel, which contains at least the identity by the above, so is non-vanishing.

Let us now denote the set of appearing momenta for all the $\mathfrak{u}(1) \times \mathfrak{u}(1)$-primary states as $Q$, counted without multiplicity. By the above discussion, $Q$ is additive, i.e.\ $Q+Q = Q$ and invariant under joint inversion of $P$ and $\bar P$, i.e.\ $-Q=Q$. Thus, $Q$ is a discrete subgroup of $\RR^2$.
Since the density of states is fixed to be $\rho_R$, we also know that the set of $(|P|,|\bar P|)$ for $(P,\bar P) \in Q$ agrees with the same set for $(P,\bar P) \in \Lambda_R$.
We claim that this implies that $Q=\Lambda_R$ or $Q=\Lambda_{R^{-1}}$. Indeed, write $\sigma$ for the reflection $(P,\bar P) \mapsto (P,-\bar P)$ as above, so that $\sigma \Lambda_R=\Lambda_{R^{-1}}$. Both $\Lambda_R$ and $Q$ are already invariant under $(P,\bar P) \mapsto (-P,-\bar P)$. Therefore, matching the set $(|P|,|\bar P|)$ says that
\be
Q \cup \sigma Q=\Lambda_R \cup \Lambda_{R^{-1}}\ . \label{eq:Q matching}
\ee
All four elements in this equality are groups. To continue, we first use the following elementary group theoretic fact: For $A,\,B,\,K \subset G$ subgroups and $K \subset A \cup B$, either $K\subset A$ or $K \subset B$. Suppose to the contrary that we could pick $a \in K \setminus B \subset A$ and $b \in K \setminus A \subset B$. But $ab \not \in A$ (otherwise $b=a^{-1}(ab) \in A$) and $ab \not \in B$ (otherwise $a=(ab)b^{-1} \in B$), a contradiction.

Now \eqref{eq:Q matching} says in particular that $Q \subset \Lambda_R \cup \Lambda_{R^{-1}}$, so $Q \subset \Lambda_R$ or $Q \subset \Lambda_{R^{-1}}$. Without loss of generality, suppose that $Q \subset \Lambda_R$ (otherwise we replace $R \to R^{-1}$ in the following argument). Similarly, $\Lambda_R \subset Q$ or $\Lambda_R \subset \sigma Q$. In the former case, we would conclude that $Q=\Lambda_R$ and would be done. In the latter case, $Q \subset \Lambda_R \subset \sigma Q$ and in particular $Q \subset \sigma Q$. But since $\sigma$ is an involution, we can apply it to the inclusion to also conclude that $\sigma Q \subset Q$ and thus $Q=\sigma Q=\Lambda_R$ as desired.

This tells us the field content of the theory in terms of $\mathfrak{u}(1) \times \mathfrak{u}(1)$ representations, not just in terms of Virasoro representations.

To fully characterize the theory, we also need to determine the sphere three-point function, i.e.\ the OPE coefficients. In fact the $\mathfrak{u}(1) \times \mathfrak{u}(1)$ Ward identities allow one to determine sphere $n$-point functions up to an overall normalization. This is possible because one can write $L_{-1}=\frac{1}{2}(j \,j)_{-1}$ thanks to the identification $T=T_j$. The LHS can be evaluated on a vertex operator in terms of the coordinate derivative $\partial_{z_i}$, while the right hand side can be fully evaluated in terms of the momenta. We thus obtain $n$ first-order differential equations for the coordinate dependencies of the correlator (the abelian version of the Knizhnik-Zamolodchikov equation), as well their antiholomorphic avatars. They determine the sphere $n$-point function, up to an overall constant, to be of the form
\be
   \Big\langle\prod_{i=1}^n V_{P_i,\bar P_i}(z_i,\bar z_i)\Big\rangle
   = C(\{P_i,\bar P_i\}) \prod_{1\le i<j\le n}
   (z_i-z_j)^{2P_iP_j}(\bar z_i-\bar z_j)^{2\bar P_i\bar P_j}\ , \label{eq:free boson n point function}
\ee
provided momentum conservation holds.
Consider the special case of the sphere four-point function \eqref{eq:sphere four-point function}. The OPE in the $V_{P_1,\bar P_1} \times V_{-P_1,-\bar P_1}$ channel produces only the identity by momentum-conservation, and if we canonically normalize the two-point function, this shows that the corresponding prefactor $C(\{P_i,\bar P_i\})$ in \eqref{eq:free boson n point function} is unity. Decomposing in the other channel then shows that the structure constants are unity when momentum conservation is satisfied. This characterizes the theory completely and identifies it with the free boson theory.

\subsection{\texorpdfstring{$\ZZ_2$}{Z2} orbifold of the free boson} \label{subsec:Z2 orbifold free boson}
Next, we want to show that a theory with $\rho_R^\mathrm{orb}=\frac{1}{2}(\rho_R+2 \rho_2-\rho_1)$ as in \eqref{eq:orbifold solution} is necessarily a $\ZZ_2$-orbifold of the free boson. We denote the theory at hand $\mathcal{T}$. Our strategy is the following. We first identify the chiral algebra of the theory and show that, \emph{as an algebra}, it contains the $\ZZ_2$-orbifold of the Heisenberg algebra. This implies that this algebra's fusion rules of are $\ZZ_2$-graded (if we already knew that the theory was the $\ZZ_2$-orbifold of the free boson, the $\ZZ_2$ odd states would be the twisted sector states). The whole theory then inherits this $\ZZ_2$ symmetry. We can thus consider the gauged theory $\mathcal{T}/\ZZ_2$. Its torus partition function can be computed to be $\rho_R$ and, by the previous subsection, we thus know that $\mathcal{T}/\ZZ_2$ is the free boson theory. But gauging in two dimensions is invertible and this statement also shows that $\mathcal{T}$ is a $\ZZ_2$-orbifold of the free boson. The $\ZZ_2$ symmetry acting on $\mathcal{T}$ is then identified with the quantum symmetry of the orbifold.

For $R=1$, we have $\rho=\rho_2$, which we already know to be the free boson at radius $R=2$, which indeed can be realized as a $\ZZ_2$-orbifold of the free boson at $R=1$. Therefore, we can assume that $R \ne 1$ in the following.
Let us denote the chiral algebra in the following by $\mathcal{A}$.
\paragraph{Chiral fields.} Let us first work out the decomposition of $\mathcal{A}$ into Virasoro representations, which we denote by $\mathcal R^\Vir_h$. This is just a matter of computing $D_{P,0}$ from $\rho=\frac{1}{2}(-\rho_1+2\rho_2+\rho_R)$. For generic $R$, this gives in terms of Virasoro representations
\be
\mathcal{A} \cong \bigoplus_{n=0}^\infty \mathcal{R}^\Vir_{h=(2n)^2}\ , \label{eq:Z2 orbifold chiral algebra decomposition}
\ee
i.e.\ it contains all degenerate Virasoro representations with $P=2n$ even. For particular values of $R$, the algebra can be bigger, which we treat below. We first focus on the general case. The lowest additional state beyond Virasoro appears for $h=4$. One can take the following brute-force route to determine $\mathcal{A}$ at generic radius as a VOA. Make an ansatz of the OPE of the spin-4 field $W_4$ with itself. That does not contain yet the next Virasoro primary state of spin $16=4^2$ in its singular part of the OPE and thus closes on itself. In terms of primary fields, the $W_4W_4$ OPE thus contains only the identity $\id$ and $W_4$ itself. We normalize $W_4$ by fixing the two-point function of $W_4$ to be the convenient value $\frac{3}{200}$. This means that there is only one free structure constant $c_{444}$. Imposing the Jacobi identity requires then that $c_{444}^2=1$, which is the correct value for the $\ZZ_2$-invariant part of the Heisenberg algebra.\footnote{In practice, we did this using Thielemans' \texttt{Mathematica} package~\cite{Thielemans:1991uw}. The normalization of $\frac{3}{200}$ is chosen precisely such that the Jacobi identity implies $c_{444}^2=1$.} We shall call the $\ZZ_2$-invariant part of the Heisenberg algebra by its standard name $M(1)^+$ in the following.

It is known that $M(1)^+$ is generated by $T$ and $W_4$ (meaning that the primaries for $n \ge 2$ in \eqref{eq:Z2 orbifold chiral algebra decomposition} can be realized in terms of normal-ordered products thereof)~\cite[Theorem 2.7]{Dong:1997id}. Thus, the $W_4W_4$ OPE characterizes the algebra completely. Thus, for generic values of $R$, we see that $M(1)^+ \subset \mathcal{A}$. But since it already accounts for the decomposition \eqref{eq:Z2 orbifold chiral algebra decomposition}, it is equal to it.

It remains to show that for non-generic $R$, we also have $M(1)^+ \subset \mathcal{A}$. For this, it helps to notice the following trick. Let
\be
\mathcal{A}_\mathrm{deg} \subset \mathcal{A} \label{eq:Adeg}
\ee
denote the subalgebra of the chiral algebra consisting of degenerate Virasoro representations, i.e.\ those for which $h$ is a perfect square. Since degenerate representations only fuse to degenerate representations, this is indeed a subalgebra. For generic $R$, $\mathcal{A}=\mathcal{A}_\mathrm{deg}$ thanks to \eqref{eq:Z2 orbifold chiral algebra decomposition}. Since we assume $R \ne 1$ and $R \ge 1$, the only case in which another degenerate field could appear in the $W_4W_4$ OPE is $R=2$, in which case the OPE contains an additional spin-4 field. (At $R=1$ the chiral algebra would contain the Heisenberg algebra and thus in particular also its $\ZZ_2$-singlet sector $M(1)^+$; the $R=\frac{1}{2}$ case is T-dual to $R=2$.) Let us call the two spin-4 fields in that case $W_4^i$ for $i=1,2$. We can assume that their two-point function is normalized as $W_4^i W_4^j \supset \frac{3}{200} \delta^{ij} \id$. The OPE has the free constants $c_{444}^{ijk}$ with $c_{444}^{ijk}=c_{444}^{jik}$.\footnote{The Jacobi identity implies in fact that $c_{444}^{ijk}$ is totally symmetric in $ijk$.} We can again impose the Jacobi identities on this ansatz with the result
\be
c_{444}^{111}=-c_{444}^{221}=-c_{444}^{122}=\cos(\varphi)\ , \qquad c_{444}^{112}=c_{444}^{121}=-c_{444}^{222}=\sin(\varphi)\ ,
\ee
where $\varphi \in [0,2\pi)$ is an angle. The freedom of this angle simply corresponds to the basis choice of $W_4^1$ and $W_4^2$. Rotating the basis allows us to set $\varphi=0$, which shows in particular that the algebra generated by $T$ and $W_4^1$ is a subalgebra of $\mathcal{A}$ that is again isomorphic to $M(1)^+$.

\paragraph{A more conceptual route.} Above, we established that $M(1)^+ \subset \mathcal{A}$ for any radius. The derivation was based on brute force. Such a derivation will be computationally very difficult for the exceptional theories and thus we explain also a more conceptual route towards the same result. For this, we may again restrict to the subalgebra $\mathcal{A}_\mathrm{deg}$ as in \eqref{eq:Adeg}. We will use the following theorem.
\begin{theorem} \label{thm:Carpi}
    Let $\mathcal{A}$ be a unitary chiral algebra with $c=1$ and $\mathcal{A}_\mathrm{deg} \subset \mathcal{A}$ its subalgebra consisting of degenerate Virasoro representations. Then $\mathcal{A}_\mathrm{deg} \cong \mathfrak{su}(2)_1^\Gamma$, where $\Gamma \subset \SO(3)\cong\Aut(\mathfrak{su}(2)_1)$ is a closed subgroup, i.e.\ $\mathcal A_\mathrm{deg}$ is a singlet sector of some $\mathfrak{su}(2)_1$ orbifold.
\end{theorem}
This theorem was proven in the recent preprint \cite[Theorem 5.6]{Carpi:2026itu} and the rational case was also proven in the recent preprint~\cite{Gannon:2026ihc}. In fact,~\cite{Gannon:2026ihc} classifies \emph{all} strongly rational (and pseudo-unitary) chiral algebras $\mathcal{A}$ with $c=1$, not just their degenerate subalgebras $\mathcal{A}_\mathrm{deg}$, while~\cite{Carpi:2026itu} classifies all  (simple) degenerate chiral algebras $\mathcal{A}_\mathrm{deg}$.

We will give a short proof for completeness. The proof relies on a few standard facts about chiral algebras. First, the degenerate modules $\mathcal{R}_{h=n^2}^\Vir$ obey the tensor product rules of $\SO(3)$ representations (rigorously proven in~\cite[Theorem 3.3]{Milas:2000qk})
\be
\mathcal{R}^\Vir_{n_1^2} \otimes \mathcal{R}^\Vir_{n_2^2} \cong \bigoplus_{n=|n_1-n_2|}^{n_1+n_2} \mathcal R^\Vir_{n^2}\ . \label{eq:Virasoro degenerate fusion rules}
\ee
Moreover, these modules generate a semisimple symmetric tensor category equivalent to $\Rep(\SO(3))$, with $\mathcal{R}^\Vir_{n^2}$ corresponding to the spin-$n$ representation $V_n$~\cite[Example 4.12]{McRae:2018wpu}.

Now, a chiral algebra that contains $\Vir_{c=1}$ and decomposes into only degenerate representations is the same data as a commutative associative algebra $A$ in this category with a one-dimensional space of invariants~\cite{Huang:2014ixa, Creutzig:2020smh}. The content of this statement is just bookkeeping of the OPEs. Write
\be
\mathcal{A}_\mathrm{deg} \cong \bigoplus_{n \ge 0} A_n \otimes \mathcal{R}^\Vir_{n^2}\ ,
\ee
where the multiplicity space $A_n$ records how many multiplets of currents of degenerate weight $h=n^2$ the algebra contains. Because of the fusion rules \eqref{eq:Virasoro degenerate fusion rules}, the content of the OPE is therefore a set of structure constants mapping $A_{n_1} \times A_{n_2}$ to $A_n$ for every $n$ appearing in the tensor product of the spin-$n_1$ and spin-$n_2$ representations, i.e.\ precisely an $\SO(3)$-equivariant product on $A=\bigoplus_{n \ge 0} A_n \otimes V_n$. Associativity of the OPE translates to associativity of this product and mutual locality of the currents translates to commutativity of $A$. Finally, since the vacuum is unique, we have a one-dimensional space of invariants, $A^{\SO(3)}=A_0 \cong \CC$. Conversely, we can reconstruct $\mathcal{A}_\mathrm{deg}$ from $A$.

It remains to classify the possible algebras $A$. Notice that a large class of such algebras is obtained by considering the space of functions $\CC[\SO(3)/\Gamma]$, for $\Gamma$ a closed subgroup of $\SO(3)$.
$\SO(3)$ acts by multiplication from the left and the algebra structure follows from the multiplication of functions.
In fact, thanks to the structure theorem~\cite[Proposition 7.1]{Pinzari:2007} and unitarity, $A$ is guaranteed to be of this form, i.e.\footnote{To apply the proposition, we also need to know that $A$ is a commutative unital C$^*$ algebra. Commutativity and the unit are ensured by the OPE, while unitarity of the chiral algebra provides a $*$-involution together with the positive definite inner product $\langle a,b\rangle=\omega(a^*b)$, where $\omega:A \to A_0=\CC$ is the projection onto the invariants. Multiplication by $a$ is then bounded on the Hilbert space completion of $A$. This defines the C$^*$ structure on the completion.}
\be
A \cong \CC[\SO(3)/\Gamma]\ .
\ee
This is precisely the algebra associated to the singlet algebra $\mathfrak{su}(2)_1^\Gamma$.
To see this, consider first the case where $\Gamma$ is the trivial group.  It is simple to work out the decomposition $\mathfrak{su}(2)_1 \cong \bigoplus_{n=0}^\infty V_n \otimes \mathcal{R}_{h=n^2}^\Vir$ in terms of $\mathfrak{su}(2) \times \Vir_{c=1}$ representations, for example by using the fact that the vacuum character of $\mathfrak{su}(2)_1$ equals $\frac{\vartheta_3(z|2\tau)}{\eta(\tau)}$, where $z$ is a fugacity for the Cartan generator $J^3_0$ of $\mathfrak{su}(2)$, see e.g.~\cite[Sect.~14.4]{DiFrancesco:1997nk}. Thus, the multiplicity spaces are $A_n = V_n$ in this case, and $A=\bigoplus_{n \ge 0} V_n \otimes V_n \cong \CC[\SO(3)]$ is the statement of the Peter--Weyl theorem. The left action on $\CC[\SO(3)]$ can be taken to be the one turning $A$ into an $\SO(3)$ equivariant algebra, while the right action is the $\SO(3)$ action acting on the multiplicity spaces, i.e.\ the action of $\Aut(\mathfrak{su}(2)_1) \cong \SO(3)$. Taking invariants on both sides gives the claim
\be
\mathfrak{su}(2)_1^\Gamma \quad \longleftrightarrow \quad \CC[\SO(3)]^\Gamma=\CC[\SO(3)/\Gamma]\ .
\ee
Since $\mathcal{A}_\mathrm{deg}$ can be recovered from $A$, the theorem follows.

\paragraph{The possible chiral algebras $\mathcal{A}_\mathrm{deg}$.} There is only a small list of possible closed subgroups of $\SO(3)$ and they give the possible degenerate chiral algebras $\mathcal A_\mathrm{deg}$. Since $\mathfrak{su}(2)_1 \cong \bigoplus_{n=0}^\infty V_n \otimes \mathcal{R}_{h=n^2}^\Vir$, one can then determine the field content of all possible chiral algebras $\mathcal{A}_\mathrm{deg}$ by computing the number of singlet representations of $\Gamma$ inside the $(2n+1)$-dimensional representation of $\SO(3)$. Let us denote this number by $a_\Gamma(n)$. To express $a_\Gamma(n)$ for the different subgroups, it is convenient to specify the generating function (the Molien series) $f_\Gamma(x) = \sum_{n=0}^\infty a_\Gamma(n)x^n$, which is listed in Table~\ref{tab:singlet numbers} for the different subgroups.
\begin{table}[htbp]
    \centering
    \renewcommand{\arraystretch}{1.3}
    \begin{tabular*}{\linewidth}{@{\extracolsep{\fill}}l|*{4}{c}}
        \toprule
        $\Gamma$ & $\{1\}$ & $\ZZ_m$ & $\mathrm{D}_m$ & $\mathrm{T}$ \\
        \hline
        \rule[-1.3em]{0pt}{3.3em}$f_\Gamma(x)$ & $\dfrac{1+x}{(1-x)^2}$ & $\dfrac{1+x^m}{(1-x)(1-x^m)}$ & $\dfrac{1+x^{m+1}}{(1-x^2)(1-x^m)}$ & $\dfrac{1+x^6}{(1-x^3)(1-x^4)}$ \\
        \bottomrule
    \end{tabular*}

    \vspace{1.1em}

    \begin{tabular*}{\linewidth}{@{\extracolsep{\fill}}l|*{5}{c}}
        \toprule
        $\Gamma$ & $\mathrm{O}$ & $\mathrm{I}$ & $\SO(2)$ & $\Orth(2)$ & $\SO(3)$ \\
        \hline
        \rule[-1.3em]{0pt}{3.3em}$f_\Gamma(x)$ & $\dfrac{1+x^9}{(1-x^4)(1-x^6)}$ & $\dfrac{1+x^{15}}{(1-x^6)(1-x^{10})}$ & $\dfrac{1}{1-x}$ & $\dfrac{1}{1-x^2}$ & $1$ \\
        \bottomrule
    \end{tabular*}
    \caption{The Molien series $f_\Gamma(x)=\sum_{n\ge0}a_\Gamma(n)\,x^n$, whose coefficient $a_\Gamma(n)$ counts the $\Gamma$-singlets in the spin-$n$ representation of $\SO(3)$, for the closed subgroups $\Gamma\subset\SO(3)$.}
    \label{tab:singlet numbers}
\end{table}

The main point of the table is that the functions $f_\Gamma(x)$ are distinct for the different $\Gamma$-subgroups.\footnote{The only apparent exception is the case $f_{\ZZ_2}(x)=f_{\mathrm{D}_1}(x)$, but in this case the corresponding groups are also isomorphic.} This means that we can uniquely reconstruct $\mathcal{A}_\mathrm{deg}$ as an algebra from knowing the decomposition of $\mathcal{A}_\mathrm{deg}$ in terms of Virasoro representations.

In the present context, the decomposition \eqref{eq:Z2 orbifold chiral algebra decomposition} tells us that $\Gamma=\Orth(2)$ for generic $R$. For special values of $R$, the multiplicities of Virasoro representations in $\mathcal{A}_\mathrm{deg}$ can only get bigger. According to the table, the only such possibilities are those that are subgroups of $\Orth(2)$. This means that $\mathcal{A}_\mathrm{deg}$ contains as a subalgebra the $\ZZ_2$-singlet sector of the Heisenberg algebra.

\paragraph{A $\ZZ_2$ symmetry.} We have established that the theory at hand at least contains the $\ZZ_2$-singlet sector of the Heisenberg algebra as a chiral algebra. Therefore, the fusion rules carry a $\Rep(\ZZ_2) \cong \ZZ_2$-symmetry under which twisted sector representations are odd. This is established rigorously in the context of VOAs in~\cite{Abe:1999rm}. Since the theory is unitary, the Hilbert space decomposes as a direct sum of irreducible $M(1)^+$-representations, and since the fusion rules of $M(1)^+$-representations are $\ZZ_2$-graded, the grading is multiplicative in the OPE, i.e.\ an automorphism of the theory. Therefore, all correlation functions of the theory respect the $\ZZ_2$-symmetry.

Next, we will show that all the characters of $M(1)^+$ are linearly independent. This will allow us to determine the field content of the theory in terms of $M(1)^+$-representations. The irreducible $M(1)^+$-modules are $M(1)^\pm$, the one-parameter family $M(1,P) \cong M(1,-P)$ with $P>0$, and the two twisted-sector modules $M(1)(\theta)^\pm$~\cite{Dong:1998rk}. The characters are
\begin{subequations}
\begin{align}
\chi_{M(1)^\pm}(\tau)&=\frac{q^{-\frac{1}{24}}}{2} \bigg(\prod_{n=1}^\infty (1-q^n)^{-1}\pm \prod_{n=1}^\infty (1+q^n)^{-1}\bigg)\ , \\ \chi_{M(1,P)}(\tau)&=q^{P^2-\frac{1}{24}} \prod_{n=1}^\infty (1-q^n)^{-1}\ , \\
\chi_{M(1)(\theta)^\pm}(\tau)&=\frac{q^{\frac{1}{48}}}{2}\bigg(\prod_{n=1}^\infty (1-q^{n-\frac{1}{2}})^{-1} \pm \prod_{n=1}^\infty (1+ q^{n-\frac{1}{2}})^{-1}\bigg)\ ,
\end{align}
\end{subequations}
with $q=\e^{2\pi i \tau}$ as usual. One can check that any finite subset of these characters is linearly independent.

Thus, the field content as $M(1)^+$-representations agrees with the one of the $\ZZ_2$-orbifold of the free boson. In particular, we can write $\rho^\mathrm{orb}_R=\rho_{R,+}^\mathrm{orb}+\rho_{R,-}^\mathrm{orb}$, where the $+$ and $-$ signs denote the untwisted and twisted sector contributions which are even and odd under the diagonal $\ZZ_2$-symmetry.

Thus we conclude that the theory $\mathcal{T}$ under consideration indeed carries a $\ZZ_2$-symmetry. Since $\mathrm{H}^3(\ZZ_2,\Uone)=\ZZ_2$, the $\ZZ_2$-action could potentially be anomalous. However, this putative anomaly can already be detected at the level of the torus partition function and manifests as a failure of modular invariance of the gauged torus partition function, see e.g.~\cite{Bhardwaj:2017xup}. Carrying out the gauging on the torus partition function is however guaranteed to yield the torus partition function of the free boson. Indeed, the density of states of the trace with $\ZZ_2$-insertion is given by $\rho_{R,+}^\mathrm{orb}-\rho_{R,-}^\mathrm{orb}$ and the twisted sector can then be computed by applying modular transformations. On the level of the partition function, this gauging is precisely the gauging of the $\ZZ_2$ quantum symmetry of the orbifold theory, which recovers the free boson partition function~\cite{Vafa:1986wx}. Thus $Z_{\TT^2}[\mathcal{T}/\ZZ_2]=Z_{\TT^2}[\mathrm{S}^1_R]$.

Finally, we use the result of Section~\ref{subsec:free boson} to conclude that $\mathcal{T}/\ZZ_2 \cong \mathrm{S}^1_R$, the free boson theory. Using invertibility of the gauging shows that
\be
\mathcal{T} \cong \mathrm{S}^1_R/\ZZ_2 \label{eq:T Z2 orbifold}
\ee
is indeed a $\ZZ_2$-orbifold of the free boson as desired.
Let us also notice that, up to conjugation, there are three non-anomalous $\ZZ_2$-actions on the free boson theory by which we can orbifold. At generic radius the symmetry group is $(\Uone\times\Uone)\rtimes\ZZ_2$, generated by the momentum and winding shifts and by the reflection~\cite[eq.~(3.2)]{Thorngren:2021yso}. The product of the two half-shifts is anomalous, since its twisted sector has spins in $\frac{1}{4}+\frac{1}{2}\ZZ$. The first non-anomalous action is a half-shift of the circle and the orbifold results in $\mathrm{S}^1_R/\ZZ_2 \cong \mathrm{S}^1_{R/2}$. The second one is the half-shift in the T-dual frame, which results in $\mathrm{S}^1_R/\ZZ_2 \cong \mathrm{S}^1_{2R}$~\cite{Ginsparg:1987eb}. The third one is the reflection giving the partition function of interest. The gauging is also unique in the sense that no discrete theta-angles can be introduced because $\mathrm{H}^2(\ZZ_2,\Uone)=0$. This shows that the orbifold in \eqref{eq:T Z2 orbifold} is uniquely specified.

\subsection{Exceptional theories}
It remains to show that the exceptional cases can also be uniquely realized as $\SU(2)_1/\Gamma$ with $\Gamma=\mathrm T,\,\mathrm O,\,\mathrm I$ the tetrahedral, octahedral and icosahedral groups. The strategy is the same as for the $\ZZ_2$-orbifold of the free boson. We will first determine the chiral algebra and then show that it gives rise to a non-anomalous, non-invertible $\Rep(\Gamma)$ symmetry of $\mathcal T$, realized by topological defect lines. Gauging this symmetry leads to $\SU(2)_1$, whose uniqueness was already established in Section~\ref{subsec:free boson}. Therefore the three exceptional theories can be realized as $\SU(2)_1/\Gamma$. Another difference from the $\ZZ_2$-orbifold is that the inverse gauging is potentially ambiguous because $\mathrm{H}^2(\Gamma,\Uone)\cong\ZZ_2$ for all three groups. We will explain at the end that the two choices lead to equivalent theories.

Let us also notice that the tetrahedral and octahedral groups are solvable. We have $\mathrm T\cong\mathrm A_4$, with the Klein four-group $\mathrm V_4$ as a normal subgroup, and $\mathrm O\cong\mathrm S_4$, with $\mathrm A_4$ as a normal subgroup. The quotient groups are $\mathrm A_4/\mathrm V_4\cong\ZZ_3$ and $\mathrm S_4/\mathrm A_4\cong\ZZ_2$, respectively. As a consequence of the $\Rep(\mathrm T)$ (resp.\ $\Rep(\mathrm O)$) symmetry, we also have an ordinary $\ZZ_3$ (resp.\ $\ZZ_2$) symmetry that we could gauge. These gaugings lead to $\SU(2)_1/\mathrm{V}_4$ and $\SU(2)_1/\mathrm T$, respectively~\cite{Thorngren:2021yso}.\footnote{The former can be identified with the $\ZZ_2$-orbifold of the free boson at radius $R=2$. Indeed, $\mathrm V_4\cong\ZZ_2\times\ZZ_2$; at the self-dual radius one generator can be taken to act by a half-shift and the other by a reflection. Performing first the half-shift orbifold leads to a free boson at $R=\frac12$, or equivalently at $R=2$, and then one performs the reflection orbifold.}
This gives an alternative treatment of the tetrahedral and octahedral cases using only ordinary symmetries. For the icosahedral theory this is not an option, and we therefore treat the three theories uniformly.

\paragraph{Chiral fields.} The first step consists of identifying the chiral algebra $\mathcal A$. As one can see from the density of states \eqref{eq:exceptional solutions}, $\frac12(-\rho_1+\rho_2+\rho_3+\rho_i)$ for $i=3,\,4,\,5$ for the tetrahedral, octahedral and icosahedral cases, the chiral algebra only consists of degenerate Virasoro primary fields, i.e.\ $\mathcal A=\mathcal A_{\mathrm{deg}}$. We can then use Theorem~\ref{thm:Carpi} to conclude that $\mathcal A=\mathcal A_{\mathrm{deg}}\cong\mathfrak{su}(2)_1^\Gamma$, since the spectrum determines the group $\Gamma$ completely. Let us also notice that the brute-force route towards this result that we followed in Section~\ref{subsec:Z2 orbifold free boson} is in principle also possible, but hardly practical: for the icosahedral case, the lowest higher-spin field appears at $h=36$ and its OPE has a very large number of terms.

\paragraph{A $\boldsymbol{\Rep(\Gamma)}$ fusion subcategory.} For the $\ZZ_2$-orbifold, we proceeded by reading off the field content of the theory in terms of $\mathcal A$-modules. This required explicit knowledge of the irreducible characters of the algebra, which will not be needed in the following argument. We instead proceed by quantum Schur--Weyl duality.

We can consider $\mathfrak{su}(2)_1$ both as an $\mathcal A$-module and a $\Gamma$-module. In fact, we have the decomposition
\be
\mathfrak{su}(2)_1\cong\bigoplus_{\varrho\in\Irr(\Gamma)}R_\varrho^*\otimes\mathcal A_\varrho\ ,
\ee
where $R_\varrho$ is the representation space of the irreducible representation $\varrho$ of $\Gamma$ and $R_\varrho^*$ is associated to the dual representation $\varrho^*$. This defines the $\mathcal A$-modules $\mathcal A_\varrho$, with $\mathcal A_0\cong\mathcal A$ for the trivial representation that we denote by 0. Furthermore, the $\mathcal A_\varrho$ are irreducible and pairwise non-isomorphic~\cite{Dong:1996as}.

The result of~\cite{McRae:2018wpu} shows that the modules $\mathcal A_\varrho$ generate a symmetric tensor subcategory equivalent to $\Rep(\Gamma)$. This equivalence can be written as
\be
\Phi:\Rep(\Gamma)\longrightarrow\Rep(\mathcal A)\ ,
\qquad
\Phi(R):=\bigl(R\otimes\mathfrak{su}(2)_1\bigr)^\Gamma\ ,
\qquad
\Phi(R_\varrho)\cong\mathcal A_\varrho\ ,
\ee
where the superscript $\Gamma$ denotes the $\Gamma$-singlet part.
Consequently,
\be
\mathcal A_\varrho\otimes\mathcal A_\sigma
\cong\bigoplus_{\tau\in\Irr(\Gamma)}
\tensor{N}{_{\varrho\sigma}^{\tau}}\,\mathcal A_\tau\ ,
\qquad
\tensor{N}{_{\varrho\sigma}^{\tau}}
=\dim\Hom_\Gamma(R_\varrho\otimes R_\sigma,R_\tau)\ .
\ee
Note that these are far from being all the modules of $\mathcal A$: in particular, the analogs of twisted-sector modules are not included in this subcategory.

Under $\Phi$, the regular representation of $\Gamma$ realized as the space of functions $\Fun(\Gamma)$ on $\Gamma$ is mapped to the original chiral algebra extension,
\be
\Phi\big(\Fun(\Gamma)\big)=\big(\Fun(\Gamma) \otimes \mathfrak{su}(2)_1\big)^\Gamma \cong \mathfrak{su}(2)_1 \cong \bigoplus_{\varrho \in \Irr(\Gamma)} \dim R_\varrho\, \mathcal{A}_\varrho \ . \label{eq:Phi Fun Gamma}
\ee
The identification with $\mathfrak{su}(2)_1$ holds as an algebra extension, not just as an $\mathcal A$-module.

\paragraph{$\boldsymbol{\alpha}$-induction.} For this step we use the stronger fact that $\mathcal A$ is strongly rational. For the tetrahedral and octahedral cases this follows from the regularity theorem for fixed points by finite solvable groups~\cite{Carnahan:2016guf}, see also~\cite{Miyamoto:2015, McRae:2021yyb}, while strong rationality of the icosahedral fixed-point algebra is proved in the recent preprint~\cite{Xu:2026aa}.

Decompose the Hilbert space as $\mathcal A$ and $\overline{\mathcal A}$ representations as follows,
\be
\mathcal H=\bigoplus_{i,j}M_{ij}\,\mathcal A_i\boxtimes\overline{\mathcal A}_j\ ,
\qquad \mathcal A_0=\mathcal A\ .
\ee
Since $\mathcal{A}$ is strongly rational, this sum is finite and $\mathcal{T}$ is a rational CFT. Since $\mathcal A$ is the full chiral algebra, we have
\be
M_{i0}=M_{0i}=\delta_{i0}\ .
\label{eq:no chiral extension exceptional}
\ee
$\alpha$-induction assigns to every $\mathcal A$-representation $\mathcal A_i$ a topological defect line $\alpha^+(\mathcal A_i)$ of $\mathcal T$, in such a way that fusion of defects reproduces fusion of representations~\cite{Frohlich:2003hm,Frohlich:2006ch}. This assignment can fail to be injective and it does so precisely when the chiral algebra of $\mathcal{T}$ is bigger than $\mathcal{A}$ \cite[Prop.~2.36]{Frohlich:2003hm}. In our case, the chiral algebra of $\mathcal{T}$ is exactly $\mathcal{A}$ and thus $\alpha$-induction is fully faithful and the defects
\be
\mathcal L_\varrho:=\alpha^+(\mathcal A_\varrho)
\ee
have the same fusion rules and the same junction spaces as the representations $\varrho$ of $\Gamma$, and therefore realize a $\Rep(\Gamma)$-symmetry of $\mathcal T$. The regular algebra $\Phi(\Fun(\Gamma))$ is mapped to the gaugeable defect
\be
\mathcal L_{\Phi(\Fun(\Gamma))}=\bigoplus_{\varrho\in\Irr(\Gamma)}\dim R_\varrho\, \mathcal L_\varrho\ . \label{eq:gaugable defect}
\ee
Its canonical multiplication and junctions define a non-anomalous gauging~\cite{Frohlich:2009gb,Bhardwaj:2017xup}.

\paragraph{Gauging $\boldsymbol{\Rep(\Gamma)}$.}
We can therefore consider the gauged theory
\be
\mathcal{T}':=\mathcal{T}/\Rep(\Gamma)\ .
\ee
Gauging $\Rep(\Gamma)$ does not merely produce a new modular invariant for $\mathcal{A}$, it extends the chiral algebra.
The local fields of $\mathcal{T}'$ are the states of the $\mathcal{L}_\varrho$-twisted sectors of $\mathcal{T}$ and since we gauge \eqref{eq:gaugable defect}, they appear with multiplicity $\dim R_\varrho$. The trivial representation gives the untwisted sector. The multiplicity of the holomorphic representation $\mathcal{A}_i \boxtimes \overline{\mathcal{A}}_0$ in the defect Hilbert space of $\mathcal{L}_\varrho$ is given by the number of topological junctions between $\mathcal{L}_i \otimes \mathcal{L}_\varrho$ and the trivial defect~\cite[eq.~(2.15)]{Frohlich:2006ch}. By injectivity of $\alpha$-induction, this equals $\dim \Hom_{\Rep(\mathcal{A})}(\mathcal{A}_i \otimes \mathcal{A}_\varrho,\mathcal{A}_0)$. Since $\mathcal{A}_\varrho \otimes \mathcal{A}_{i} \supset \mathcal{A}_0$ only for $i=\varrho^*$, the $\mathcal L_\varrho$-twisted sector contains exactly one holomorphic multiplet, transforming in $\mathcal A_{\varrho^*}$. The holomorphic fields of $\mathcal{T}'$ therefore assemble into $\bigoplus_{\varrho \in \Irr(\Gamma)}\dim R_\varrho\, \mathcal A_\varrho=\Phi(\Fun(\Gamma))$, which defines an extended chiral algebra. By \eqref{eq:Phi Fun Gamma}, the extended chiral algebra is therefore $\mathfrak{su}(2)_1$. A similar conclusion applies to the anti-holomorphic fields.

$\mathfrak{su}(2)_1$ contains in particular the Heisenberg algebra. The only possible partition function among the list \eqref{eq:free boson and spurious solution}, \eqref{eq:orbifold solution} and \eqref{eq:exceptional solutions} with a Heisenberg chiral algebra is the free boson partition function, given that we already ruled out the existence of the spurious solution in Section~\ref{subsec:spurious solution}. The uniqueness theorem established in Section~\ref{subsec:free boson} thus implies that $\mathcal{T}'$ must be a free boson theory and, since the algebra enhances to $\mathfrak{su}(2)_1$, it must sit at the self-dual radius, i.e.
\be
\mathcal{T}'\cong\SU(2)_1\ .
\ee

\paragraph{Gauging $\boldsymbol{\Gamma}$.} Gauging $\Rep(\Gamma)$ is invertible and the dual symmetry of $\mathcal{T}$ becomes the symmetry $\Gamma$ acting on $\SU(2)_1$~\cite{Bhardwaj:2017xup}. Since we know the action of $\Gamma$ on the left- and right-moving chiral algebra, this determines the action of $\Gamma$ on the full theory, up to a relative automorphism of $\Gamma$ twisting the action on the right-movers with respect to the left-movers. For the tetrahedral and octahedral case, there is a unique 3-dimensional representation mapping into $\SO(3)$ and thus the action is unique. For the icosahedral case, there are two inequivalent 3-dimensional representations $\mathbf{3}$ and $\mathbf{3'}$ and one can in principle let $\mathrm{I}$ act in the representation $\mathbf{3}$ on the left-movers and in the representation $\mathbf{3'}$ on the right-movers.

However, that orbifold cannot lead to a theory $\mathcal{T}$ with density of states $\rho^\mathrm{ico}$.\footnote{In fact, that orbifold is anomalous as can be seen from attempting to compute the density of states.} To see this, suppose for the sake of contradiction that it would.
Recall that the irreducible representations of $\mathrm{I}$ are $\varrho \in \{\mathbf{1},\mathbf{3},\mathbf{3'},\mathbf{4},\mathbf{5}\}$.
The states in the untwisted sector of the orbifold decompose in terms of $\mathcal{A} \boxtimes \overline{\mathcal{A}}$-representations as
\be
\bigoplus_\varrho \mathcal{A}_\varrho \boxtimes \overline{\mathcal{A}}_{\phi(\varrho)}\ ,
\ee
where $\phi$ swaps $\mathbf{3}$ to $\mathbf{3'}$ and fixes the other irreps. From the branching rule $\SO(3) \longrightarrow \mathrm{I}$, one can see that the primary states of $\mathcal{A}_{\mathbf{3}}$ have $h=1$ (since they arise from the currents of $\mathfrak{su}(2)_1$ themselves), while the primary states of $\mathcal{A}_{\mathbf{3'}}$ have $h=9$. Thus, the existence of the term $\mathcal{A}_{\mathbf{3}} \boxtimes \overline{\mathcal{A}}_{\mathbf{3'}}$ in the untwisted sector means that the resulting orbifold partition function has Virasoro primary states with $(h,\bar h)=(1,9)$ and $(h,\bar h)=(9,1)$. However, it is easy to check that the density of states $\rho^\mathrm{ico}$ does not contain such a Virasoro primary state. Therefore, that orbifold cannot lead to a theory with the correct density of states.

The only remaining freedom in the gauging is to include a discrete torsion class in $\mathrm{H}^2(\Gamma,\Uone)$. For $\Gamma=\mathrm{A}_4,\, \mathrm{S}_4,\, \mathrm{A}_5$, we have $\mathrm{H}^2(\Gamma,\Uone) \cong \ZZ_2$. As discussed in~\cite{Thorngren:2021yso}, the two different choices of gauging are related by a chiral $\ZZ_2$ rotation of the $\SU(2)_1$ theory and lead to equivalent CFTs. There is thus only one possible gauging and thus
\be
\mathcal{T} \cong \SU(2)_1/\Gamma\ .
\ee
This finally finishes the full classification and proves Theorem~\ref{thm:c1 classification}.

\section{Discussion}
In this paper, we have demonstrated that Ginsparg's proposed classification of unitary $c=1$ CFTs given in Theorem~\ref{thm:c1 classification} is complete.
Along the way, we have developed some useful technology, especially with regard to the modular bootstrap. The proof hinged on two main techniques: (i) leveraging S-modular invariance of $\rho$ to precisely quantify the growth of states and (ii) applying Cohen's theorem for the Bohr compactification $\mathrm{b}\RR^2$ to conclude that $\rho$ must have a lattice-like structure. From there, the techniques are relatively standard to constrain $\rho$ further into the different admissible forms.

\begin{figure}[htbp]
\centering
% strand colours (Okabe--Ito palette)
\definecolor{strandUnitary}{HTML}{0072B2}
\definecolor{strandVacuum}{HTML}{009E73}
\definecolor{strandNonUnitary}{HTML}{E69F00}
\definecolor{strandCeffOne}{HTML}{D55E00}
\definecolor{strandCless}{HTML}{CC79A7}
\begin{tikzpicture}[
  font=\scriptsize,
  station/.style={circle, draw=black!55, fill=black!3, line width=0.6pt,
                  minimum size=2*\pRad cm, inner sep=0pt, align=center},
  detail/.style={anchor=north, align=center, inner sep=1pt, text=black!80},
  strand/.style={line width=1.1pt, draw=#1, line cap=round,
                 -{Stealth[length=2mm, width=1.6mm, round]}},
]
  % ---------------- geometry ----------------
  \def\pSep{2.58}    % distance between the centers of neighbouring nodes
  \def\pRad{1.07}    % radius of the nodes
  \def\pDel{0.11}    % distance between parallel strands
  \def\pWaist{0.4}   % height of the innermost strands between two nodes, minus \pDel
  % A strand at level l runs around a node on a circle of radius \pRad+l*\pDel and,
  % between two nodes, around a "waist" circle of radius \pQ-l*\pDel centred at height \pY
  % and tangent to both. Strands at different levels are then exactly parallel.
  \pgfmathsetmacro\pQ{((\pSep/2)^2+\pWaist^2-\pRad^2)/(2*(\pRad-\pWaist))}
  \pgfmathsetmacro\pY{\pQ+\pWaist}
  \pgfmathsetmacro\pTheta{atan(\pY/(\pSep/2))}
  % \pStrand{colour}{side: 1 above, -1 below}{level}{first node}{last node}
  \newcommand\pStrand[5]{%
    \begin{scope}[yscale=#2]
      \pgfmathsetmacro\sR{\pRad+#3*\pDel}%
      \pgfmathsetmacro\sQ{\pQ-#3*\pDel}%
      \pgfmathsetmacro\sH{\pWaist+#3*\pDel}%
      \pgfmathsetmacro\sDx{sqrt(\pRad^2-\sH^2)}%
      \pgfmathtruncatemacro\sFirst{#4+1}%
      \pgfmathtruncatemacro\sLast{#5-1}%
      \ifnum\sFirst>\sLast
        \draw[strand=#1] (#4*\pSep+\sDx,\sH) -- (#5*\pSep-\sDx,\sH);
      \else
        \draw[strand=#1] (#4*\pSep+\sDx,\sH) -- (#4*\pSep+\pSep/2,\sH)
          \foreach \pIter in {\sFirst,...,\sLast}{
            arc[start angle=270, end angle=360-\pTheta, radius=\sQ]
            arc[start angle=180-\pTheta, end angle=\pTheta, radius=\sR]
            arc[start angle=180+\pTheta, end angle=270, radius=\sQ]
          }
          -- (#5*\pSep-\sDx,\sH);
      \fi
    \end{scope}}
  % ---------------- nodes ----------------
  \node[station] at (1*\pSep,0) {Axioms};
  \node[station] at (2*\pSep,0) {Upper\\ bounded\\ density};
  \node[station] at (3*\pSep,0) {Finite sum\\ of Narain\\ combs};
  \node[station] at (4*\pSep,0) {$m_R\ge 0$\\ for $R\neq 1$};
  \node[station] at (5*\pSep,0) {Classification\\ of partition\\ functions};
  \node[station] at (6*\pSep,0) {Classification\\ of CFTs};
  % ---------------- strands ----------------
  \pStrand{strandUnitary}{1}{2}{1}{6}
  \pStrand{strandVacuum}{1}{1}{1}{5}
  \pStrand{strandCeffOne}{-1}{1}{1}{3}
  \pStrand{strandCless}{-1}{2}{1}{3}
  \pStrand{strandNonUnitary}{-1}{3}{1}{4}
  % second branch of the c_eff=1 strand: leaves "Upper bounded density" at angle -\pBranch
  % and merges tangentially into the first branch, \pMerge to the right of its waist
  \def\pBranch{22}
  \def\pMerge{0.08}
  \pgfmathsetmacro\pBranchH{\pWaist+\pDel}
  \draw[line width=1.1pt, draw=strandCeffOne, line cap=round]
    ({2*\pSep+\pRad*cos(\pBranch)},{-\pRad*sin(\pBranch)})
    to[out=-\pBranch, in=180] (2*\pSep+\pSep/2+\pMerge,-\pBranchH);
  % ---------------- details below the nodes ----------------
  \pgfmathsetmacro\yDetail{-\pRad-3*\pDel-0.1}
  \node[detail] at (2*\pSep,\yDetail) {$|\rho|(B_R)\le C_\rho R^2$};
  \node[detail] at (3*\pSep,\yDetail) {$\rho=\frac{1}{2}\sum_R m_R\,\rho_R$\\[1pt] $m_R\in\ZZ$};
  \node[detail] at (5*\pSep,\yDetail) {Ginsparg's list\\[1pt] + spurious solution};
  \node[detail] at (6*\pSep,\yDetail) {Ginsparg's list};
  % ---------------- legend ----------------
  \pgfmathsetmacro\yLegend{\yDetail-1.15}
  \newcommand\pLegendEntry[4]{% x, y, colour, text
    \draw[strand=#3] (#1,#2) -- ++(0.7,0);
    \node[anchor=west, inner sep=0pt] at (#1+0.85,#2) {#4};}
  \pLegendEntry{\pSep-\pRad+0.75}{\yLegend}{strandUnitary}{Unitary $c=1$}
  \pLegendEntry{\pSep-\pRad+0.75}{\yLegend-0.38}{strandVacuum}{Non-unitary $c=1$, $h,\bar h\ge 0$, unique vacuum}
  \pLegendEntry{\pSep-\pRad+0.75}{\yLegend-0.76}{strandNonUnitary}{Non-unitary, $c=1$, $h,\bar h\ge 0$}
  \pLegendEntry{\pSep-\pRad+8.4}{\yLegend}{strandCeffOne}{Non-unitary $c_\mathrm{eff}=1$}
  \pLegendEntry{\pSep-\pRad+8.4}{\yLegend-0.38}{strandCless}{Unitary $c<1$, or non-unitary $c_\mathrm{eff}<1$}
\end{tikzpicture}
\caption{Validity of the different parts of our argument with modified assumptions (non-unitarity or $c<1$). Arrows point from the assumptions to the valid conclusions in each case. `Unitary $c=1$' is our main case where the full argument applies. The first node `Axioms' refers to those of \ref{axiom:S invariance}--\ref{axiom:vacuum} that apply: \ref{axiom:S invariance}--\ref{axiom:integrality} are satisfied in all cases; positivity \ref{axiom:positivity} is satisfied for all of the $c=1$ cases; and unique vacuum \ref{axiom:vacuum} is only satisfied for the top two cases.}
\label{fig:validity}
\end{figure}
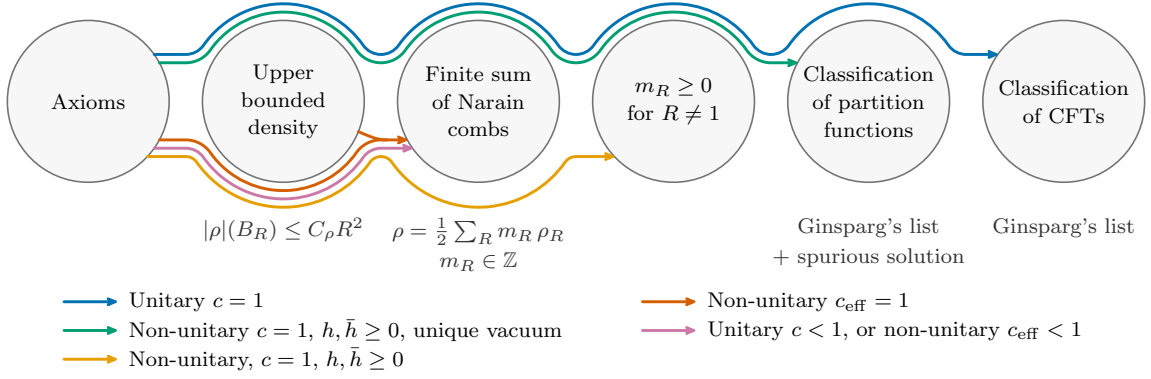

We now make some further remarks and discuss some possible future generalizations of these techniques. Figure~\ref{fig:validity} summarizes which steps of our argument remain valid for non-unitary theories and for $c<1$.

\paragraph{Non-unitary CFTs.} Many of the results of this paper also apply to non-unitary theories with $c=1$. In order for our bootstrap techniques to apply, we merely need to assume that all states have $h,\bar h\ge 0$. Null vectors might not decouple for non-unitary theories, but this only improves the positivity properties of the degeneracies. If we additionally assume a unique vacuum, our classification of partition functions still holds: $Z$ must be one of Ginsparg's partition functions or the spurious solution, i.e.\ \eqref{eq:free boson and spurious solution}, \eqref{eq:orbifold solution} or \eqref{eq:exceptional solutions}. Even without a unique vacuum, the density of states still has to be a finite linear combination of free boson densities of the form \eqref{eq:rho integer free boson decomposition} with $m_1\in \ZZ$ and $m_{R\neq 1}\in \NN$.

Many of the techniques also carry over to the case of $c_\mathrm{eff}=c-24h_\mathrm{min}=1$, which still ensures only a polynomial growth of density of states. However, null-vectors are now located at $P_{r,s}=\frac{1}{2}|\beta r-\beta^{-1} s|$ with $r,\, s \in \NN$ and $c=1-6(\beta-\beta^{-1})^2$, as opposed to the half-integer locations for $c=1$. Thus our proof of upper bounded density does not apply, although it seems plausible to us that it can be adapted, at least in the case of $\beta^2 \in \QQ$ where the set of degenerate weights remains discrete. Assuming upper bounded density, however, the results of Section~\ref{sec:modular bootstrap crystalline measures} still apply (except for the last step in Subsection~\ref{subsec:Kiritsis} that relies on positivity in the sense of $c=1$ CFTs). This again implies that the density of states is a linear combination of free boson densities of the form \eqref{eq:rho integer free boson decomposition}, now with $m_R\in \ZZ$.
Naturally, the results of Section~\ref{sec:uniqueness} completely fail without unitarity.

\paragraph{The case of $\boldsymbol{c<1}$.} CFTs with $c<1$, or more generally $c_\mathrm{eff}<1$, are Virasoro minimal models. These are RCFTs and are of course fully classified. It is, however, interesting to note that the argument of this paper also applies to them. Parametrize conformal weights in terms of Liouville momenta as $h=\frac{c-1}{24}+P^2$, and similarly for the right-movers, and define a density of states $\rho$ by decomposing into Fock characters as in \eqref{eq:definition density of states}.  The density of states of a single product of left- and right-moving degenerate Virasoro characters satisfies \eqref{eq:bounded density} and, since there are only finitely many characters, the density of states defined in this way also has upper bounded density. Thus, only the positivity axiom~\ref{axiom:positivity} and the vacuum axiom~\ref{axiom:vacuum} fail. The latter fails because the vacuum has $P=\bar P=\sqrt{\frac{1-c}{24}} \ne 0$. This reduced set of axioms is still sufficient for all the arguments in Section~\ref{sec:modular bootstrap crystalline measures} to apply and to again obtain the conclusion \eqref{eq:rho integer free boson decomposition} with $m_R \in \ZZ$. Thus, it follows that the minimal model density of states can be expressed as finite linear combinations of free boson partition functions with half-integer coefficients. This is indeed the case, e.g.\ for the $\mathcal{M}(p,p')$ minimal model with the A-modular invariant, one has
\be
\rho^{\mathcal{M}(p,p')}=\frac{1}{2}\big(\rho_{\sqrt{pp'}}-\rho_{\sqrt{p/p'}}\big)\ ,
\ee
see e.g.~\cite{Kapec:2020xaj}. This relation carries over to the partition function. Similar formulas apply for other modular invariants.

\paragraph{Role of integrality.} An important role was played in this work by integrality of the degeneracies. This was crucially used to prove boundedness of the density of states, and to prove the classification theorem of crystalline measures. Integrality has historically been a difficult property to implement in the modular bootstrap, particularly in numerical studies, because it breaks the convexity of the problem.\footnote{We point out that some recent studies have made use of integrality~\cite{Fitzpatrick:2023lvh,Chiang:2023qgo}, but it certainly remains relatively less explored. }

\paragraph{Genericity of extended chiral algebras.} There is a prevailing lore in the physics community that ``generic'' unitary 2d CFTs with discrete spectrum and $c \ge 1$ should only have Virasoro symmetry and no extended symmetry. We find it interesting to remark that this expectation fails for $c=1$, where all the examples of Theorem~\ref{thm:c1 classification} possess an extended chiral algebra. Moreover, perhaps surprisingly, rational CFTs are \emph{dense} in the moduli space of all CFTs at $c=1$ (although they have zero measure).

\paragraph{The power of the modular bootstrap.} For most of the paper, we only imposed the modular bootstrap axioms and have shown that the possible partition functions are already uniquely classified by them, except for the spurious solution. Demanding the full CFT axioms including crossing symmetry was not necessary. This is perhaps surprising and shows how strong the modular bootstrap axioms really are. Technically, the main difficulty is usually to impose all the axioms simultaneously in an effective way.

\paragraph{Non-discrete theories.} We should of course mention that the discreteness assumption is very important in our analysis. There are known unitary non-discrete CFTs with $c=1$ beyond the non-compact free boson and its $\ZZ_2$-orbifold, namely timelike Liouville theory~\cite{Ribault:2015sxa} and Runkel-Watts theory~\cite{Runkel:2001ng}. Interestingly, the latter two theories share the torus partition function of the non-compact free boson, so we expect the modular bootstrap approach to become much less powerful if the discreteness assumption is removed.

\paragraph{$\boldsymbol{\mathcal{N}=1}$ SCFTs with $\boldsymbol{c=\frac{3}{2}}$.} A similar proposed classification exists for unitary $\mathcal{N}=(1,1)$ SCFTs with central charge $c=\frac{3}{2}$ and discrete spectrum~\cite{Dixon:1988ac}, where five continuous families and six isolated theories are found. Since characters of the $\mathcal{N}=1$ Virasoro algebra already account for the asymptotic growth of states, the Fock density of states plausibly also defines a tempered measure and methods similar to those in Section~\ref{sec:modular bootstrap crystalline measures} could plausibly show that the density of states is lattice-like, which could lead to a full proof.

\paragraph{$\boldsymbol{\mathcal{N}=2}$ SCFTs with $\boldsymbol{c=3}$.} In a similar spirit, all known unitary discrete $c=3$ $\mathcal{N}=2$ SCFTs are obtained as orbifolds of $\mathbb{T}^2$ sigma models~\cite{Aspinwall:1994ay, Dulat:2000ae}, but to our knowledge completeness of that list was not claimed. This is another instance where the $\mathcal{N}=2$ characters already account for the full growth of states and the density of superprimary states becomes a tempered measure. Thus the methods of this paper plausibly apply to give a full classification.

\paragraph{Other theories with $\boldsymbol{c=c_\mathrm{current}}$.} The $\mathcal{N}=0,\, 1$ and $2$ Virasoro cases all satisfy $c=c_\mathrm{current}$, where $c_\mathrm{current}$ measures the Cardy growth of states supplied by the characters of the chiral algebra. In all such cases, the density of primary states with respect to the extended algebra should define a tempered measure and the techniques of this paper may be useful to classify the corresponding theories.

\paragraph{Other classifications.} Other, more surprising classifications have been proposed for $c>c_\mathrm{current}$. For example, the moduli space of $\mathcal{N}=(4,4)$ SCFTs with $c=6$ is expected to fall into two components: the moduli space of the $\mathbb{T}^4$ sigma model and of the $\mathrm{K3}$ sigma model~\cite{Nahm:1999ps}. However, in this case, $6=c>c_\mathrm{current}=3$ and thus we do not expect the techniques of this paper to be applicable. This is in accordance with the expectation that the spectrum of a generic $\mathrm{K3}$ sigma model is very far from lattice-like. Similarly, a catalog of all known CFTs possibly without extended chiral algebra with $c=2$ has been compiled in~\cite{Dulat:2000xj} and conjectured to be complete.

\paragraph{Extensions to $\boldsymbol{c>1}$.} It would of course be interesting to generalize these techniques to learn about the more challenging setting of $c>1$ CFTs. This would be technically much more difficult because (i) at $c>1$, there are also \emph{imaginary} values of $P,\bar P$ (including the identity operator), so one would have to allow $\rho$ to be a generalized distribution such as an ultradistribution~\cite{Gelfand:1968} or a hyperfunction as in~\cite[Appendix A]{Maxfield:2019hdt} and (ii) there is an exponential Cardy growth of states at large dimension. The corresponding analogs of crystalline measures have not been studied. This matches the fact that we expect a much richer and more complicated set of physical theories.

\section*{Acknowledgments}
We would like to thank Nathan Benjamin and Scott Collier for useful discussions and/or comments on the draft. LE and NL are supported by the European
Research Council (ERC) under the European Union’s Horizon 2020 research and
innovation programme (grant agreement No 101115511). We acknowledge use of
GPT 5.6 (OpenAI) and Opus 5, Fable 5 (Anthropic) to find various arguments presented in this paper, as an adversarial referee, and to assist in drawing figures. The authors take the full responsibility for the contents of the paper.

\bibliographystyle{JHEP}
\bibliography{bib}

\end{document}